\documentclass[%
 reprint,
 amsmath,amssymb,
 aps,
 rmp,
 showkeys
 include graphics
]{revtex4-2}

\usepackage{graphicx}
\usepackage{dcolumn}
\usepackage{bm}

\usepackage{aas_macros}
\usepackage{xcolor}
\usepackage{xspace}
\usepackage{totcount}
\usepackage{longtable}
\usepackage{comment}
\usepackage[normalem]{ulem}
\usepackage{tabularx}

\newcommand{\HIDECOMMENTS} 

\ifdefined\HIDECOMMENTS
    \newcommand{\note}[1]{\textcolor{blue}{}}
    \newcommand{\MC}[1]{\textcolor{red}{}}
    \newcommand{\RT}[1]{\textcolor{cyan}{}}
    \newcommand{\AG}[1]{\textcolor{black}{#1}} 
    \newcommand{\SG}[1]{\textcolor{orange}{}}
\else

\ifdefined\BOLDCOMMENTS
    \newcommand{\note}[1]{\textbf{#1}}
    \newcommand{\MC}[1]{\textbf{#1}}
    \newcommand{\RT}[1]{\textbf{#1}}
    \newcommand{\AG}[1]{\textbf{#1}} 
    \newcommand{\SG}[1]{\textbf{#1}}
\else
    \newcommand{\note}[1]{\textcolor{blue}{ #1}}
    \newcommand{\MC}[1]{\textcolor{red}{ #1}}
    \newcommand{\RT}[1]{\textcolor{cyan}{ #1}}
    \newcommand{\AG}[1]{\textcolor{magenta}{ #1}} 
    \newcommand{\SG}[1]{\textcolor{orange}{ #1}}
\fi

\fi

\newcommand{\asca}{\emph{ASCA}\xspace}
\newcommand{\ASCA}{\emph{ASCA}\xspace}

\newcommand{\xmm}{\emph{XMM-Newton}\xspace}
\newcommand{\XMM}{\emph{XMM-Newton}\xspace}

\newcommand{\chandra}{\emph{Chandra}\xspace}
\newcommand{\Chandra}{\emph{Chandra}\xspace}

\newcommand{\suzaku}{\emph{Suzaku}\xspace}
\newcommand{\Suzaku}{\emph{Suzaku}\xspace}

\newcommand\nustar{\emph{NuSTAR}\xspace}

\newcommand\swift{\emph{Swift}\xspace}
\newcommand\Swift{\emph{Swift}\xspace}
\newcommand\SWIFT{\emph{Swift}\xspace}

\newcommand\integral{\emph{INTEGRAL}\xspace}
\newcommand\INTEGRAL{\emph{INTEGRAL}\xspace}

\newcommand\ixpe{\emph{IXPE}\xspace}
\newcommand\IXPE{\emph{IXPE}\xspace}

\newcommand\Xrism{\emph{XRISM}\xspace}

\newcommand\Athena{\emph{NewAthena}\xspace}

\newcommand\erosita{\emph{eROSITA}\xspace}
\newcommand\eROSITA{\emph{eROSITA}\xspace}
\newcommand\eRosita{\emph{eROSITA}\xspace}

\newcommand\Uhuru{\emph{Uhuru}\xspace}
\newcommand\Arielfive{\emph{Ariel~5}\xspace}
\newcommand\HEAOone{\emph{HEAO~1}\xspace}
\newcommand\Einstein{\emph{Einstein}\xspace}

\newcommand\ROSAT{\emph{ROSAT}\xspace}
\newcommand\BeppoSAX{\emph{BeppoSAX}\xspace}

\newcommand\GINGA{\emph{Ginga}\xspace}

\newcommand\Granat{\emph{Granat}\xspace}

\newcommand{\Fermi}{\emph{Fermi}\xspace}
\newcommand{\fermi}{\emph{Fermi}\xspace} 
\newcommand{\FERMI}{\emph{Fermi}\xspace}
\newcommand{\hess}{H.E.S.S.\xspace}
\newcommand{\HESS}{H.E.S.S.\xspace}
\newcommand{\Magic}{MAGIC\xspace}
\newcommand{\MAGIC}{MAGIC\xspace}
\newcommand{\Veritas}{VERITAS\xspace}

\newcommand{\herschel}{\emph{Herschel}\xspace}
\newcommand{\spitzer}{\emph{Spitzer}\xspace}

\newcommand{\sgra}{Sgr~A$^\star$\xspace}

\newcommand{\hessGC}{HESS$~$J1745$-$290\xspace}
\newcommand{\GCPWN}{G359.95$-$0.04\xspace}

\newcommand{\HII}{H\,\textsc{ii}\xspace}

\newcommand{\Ka}{~K$\alpha$\xspace}
\newcommand{\Kb}{~K$\beta$\xspace}

\renewcommand{\deg}{$^\circ$\xspace} 
\newcommand{\arcmin}{$^\prime$\xspace}
\newcommand{\arcsec}{$^{\prime\prime}$\xspace}

\newcommand{\ergs}{erg~s$^{-1}$\xspace}

\newcommand{\Msol}{M$_\odot$\xspace}

\newcommand{\Sec}{\S~}
\newcommand{\Fig}{Fig.~}

\begin{document}

\preprint{APS/123-QED}

\title{High Energy Emission from the Galactic Center}

\author{Andrea Goldwurm}
 \affiliation{%
 Universit\'e Paris Cit\'e, CNRS, CEA, Astroparticule et Cosmologie, F-75013 Paris, France }
 \altaffiliation[Also at]{
 CEA Paris-Saclay, IRFU/D\'epartement d’Astrophysique,
 91191 Gif-sur-Yvette, France}
 \email{andrea.goldwurm@cea.fr}
 
\author{Maïca Clavel}
\affiliation{
Universit\'e Grenoble Alpes, CNRS, \\
IPAG, 38000 Grenoble, France
}%


\author{Stefano Gabici}
\author{Régis Terrier}
 \affiliation{%
 Universit\'e Paris Cit\'e, CNRS, Astroparticule et Cosmologie, F-75013 Paris, France
}%

\date{15 Feb 2026; Accepted 19 Dec 2025 by Rev.Mod.Phys.; Exp. Vol.98 Iss.2 Apr-Jun 2026; \href{https://doi.org/10.1103/9nww-fclb}{doi:10.1103/9nww-fclb}} 

%
\begin{abstract} 
The center of the Galaxy is a prominent source in X-rays and gamma-rays. 
The study of its high-energy (HE) emission is crucial in understanding the
physical phenomena taking place in this dense and extreme environment,
where the closest supermassive black hole (SMBH) to us, \sgra, is lurking 
nearly invisible, today, in most of the energy spectrum.
These phenomena are probably common to other galactic nuclei and 
may explain the feedback processes between nuclear regions and galaxies, 
so important for the overall evolution of the Universe.
The Galactic center HE emission is very complex and consists of both thermal 
and non thermal radiation produced by compact and extended sources, 
surrounded by more diffuse components.
All these objects and media are interacting with each other in the narrow 
and dense Central Molecular Zone (CMZ). 
Some of them also show relevant extensions towards the Galactic poles,
indicating energetic outflows that seem to link the center 
to the recently observed large Galactic polar structures. 
In spite of the fundamental advances obtained 
in the last twenty five years with the most sensitive
X-ray and gamma-ray observatories,
several questions remain open to investigations.
We review here the main observational results and the open issues on
the high-energy diagnostics of the Galactic nuclear activity,
focusing on processes that take place in the CMZ, and in particular discussing
the role of the present and past SMBH activities in powering 
this region and possibly the whole Galaxy.
\end{abstract}


\maketitle

\tableofcontents


\newpage
\section{The multi-wavelength view of the Galactic center} \label{sec:Intro}

\begin{figure*}
	\centering
 \includegraphics[width=\textwidth]{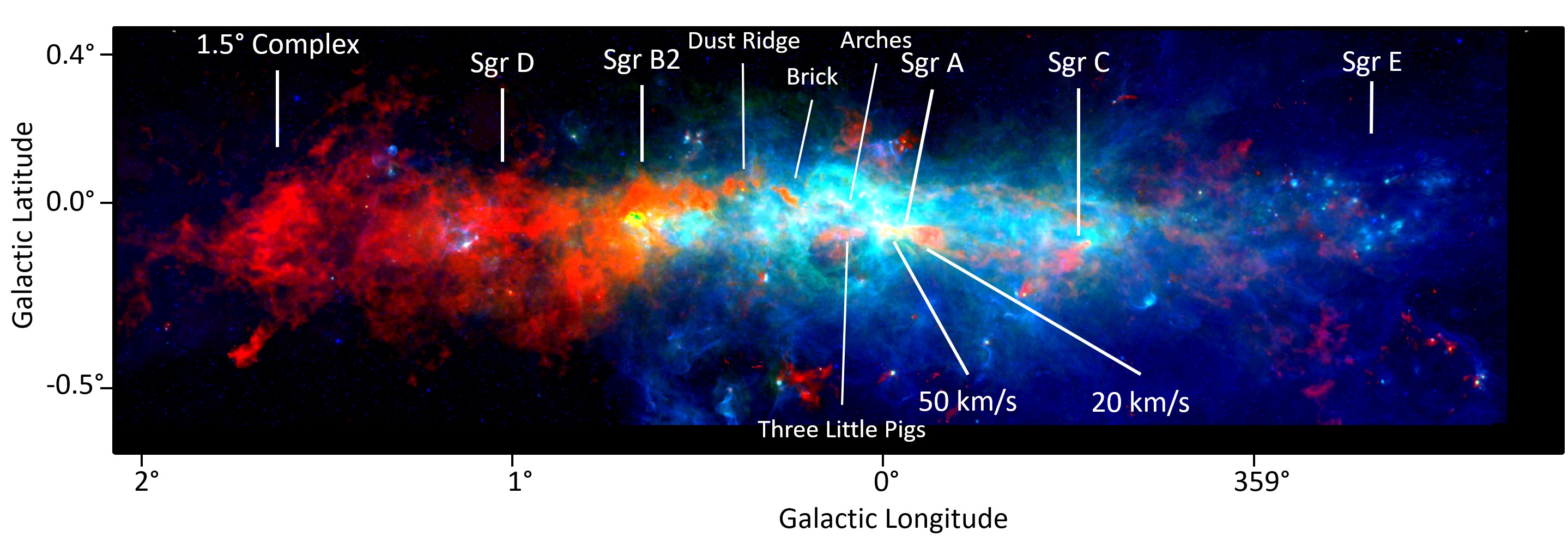}
	\caption{
	The molecular and thermal content of the Central Molecular Zone of the Galaxy
	shown by the superposition of sub-mm to mid-IR (MIR) images:
red and green show 350~$\mu$m and 70~$\mu$m emission (from \herschel) and blue shows 8~$\mu$m emission (from \spitzer).
The main molecular structures are indicated.
Note the longitude asymmetry between sub-mm-emitting cool diffuse dust associated with the dense gas (red), 
mainly distributed at positive longitudes,
and the MIR warm dust heated by recently formed high-mass stars (blue), at negative ones. 
Figure adapted from \citet[][]{batter25}.
}
	\label{fig:cmz-compIR} 
\end{figure*}

This review is dedicated to the high-energy emission from the Galactic Center and associated physical phenomena.
For Galactic Center (GC), we consider here the sky area 
of about 2$^\circ$ radius centered on the compact source Sagittarius~A$^*$ (\sgra), which is 
considered the electromagnetic counterpart of the Supermassive Black Hole (SMBH)
located at the dynamical center of the Galaxy.
For a distance of 8~kpc assumed throughout the paper (see \Sec I.A), 
this region  
corresponds to the central volume of the Galaxy of about 300~pc radius (\Fig\ref{fig:cmz-compIR}).
For high-energy (HE) emission we mean electromagnetic radiation of photon energy larger than 0.1 keV.
While the scope of the work is to present an observational view of the GC in this part of the spectrum, 
the related physical phenomena cannot be discussed without considering results obtained at other wavelengths. 
Therefore references to relevant measurements at low-energies, from radio to ultraviolet (UV)
frequencies, will be mentioned all along the paper 
and more specifically in this introduction, where we present an overview of the region.

The GC appears very similar to the majority of the galactic nuclei (GN) 
in the Universe, 
in particular the local ones. Only about 10$\%$ of the GN are currently
active and named for this as AGN, including different classes such
as  quasars, blazars and Seyfert galaxies.
The bulk of AGN is located far away, in the past, 
and the weakest ones, most commonly found in nearby galaxies,
are named low-luminosity AGN \citep{ho08}.
In this respect, even if presently rather quiet compared to other GN,
or even, probably, to other periods of the Galaxy history itself, 
overall the GC is the most dense, active and complex region of the Milky Way.
It allows us to study peculiar phenomena and physical processes in condition rarely found elsewhere, 
except in objects far away. 
Being about 100 times closer than the nearest comparable GN and 10$^5$ times closer than the closest quasar, 
the region has been largely observed and studied at all wavelengths.

The SMBH at the very center of the region represents the closest specimen of such an extreme object,
providing exquisite details on its properties, on its relation with its environment, and with the rest of the Galaxy.
The set of very high precision infrared results on \sgra environment (see \Sec\ref{subsec:SgrA}) 
makes it the best probe of the black hole solution of the Einstein's General Relativity (GR) theory,
and has been rewarded with the Nobel price of Physics in 2020 to Reinhard Genzel and Andrea Ghez \citep{genzel21, ghez21}, 
along with Roger Penrose for his theoretical contributions to the subject.

We now know that most, if not all GN contain SMBHs 
and that their interaction with the host bulges shapes and influences 
the evolution of the galaxies in the universe \citep{korho13}
even if the underlying processes and mechanisms are still not understood. 
Finally, as shown below, investigation of the GC features 
has provided evidence of past intense activity in the region, which can shed light on these feedback processes.

For all these reasons the GC is considered a real laboratory for the study of fundamental astrophysical
processes and particularly for processes at work in GN, the most violent of which generally produce
HE emission and can be studied through X and gamma-ray observations.

We will first provide a general 
description of the region, mainly based on non-HE measurements.
This chapter (\Sec\ref{sec:Intro}) provides a general overview on an extended region around the GC, 
going from the Galactic disk to the very center of the Milky 
Way. It is based on previous reviews, 
in particular those by \citet{morser96, melfal01, genzel10} 
and the more recent ones by \citet{brykra21, hensha23, morris22, ciumor26}, 
to which we refer the reader for a more complete description and bibliography on the GC.  
Only a few of the most recent publications are cited here.
In the following chapter (\Sec\ref{sec:SEAR}) we describe the historical development of the HE observations of the GC 
and the characteristics of the observatories used to survey the region in the last 25 years,
including current ones and mentioning those planned for the future. 
Then, the different components of the GC HE emission will be discussed: 
\sgra and the central 20-pc emission (\Sec\ref{sec:SGRA}), 
the X-ray point-source populations and discrete thermal and non-thermal features (\Sec\ref{sec:XRPO}), 
the different components of the X-ray diffuse emission including 
the role of Cosmic Rays (CR, \Sec\ref{sec:DXRE}) and finally the gamma-ray 
emission (\Sec\ref{sec:GRCR}).

In the conclusions (\Sec\ref{sec:Con}) we present a finding chart 
of the GC structures and sources discussed in the previous sections,
we recall the main established results 
and open questions and we list the perspectives for future observation programs
based on new HE instruments. Finally, we mention
the possible links between some of the HE GC components and their polar extensions, 
with the large Galactic polar structures of 
emission, the \fermi gamma-ray and \eROSITA X-ray bubbles,
that clearly indicate the presence of energetic outflows emanating 
from the central region of the Galaxy.

\subsection{The center of our Galaxy} \label{subsec:GC}
Our Galaxy is a barred spiral (Bbc - SBbc type), 
composed by a Galactic disk (GD), a central Galactic bulge (GB) and an extended halo.
The thin GD has a radius of 13--15 kpc and scale height of few hundred parsecs.
It contains a large fraction (0.6--0.7) of the stars of the Galaxy,
which are estimated to be about 1--$4\times10^{11}$ in total, 
for a mass of about $5\times10^{10}\rm~M_\odot$. 
It also includes most of the Galactic interstellar matter (ISM),
estimated at 10--15$\%$ of the star mass, 
in the form of diffuse gas and dust.
Stars, particularly the young ones, and ISM show important concentrations along four main spiral arms 
surrounding the GD center. 
The Sun is in the close periphery of the disk, 
in the inner edge of the Orion-Cygnus (or Local) Arm, 
moving with a velocity of 220~km~s$^{-1}$ along a circular Galactic orbit with period of 210 Myr, 
at an estimated distance of 8~kpc from the Galaxy dynamical center\footnote{
Current Sun-GC distance measures are in the range 7.9--8.2~kpc \citep[][]{do19a,gravit19}.
We adopt here, for simplicity, the value of 8~kpc, which defines the scaling relation 
between measured angles in the sky and linear sizes at the GC throughout the paper:
$1''\approx0.04\rm~pc$, $1'\approx2.33\rm~pc$, $1^\circ\approx140\rm~pc$.}.
From Earth, the Galaxy is therefore seen edge-on and from inside the rotating GD,
whose glowing in the sky produces the bright lane appearance of 
the Milky Way (MW).

Towards the disk center, a central peanut-shaped Bulge, mainly composed of old stars, 
sticks out from the Galactic plane, 
in the projected central 40$^\circ$--60$^\circ$ of the MW, 
showing slight asymmetry, with a more prominent side towards positive longitudes.
This view is actually due to a central stellar bar, composed by a fat ellipsoidal structure, 
stretching 2-5~kpc out from the center on both sides 
\citep[][see  also  their  Fig.~1  for a comprehensive view of the MW]{brykra21}
seen from the Sun with a tilt of its major axis of $\approx 20$--$30^\circ$ from the 
line of sight (LoS) towards positive longitudes.
The total bulge and bar stellar mass is estimated to be $2\times10^{10}\rm~M_\odot$.
Even if these estimations and interpretations 
may still evolve with future surveys and refined modellings, 
it is clear that the bar is the main component of the observed GB and its gravitational potential 
dominates the stellar and ISM dynamics in the central regions.
In fact its presence allows relevant inflow of matter from the GD to the center 
which otherwise will be hampered by the large angular momenta.
An extended spherical halo containing very old stellar populations, globular clusters
and dark matter (DM) envelops the disk and bulge.
Including the halo, the total mass of the MW
was recently revised by \textit{Gaia} down to $2\times10^{11}\rm~M_\odot$ 
with a ratio of DM to baryonic mass of 3 \citep{Jiao23}.

Since, from the Sun, the GC is seen through the whole GD, the dust of the ISM absorbs the radiation with
an extinction of $\approx$~30 mag in A$_V$ which totally blocks visible and UV light. 
Thus, the GC region can only be observed from radio to infrared (IR) frequencies and again at HE. 
However the peculiar location of the GC was in-fact remarked already in the 1920s, 
well before radio, IR or HE astronomy developed, based on stellar distributions and motions.
It remained not well positioned until the detection of the radio source Sgr~A 
and its association in the 1950s--1960s with the GC, 
which also led to the redefinition of the Galactic coordinates,
with the longitude $l=0$ and latitude $b=0$ now at $\approx 4$\arcmin from \sgra.

\subsection{The Galactic bar potential and matter inflow} \label{subsec:Bar}

The MW spiral structure is interrupted inwards
by the bulge, at a galactocentric radius $R$ of about 3~kpc (20$^\circ$).

The ISM distribution shows a density increase from the outskirts of the GD 
toward the center and reaches a maximum at $R\approx 4$~kpc,
forming the so called 4-kpc molecular ring. 
For smaller galactocentric distances the gas density decreases rapidly 
to low values creating the region of ISM avoidance 
(3--1.5~kpc) where the stellar bar potential does not allow for stable orbits.
The tri-axial bar potential indeed forces matter to follow the so-called x1 and x2 closed
orbits
which are related to the specific resonance points of the underlying potential (see also \Fig\ref{fig:cmz-mcdist-models}).
The x1 are nested, very elliptical orbits around the bar major axis 
while the x2 are more oval trajectories aligned to the bar minor axis 
perpendicular to and within the x1 orbits.
The x2 orbit sizes are somehow linked to the potential orbital resonances
and even though what determines their actual dimensions is still under intense study,   
they broadly seem to correspond to the rings of the central molecular zone.
The accepted picture is that 
ISM of the GD inner edge is rapidly set into 
x1 orbits closer and closer to the bar axis, 
get shocked at the cusps of the inner orbits (where the trajectories intersect) and 
is funnelled inwards along the very inner, nearly radial, x1 orbits, 
forming the so-called bar (or dust) lanes, reaching the outer x2 orbits.
Shocks 
lead to the setting of material on the x2 orbits, allowing  
the inflow of matter from the disk to reach this central region (\Fig\ref{fig:cmz-mcdist-models}, left).
Data and simulations show that, considering loss processes, 
about 0.2--1.4~M$_\odot$~yr$^{-1}$ 
of matter is conveyed in this way
from the GD edge to the GC through the bar lanes 
\citep[][]{brykra21,hensha23,su24}.  
This overall picture accounts for most of the non-circular gas kinematic features 
observed in the radial velocity -- longitude (V$_R$ -- $l$) 
diagrams of molecular lines like those of the CO 
within 20$^\circ$ from the center, 
it does not require the presence of expanding rings to explain these non-circular motions, 
and it is now an accepted model, even if various items are still object of intense research.

The region delimited by the 
x2 orbits ($R <$ 300~pc) 
contains the Nuclear Bulge (NB), 
made of the Nuclear Stellar Disc (NSD) 
and its inner Nuclear Stellar Cluster (NSC),
and the Central Molecular Zone (CMZ) of ISM.

\subsection{The nuclear bulge: the CMZ and the NSD} \label{subsec:CMZ}

\subsubsection{Molecular material and diffuse thermal sources} \label{subsubsec:MC}

The Central Molecular Zone  
\citep{morser96, brykra21, hensha23} 
is the 4$^\circ$~$\times$~2$^\circ$ 
narrow strip of the sky, corresponding to the central 600~pc~$\times$~300~pc region 
which contains the large reservoir of ISM in the form of neutral, 
mostly molecular, and ionized gas, along with the associated dust,
that can be observed at different wavelengths from radio to MIR
(\Fig\ref{fig:cmz-compIR}).
In the CMZ clouds 
the average density ($\approx 10^{3-4}$~cm$^{-3}$) is 
up to two orders of magnitude 
above the one observed in molecular clouds (MC) of the GD \citep{brykra21}.
Temperatures are also higher ($\gtrsim 40$~K compared with the typical $< 20$ K in the GD MC), 
as well as the turbulence and the magnetic pressure. 
The total mass of molecular gas amounts to $\approx2$--$6\times10^{7}\rm~M_\odot$ \citep{hensha23,batter25}, 
i.e.\ 3--10$\%$ of the total molecular content of the Galaxy,
and appears organized in structures distributed around the nucleus.
An inner, low velocity ($< 100\rm~km~s^{-1}$), 
component of these structures is located 
within 100~pc from the center, is composed by the denser clouds and represents the
dominant part of the CMZ mass, the so-called 100-pc molecular ring. 
As shown in \Fig\ref{fig:cmz-compIR}
the overall matter distribution appears 
asymmetric with a large fraction (from three quarters to two thirds)
of gas material at positive Galactic longitudes while the indicators of star formations, specifically 
the tracers of massive young stars and compact \HII regions, appear concentrated to negative longitudes.

\begin{figure}[ht]
	\centering
  \includegraphics[width=1.0\linewidth]{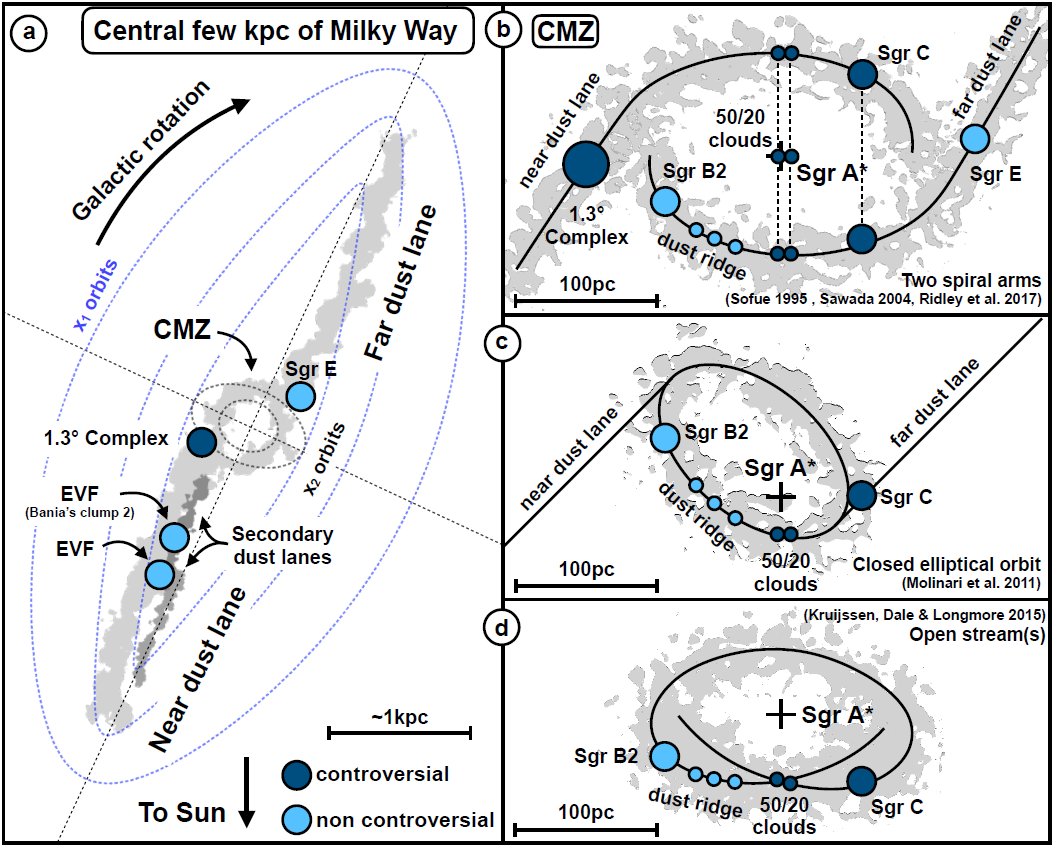}
	\caption{Top view sketches of the matter inflow along the x1 and x2 orbits of the stellar bar potential (left) and
of the different models of molecular matter distribution in the CMZ (right),
with controversial locations of individual MCs
\citep[from][]{hensha23}.
	}
	\label{fig:cmz-mcdist-models}
\end{figure}

Given our vantage point and the lack of LoS distance indicators,
the actual 3D spatial distribution and dynamics of the gas within the CMZ are however unknown.
Several models, supported by various observations, simulations and theoretical studies, 
have been proposed (\Fig\ref{fig:cmz-mcdist-models}, right): from spiral arms around the nucleus \citep{sofue95}
to a twisted, elongated closed ring not centered on \sgra\ \citep{Molina11} 
and to gas streams following open orbits \citep{kruijs15}.  
Large observational efforts
\cite{yan17}, including recent comprehensive work combining IR data from \herschel and \spitzer and 3D modelling 
\cite{batter25} and the ALMA CMZ Exploratory Survey program \cite{sofue25},
have not allowed yet to identify the most realistic of these pictures \citep[e.g.,][]{lipman25, walker25}.

Within the large
structures the gas is highly non-uniform and the densest concentrations 
form the Giant Molecular Clouds (GMC), with densities that can reach 
$10^{5-6}\rm~cm^{-3}$, temperatures 70--100~K, large inner velocities and enhanced turbulence, 
and which are threaded by strong ($\approx$ mG) magnetic fields.
The main GMCs of the CMZ, for density and volume, are the complexes corresponding to the bright 
extended radio sources known as Sgr~A, Sgr~B2, Sgr~C, Sgr~D, and Sgr~E (\Fig\ref{fig:cmz-compIR},  \ref{fig:MeerKat-CMZ-index}), 
even if some are actually composed of several clumps, close in projection but  
possibly at different LoS distances because they display different V$_R$. 

The most massive cloud, Sgr~B2, contains $\approx0.5$--$1\times10^{7}\rm~M_\odot$, 
roughly 10$\%$ of the CMZ gas, and is one of the most massive 
star-forming clouds in the entire Galaxy.
Located at $l\approx0.66^\circ$ (100~pc in projection) from \sgra, in the northern part of the large Sgr~B complex which also includes the bright \HII region Sgr~B1 in the south, 
Sgr B2 is a luminous radio and IR source.
It is composed of an extended envelope (n$_{H_{2}} = 10^{4}$~cm$^{-3}$, 20-pc radius)
and an inner high density cloud (n$_{H_{2}} = 10^{5-6}$~cm$^{-3}$, 5-pc radius), 
which contains sites of massive star formation, with ultra-compact \HII regions, proto-stars and molecular masers.
Using masers, the LoS position of Sgr~B2 was estimated to be about 130~pc in front of the sky plane defined by \sgra
but this estimate relies on kinematic assumptions.
Sgr~C is instead located on the Galactic west side of the GC (\Fig\ref{fig:cmz-compIR}) also at about 100~pc in projection
from \sgra and, even if considerably less massive ($\approx 5\times10^{5}$~M$_\odot$) than Sgr~B2,
it is a site of intense star formation. 
Extended over 40~pc and composed by several clumps, possibly not all physically connected,
its structure is complex, presenting other superposed diffuse sources: 
a SNR that seems interacting with the molecular material, 
an \HII region  
and structures of hot gas extending north, the Sgr~C Chimney. 
Further out at negative longitudes the GMC Sgr~E extends from $l\approx-1^\circ$ to $-1.4^\circ$ 
and hosts a substantial number of compact \HII regions and other star formation indicators.
Its large negative radial velocities indicate that it is approaching and is probably located 
at the interaction point of the far dust lane with the CMZ rings of the external x2 orbits. 
In a nearly symmetric position to the center, at $l\approx1.2^\circ$, 
is located another GMC, Sgr~D, which also presents a complex structure including 
a very bright \HII region, 
a radio-bright shell SNR more to the south, 
and another SNR detected in X-rays,
all these superposed to extended radio line emission of the cloud molecular material
that appears receding at 50--70 km~s$^{-1}$.
Further out there is the large l=1.5$^\circ$-complex (or 1.3$^\circ$-complex) 
showing extreme velocity features in the V$_R$ -- $l$ diagram, 
which is probably out of the 100-pc ring and linked to either the x2 orbit, the bar lane or their intersection.
The central Sgr~A radio complex is a highly composite source, including the GMC
M--0.02--0.07 (also known as the 50~km~s$^{-1}$ cloud) and M--0.13--0.08  (the 20~km~s$^{-1}$ cloud),
located east and south, respectively, of the Sgr~A core,
which also contains a massive molecular nuclear disk (see \Sec \ref{subsubsec:SgrAcomplex}).

In addition to the massive complexes, other smaller features, filaments and medium-size clumps of molecular
gas are present in the CMZ, in particular between Sgr~A and Sgr~B \citep[][\Fig\ref{fig:cmz-compIR} and \ref{fig:xmm-echoes}]{batter25}. 
Within 4\arcmin and 14\arcmin 
on the Galactic East from \sgra, roughly between the 50~km~s$^{-1}$ cloud and the Radio Arc,
lies a set of clouds,
which we refer to as the Sgr~A molecular complex (to avoid confusion with Sgr~A). 
This region includes important features that are known under different names in the literature. 
It is the case of G0.068--0.075 (the Stone) and G0.106--0.082 (the Sticks), 
two of the MC set known as the Three Little Pigs,
that we will refer to as the Bridge in this review, as well as of the MC G0.11--0.11, 
just south of the Bridge.
The Sgr~A molecular complex has been studied in great detail due to its bright variable non-thermal X-ray emission 
(\Sec\ref{sec:DXRE}). 
Another set, not visible in X-rays, includes the very dense 
IR-dark Brick cloud  
and the clumps of the Dust Ridge between the Brick and Sgr~B2, which are thought to trace part of 
the 100~pc ring. 
As already remarked, if the MC 2D distribution on the sky and radial velocity are well known, 
their 3D distribution and actual motion are not (\Fig\ref{fig:cmz-mcdist-models}).

A diffuse, warm (T $\approx$ 200~K) molecular gas of low density ($\approx$ 50~cm$^{-3}$) 
permeating the CMZ with high 
volume filling factor, was put in evidence at the end of the 1990s
by near-infrared (NIR) observations of the H$_3^+$ lines in the GC
(see \Sec\ref{subsec:CRNT}). 
This component appears not-rotating but rather expanding radially
and represents a low fraction of the CMZ molecular content.

\subsubsection{Non-thermal sources, magnetic field and outflows}
\begin{figure*}
	\centering
 \includegraphics[width= \textwidth]{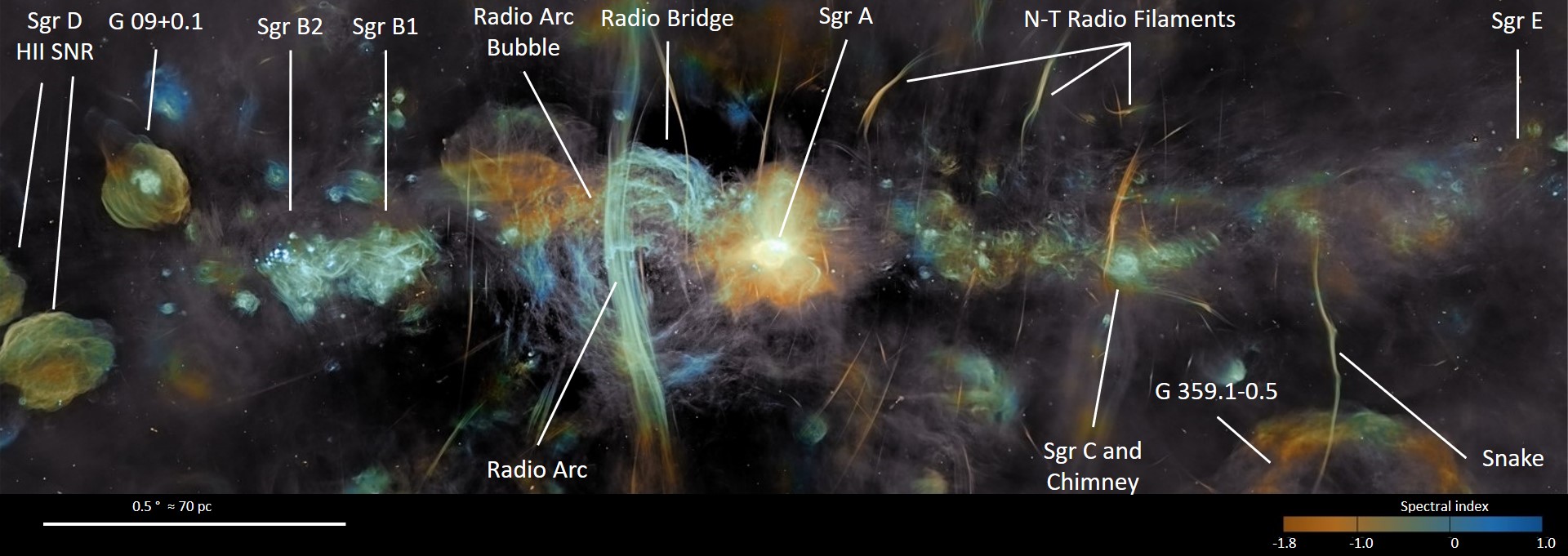}
	\caption{MeerKAT radio composite image of the GC in Galactic coordinates, 
	showing both the thermal and non-thermal radio sources of the CMZ, 
	with the color code giving the power-law spectral index. Figure adapted from \citet{heywoo22}\AG{, original image credit I. Heywood, SARAO; color processing J.C. Munoz-Mateos, ESO}.
	}
	\label{fig:MeerKat-CMZ-index}
\end{figure*}

In addition to neutral and ionized gas, traced by the thermal features, the CMZ contains a remarkable number of non-thermal 
radio sources as shown by the extraordinary MeerKAT composite GC images \citep[][\Fig\ref{fig:MeerKat-CMZ-index}]{heywoo22}.
These are either shell supernova remnants (SNRs), pulsar wind nebulae (PWNe)
or radio non-thermal filaments (NTFs). 
They demonstrate the importance of star formation (SF) in the region  
and the presence of magnetic fields \citep{morris15} and particle acceleration processes.
The most spectacular feature is the Radio Arc located at about 30--50~pc on the Galactic East 
from \sgra and extending over about 50--100~pc nearly perpendicularly to the Galactic plane 
and over about 10~pc in latitude. 
It is composed by tens of nearly aligned bright arched and narrow filaments
with non-thermal spectra which are produced by synchrotron emission of relativistic
electrons in a poloidal magnetic field.
Several hundreds other NTF have now been identified in the GC, in particular with MeerKAT \citep{yusefz22}.
The brightest and longest of them 
are also generally oriented towards the Galactic poles,
demonstrating that, out of the GMC and particularly above the plane, the GC magnetic field lines are overall poloidal
\citep[but see][for the orientation of the short filaments]{yusefz23}.

Within the intermediate-density clouds of the CMZ, instead, field lines appear, 
from IR polarization measurements of warm dust,
largely toroidal and seem to remain so 
also in the inter-cloud regions close to the plane, which prompted a modelling where 
the overall GC vertical magnetic field is dragged into a planar configuration 
by the shear of the 100-pc ring gas circular motion \citep{morris15}.
This scenario is also supported by the recent results from the large CMZ FIR survey provided by SOFIA \citep{pare25},
that confirm the presence of a pervasive vertical magnetic field also outside the NTFs and fields lines
aligned with the density structures in the dense MCs.

The field strength was estimated at a few mGauss for the Radio Arc 
and in several other locations, but lower values (100--400~$\mu$G) have been derived recently using MeerKAT data 
of detected NTFs \citep{yusefz22}.
The overall field strength 
is still controversial with estimates that range from 10--50~$\mu$G to few mG \citep{crocke10, morris15},
but, despite the lack of consensus, it is clear that the GC magnetic field 
is stronger than in the GD and it plays an important role in the dynamics and physical processes in the CMZ.

Other large-scale features and bubbles in radio/IR, extending above and below the Galactic plane, 
probably produced by large outflows and outbursts, demonstrate the extraordinary activity of the region
(see \Sec\ref{subsec:DXSC}).

\subsubsection{Massive star clusters and the nuclear stellar disc}\label{subsubsec:SFR}
The large quantity of gas in the CMZ implies a large star formation rate (SFR). 
The GC SFR, estimated roughly constant and equal to $0.07^{+0.08}_{-0.02}\rm~M_\odot~yr^{-1}$ over the last 5~Myr \citep{hensha23},
is much smaller than that in the GD, in terms of per unit of mass, 
but is very large if one considers the very small volume within which it has been estimated.
The low SFR per unit of mass is probably connected to the large 
turbulence and other dynamical phenomena 
of the region, or even to the large CR pressure and ionization.
Indeed, as mentioned above, some of the GMCs host on-going intense star formation sites (Sgr~B2, Sgr~A, Sgr~C, also
associated to young \HII regions) while others do not (Sgr~B1, the Brick, 1.5$^\circ$-complex).

The GC contains three prominent young star clusters, amongst the most massive of the MW: 
the Arches, the Quintuplet and the Young Nuclear Cluster surrounding the GC nucleus.
The Arches is located 26~pc East of \sgra in projection,
with a stellar mass of $\approx$~10$^{4}$~M$_\odot$ concentrated within a radius of 1~pc. It is mostly composed of massive Wolf-Rayet (WR) 
and O stars 
with ages between 2.5 and 3.5~Myr,
which excite 
the gas thermal emission observed in the nearby radio-bright Arched filaments.
The Quintuplet cluster is at 30~pc in projection from \sgra, east of the Arches 
and is surrounded by the nearby thermal Sickle filament which it likely powers.
Of similar mass and dimension than the Arches, it appears slightly older 
(4--5~Myr), making it unclear whether the two clusters were born from the same formation event or not.
However, they probably evolved from similar scenarios, caused by gas over-density regions
at the intersections of x1 and x2 orbits. 
The overall stellar content corresponding to the CMZ region is the so-called Nuclear Stellar Disk (NSD), 
a high-density stellar structure clearly connected to, but different from, 
the central part of the Galactic bulge/bar, and which forms an axis-symmetric rotating disk, 
of radius $\approx230$~pc and scale height $\approx45$~pc, 
made of old (10~Gyr) and middle-aged (1~Gyr) stars probably born
in the dense environment of the CMZ. 
Stars rotate 
in the same direction as the MW, 
their density increases towards the center, making a total mass of the NSD of 10$^9$~M$_\odot$,
which dominates the gravitational potential for $R \approx30$--300 pc.

\subsection{The Nucleus: the Nuclear Star Cluster and Sgr~A}
\subsubsection{The stellar content and the Young Nuclear Cluster}\label{subsubsec:YNC}

Within the inner $R$ = 30~pc (12.5$'$) and down to about 1--3~pc from where the SMBH dominates,
the gravitational potential is driven by another stellar component, %
the Nuclear Stellar Cluster (NSC).
Discovered at the end of the 1960s 
as a central prominent emission in IR, the NSC is a very dense and massive star cluster, 
quite similar to those observed in the centers of other spiral galaxies.
The NSC is more spherical than the NSD, 
it is very dense and compact,
with a huge central stellar density of 10$^{6-7}$ M$_\odot$~pc$^{-3}$ in the core, 
and an effective radius of 5~pc. 
It contains $\approx 3\times10^{7}\rm~M_\odot$ of stars mainly cold, old and of middle age ($>$ 5~Gyr).
Stellar density increases towards the center
and recent studies have shown that the derived 3D density profile 
within the influence radius of the SMBH nearly matches the predicted 
distribution expected for a relaxed star cluster with a central BH.
This cusp has been searched for several years and seems now a well established feature of the NSC.

In addition to the NSC old stars, the central parsec ($R <$ 0.5~pc, 12.5\arcsec) contains 
a remarkable cluster of about 100--200 luminous, hot, early-type stars of ages 2--8~Myr,
of total mass 10$^{4}$~M$_\odot$:
the Young Nuclear Cluster (YNC).
Mostly composed by post-main sequence blue O/B giant, super-giant, main sequence and 
WR stars,
the majority of these stars have orbital motion opposite to the MW and to the underlying NSC stars. 
A significant fraction of them (20--30\%)
shows a pattern of clockwise rotation 
and appears to lay in a rotating disk, between 1$''$ and 5$''$ (0.04--0.2~pc) from \sgra. 
The most striking concentrations of such stars were identified initially as individual IR sources:
IRS16, IRS13 and IRS7. The YNC hot stars produce $2\times10^7\rm~L_\odot$ of UV radiation 
which ionizes the gas of the central pc creating the central thermal source, Sgr~A West, and heating the dust.
The $\approx$ 30 massive post-main-sequence WR or helium emission-line stars of the cluster 
are also powerful emitters of stellar winds, 
with typical mass losses of 10$^{-5}$--10$^{-4}$ M$_\odot$~yr$^{-1}$ and terminal velocities of 500--2000~km~s$^{-1}$.
Their collective momentum and radiation pressure clear the gas out of the region, likely creating the central cavity.

\subsubsection{The Sagittarius A complex and the central cavity}\label{subsubsec:SgrAcomplex}

\begin{figure}[ht]
	\centering
 \includegraphics[width=0.35 \textwidth]{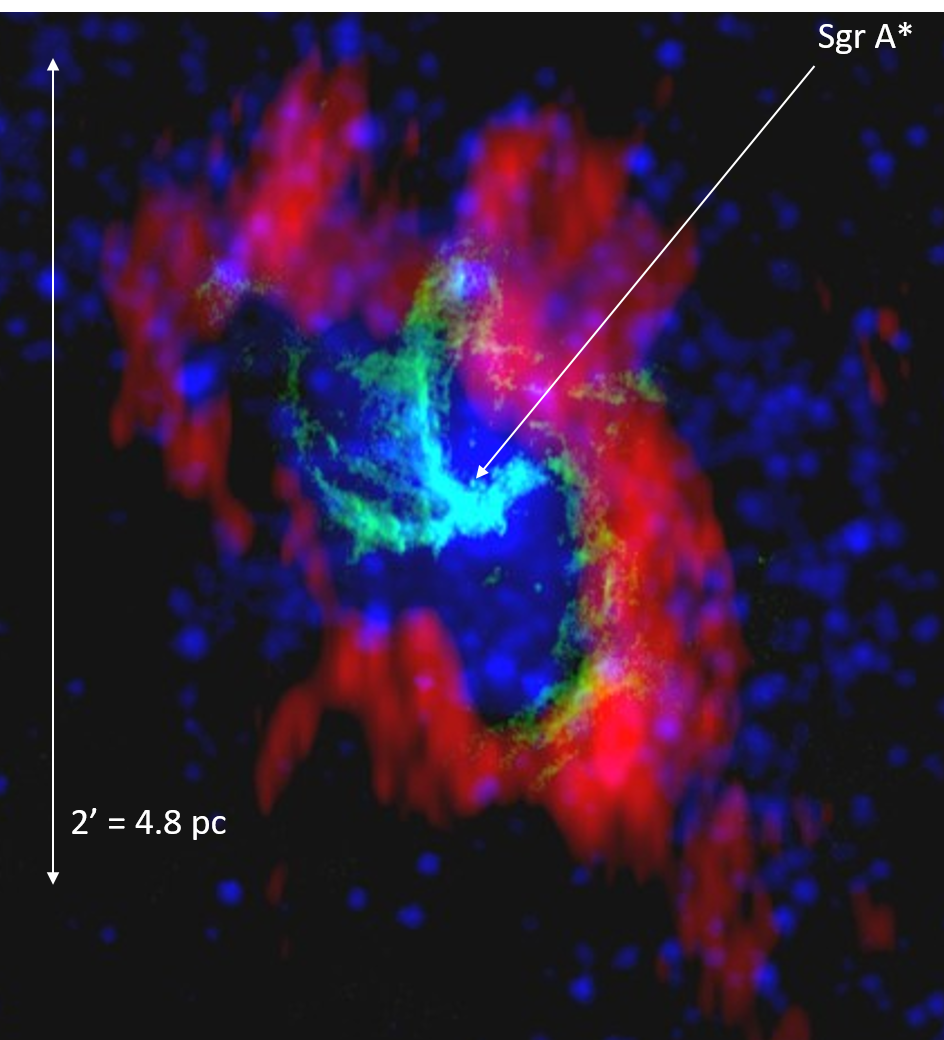}
	\caption{
	The molecular CND and the Sgr~A West mini-spiral of ionized gas
 shown by a composite image of the 3.4~mm HCN line emission (red, BIMA) and 
 the 3.6~cm radio continuum (green, VLA) from warm ionized gas, along with the star's IR emission (blue, \spitzer). 
 The image is in equatorial projection centered on \sgra (indicated by a white arrow) 
 with North to the top, East to the left and scale given by the double-sided arrow. From NRAO archives ({https://www.nrao.edu/archives/items/show/33428}).
 }
	\label{fig:cnd}
\end{figure}

The ISM component within the central $R$ = 15~pc (6.3$'$), 
spatially superposed upon, and interacting with, the stellar content described above, 
is the Sgr~A radio and molecular complex.
\citet[][recently updated by \citealp{brykra21}]{ferrie12} reviewed the wealth of published data on Sgr~A,
and provided a
detailed description of its structure and properties, 
with a global picture
presently largely accepted, apart from the location of the two GMC of the complex.
In this picture, the external belt of GMC 
that, in projection, surrounds
Sgr~A core on the East (50 km~s$^{-1}$ cloud) and South (20 km~s$^{-1}$ cloud, see also \Fig\ref{fig:cmz-compIR})
are part of the complex, the first one slightly behind and the other in front of \sgra. 
However, in other CMZ matter distribution models these two clouds are sometimes far away from \sgra
and part of the CMZ 100-pc ring (\Fig\ref{fig:cmz-mcdist-models}).
Their masses are both of $2\times10^{5}\rm~M_\odot$, with densities of $2\times10^{4}\rm~cm^{-3}$
and temperatures around 60~K.
Other molecular streams and ridges are seen to extend from the GMCs toward the
inner region connecting them to, and probably feeding, the Circum-Nuclear Disk (CND) or Ring,
an asymmetrical, clumpy, annular structure of very dense 
molecular gas and dust within $\approx1.2$--5~pc from the center, with a total mass
about one order of magnitude lower than those of the close-by GMC, from recent estimates,
and rotating rapidly, at $\approx$~100 km~s$^{-1}$, around \sgra (\Fig\ref{fig:cnd}). 

Superposed, in projection, on part of the molecular structures is Sgr~A East,
a distinct non-thermal radio source composed of a halo of diffuse emission with a triangular shape 
and an inner oval shell ($3' \times 4'$, or 7~pc $\times$ 9 pc) 
centered about 50$''$ (2~pc) east of \sgra (\Fig\ref{fig:SgrAEast}, left). 
Sgr~A East is also bright in X-rays
and is considered a mixed-morphology SNR where the observed radio synchrotron shell of the external shock 
sweeps out the inner cavity which contains hot ionized gas from both stellar ejecta and shocked ISM.

Within the CND ($R$ $<$ 1.5~pc) there is a central cavity, of much lower neutral gas density, 
carved by the powerful stellar winds of the YNC, 
\citep[but see][for the effect of stellar winds on the CND]{blank16} 
and which hosts the \HII region Sgr~A West, a thermal nebula ionized by the UV radiation
of the young stars, especially those of the IRS16 concentration.
The dense part of the ionized gas takes the form of a mini-spiral
composed of three arms and a bar (\Fig\ref{fig:cnd}), also rotating around the SMBH.
Part of the arms, particularly the western one, join the inner side of the CND ring,
and show similar circular movement, 
while the east and north arms show inward motion towards the bar and the center of the cavity.  
The contrasting processes of the SN event and the SMBH mass accretion
certainly play a role in shaping the inner 20~pc of the MW \citep{ferrie12},
and an overall picture on how matter inflow takes place has now emerged from the data \citep{brykra21}. 
The inflow 
from the GMC of the CMZ into the CND seems mediated by the molecular streamers
and then, through the mini-spiral and central cavity, the material is conveyed to the very center of the NSC, 
giving rise, in spite of the tidal stresses of the SMBH, to episodic in-situ formation of stars,
explaining the presence of the YNC in the central parsec.

\subsection{The central 0.1~pc and Sagittarius A$^\star$} \label{subsec:SgrA}

The central parsec of the Galactic Nucleus is dynamically fully dominated by the SMBH since 
the NSC enclosed mass is lower than the SMBH mass already at $R$ = 3~pc (1.25$'$). 
Deep inside the YNC inner disk all objects move at extreme velocities 
apart from one single point-like source that appears totally steady: \sgra.

\subsubsection{The S-cluster and the G objects}
At distances $R$~$< 1''$ (0.04~pc), there is another different group of young stars: 
the central, or S, or \sgra cluster,  composed of at least 50 main sequence dwarf B stars, 
less massive (8--20~M$_\odot$) and not as young ($< 15$~Myr) as the O/WR ones seen further out.
The S stars show elliptical orbits around the same focus but distributed isotropically, therefore 
indicating a different formation process than the YNC, without the intervention of a disk. 
These few tens of stars, 
have allowed the most precise estimations of the parameters of the SMBH. 
Indeed, over the last 30 years, NIR high-resolution observations conducted at the NTT, VLTI 
and Keck telescopes 
with an increasingly sophisticated instrumentation,
culminated with the upgraded Keck/NIRC2  
and VLTI/GRAVITY instruments, 
have led to extremely precise measurements 
of their orbit parameters \citep{genzel10, do19a, gravit22}. 
In particular, one of the brightest stars of the S-cluster, S2 or S0-2, 
could be followed twice over its entire 16-yr orbit and particularly at its periapse passages, 
in 2002 and 2018, when it transited with a velocity excursion of 8000~km~s$^{-1}$ at only 120~AU from the SMBH.
The exceptional data of the S2 proper motion have not only led to the determination of the mass and distance of the SMBH 
but also revealed clear GR effects.

Such NIR observations in the central 0.1~pc have also revealed the presence of several objects 
of gas/dust appearance showing rapid movement along elliptic Keplerian orbits around the SMBH. 
The first and most famous one is the G2 object, detected in 2012 
\citep{gilles12},
that was first interpreted as a $10^{-3}\rm~M_\odot$ compact ionized gas clump
and which passed at periastron in 2013, 
showing clear tidal effects on its structure, even if not a total disruption \citep{gilles13}.
After this first discovery several other G objects have been identified and are now regularly monitored 
\citep{ciurlo20}. 
The presence of such a population of dusty objects in the same region of the S-cluster has further challenged
the pure dust interpretation, 
as the hypothesis of young low-mass stars with distended gaseous-dusty envelopes 
or even binary systems seem also compatible with the data.

\subsubsection{Sagittarius~A$^\star$, the Galactic black hole}\label{subsubsec:SgrAs}

At the very center of Sgr~A West is located the remarkable source \sgra
\citep[\Fig\ref{fig:cnd}, see e.g.][and references therein]{melfal01, morris22}
detected from about 0.5~m wavelength to hard X-rays, apart from the 1~$\mu$m to 1~nm range affected by extinction.
Discovered 50 years ago \citep{balbro74},
as a distinct point-like component of Sgr~A, this bright, compact, variable and non-thermal
radio source was quickly associated to the putative SMBH of the Galaxy predicted by \citet{lynree71}
just few years before. 
The inverted non-thermal radio spectrum did not quite resemble the spectra of quasars
and the luminosity was rather low, but it pointed to a compact object of large brightness without prominent counterpart 
at the very center of the MW.

Proper motion measurements of the source over 18~yr with VLBA found no intrinsic motion down to $<1\rm~km~s^{-1}$ \citep{reibru20}.
In the inner region where everything moves very fast, this allows, under reasonable assumptions, 
to set severe lower limits on its mass. 
However the best parameters of the source are given by the NIR measurements of the motion of the S-cluster stars.
In fact the location of \sgra is fully compatible, within 1~mas (8~AU),
with the location of the focus of their orbits and the source is identified with the object that drives 
the star motions.
The estimated enclosed mass, within the S2 periapse distance, 
is of $(4.1\pm0.1)\times 10^{6}\rm~M_\odot$ at a distance of $8.1\pm0.1$~kpc,
where, following \citet{morris22}, these values (and errors) are actually the average (and range) between the Keck and VLTI results  that differs by 5 and 2.5\% in mass and 
distance, respectively.
Such a large mass compressed in such a small volume implies a minimum density of $5\times10^{15}\rm~M_\odot~pc^{-3}$ and 
cannot be stable for long,
it must collapse into a SMBH \citep[see][for discussion on alternative models]{genzel10}.
The gravitational red-shift \citep{gravit18a, do19a}
and the Schwarzschild precession of the S2 orbit \citep{gravit20a}
have already been detected in particular by GRAVITY, 
while higher-order GR effects, 
like the Lense-Thirring one which may provide information on the BH spin,
need the detection of closer (and fainter) stars with future instrumentation.

Such a SMBH 
has a Schwarzschild radius R$_S = 1.2\times10^{12}$~cm, an Eddington Luminosity of L$_E = 5\times10^{44}\rm~erg~s^{-1}$, 
and it is expected to accrete any stellar or gaseous object that would venture too close to its horizon, such as 
a fraction of the stellar-wind gas of the YNC. 
The accretion process in \sgra should produce a large amount of radiation while the source is clearly under-luminous
all over the electromagnetic spectrum.
From observations and detailed simulations 
the accretion rate from the stellar wind is estimated at 10$^{-6}$--10$^{-5}$~M$_\odot$~yr$^{-1}$ 
\citep[][]{quatae02, bagano03}
at a Bondi radius of $\approx0.12$--$0.2$~pc (3\arcsec--5\arcsec) 
giving a significant accretion power 
of L$_A \approx 3 \times 10^{39-40}\rm~erg~s^{-1}$ (assuming an efficiency of 10\%), 
with important variations induced by wind clumps, turbulence 
and shocks \citep[see][for the latest studies]{calder20, calder25}.
The radio and sub-mm \sgra fluxes (of about 1~Jy) translate into a total luminosity of less than $10^{36}\rm~erg~s^{-1}$,
some $10^{-9}$ less than the L$_E$ and $10^{-4}$ less than the estimated L$_A$ from stellar winds.
In the quiescent state, the radio to sub-mm spectrum is described by a power law with a positive index of 4/3 (in $\nu$F$_\nu$)
with a sub-mm bump that peaks at 350~$\mu$m. 
It is consistent with a self-absorbed 
thermal synchrotron emission of electrons at temperatures $\approx$ 10$^{11}$ K, 
with densities of the order of $10^{6}\rm~cm^{-3}$, in a magnetic field of 10--50~G 
\citep[][]{falmar00, yuan02}.
It becomes optically thin above $\approx 100$--300~GHz and then decreases steeply, 
becoming undetectable in the far-mid IR regimes where diffuse dust emission completely dominates measurements.

\begin{figure}[ht]
	\centering
  \includegraphics[width=\columnwidth]{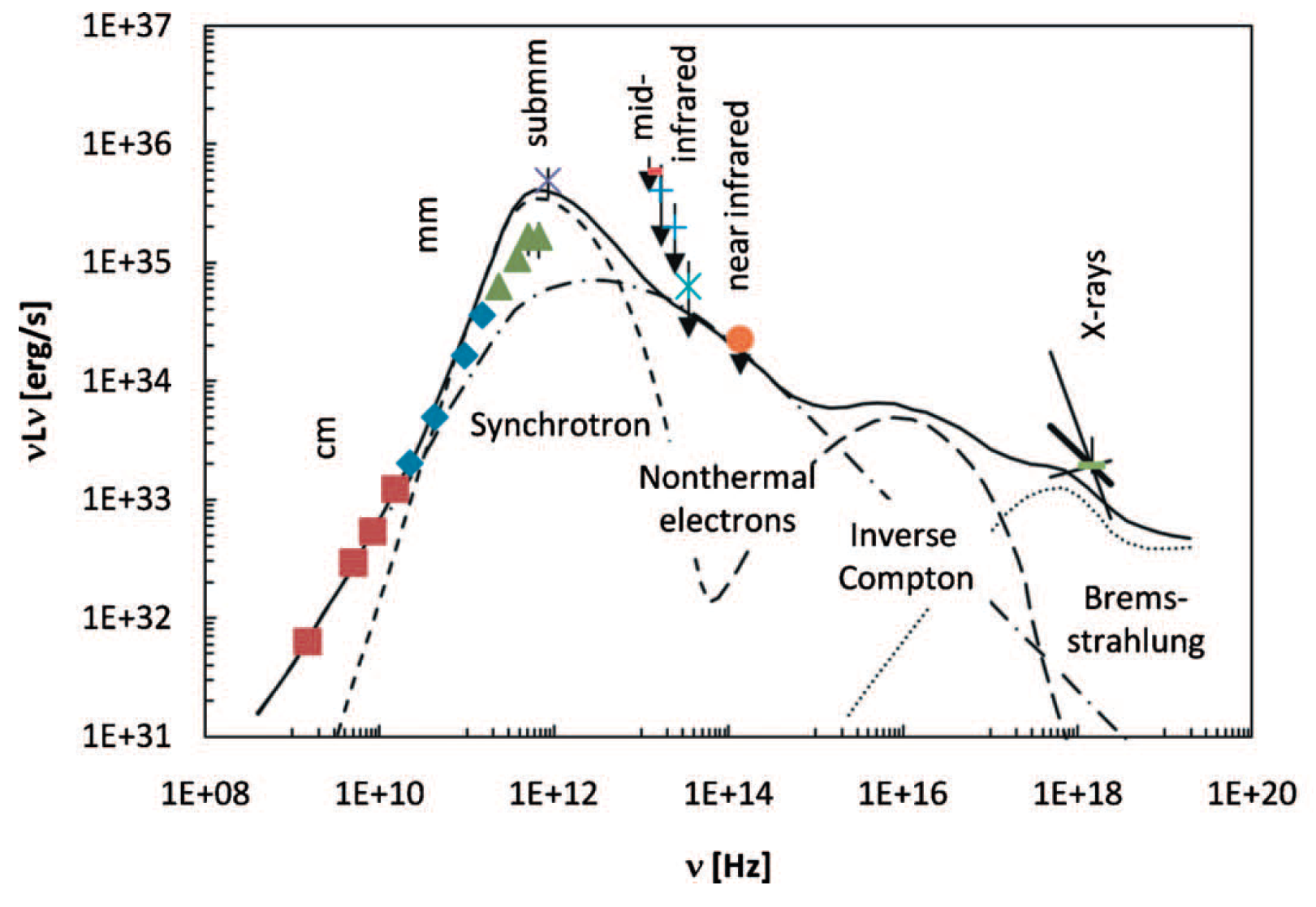}
	\caption{Sketch of the SED of the \sgra quiescent emission, from \citet[][]{genzel10}. 
	Compiled data in different bands are fitted with a RIAF model (broken and dotted lines), 
	plus a non-thermal electron component (broken-dotted line) \citep{yuan03}.
	For the most recent and complete \sgra SED data compilation see \citet{eventh22b}.
	}
	\label{fig:SgrAstar-sed}
\end{figure}

The quiescent \sgra spectral energy distribution (SED) is represented in \Fig\ref{fig:SgrAstar-sed} 
and it was modeled with advection dominated accretion flow (ADAF) 
or radiatively inefficient accretion flow (RIAF) models, 
the latter ones including outflows or convection in addition to advection.
They predict three peaks of emission: the inner thermal synchrotron one mentioned above, 
an undetected inverse Compton bump in the optical/UV range 
and a thermal bremsstrahlung emission in X-rays from the hot electrons of the entire flow. 
The ADAF/RIAF models \citep[see][and references therein]{yuanar14, narqua23}
were developed in the 1990s when it appeared clear that, at low accretion rates, 
the matter falling into BHs does not behave as predicted by the standard optically thick
and geometrically thin disks of \citet{shasun73}, 
but rather sets in a two temperature flow with ions 
and electron thermally decoupled (with ions much hotter than electrons). 
Subsequent GR magneto-hydrodynamic (MHD) simulations have confirmed these semi-analytical models 
but also revealed a large complexity of situations.  
In particular, outflows may play a major role 
and solutions depend whether the magnetic energy is dominant over the gas pressure, 
giving rise to the so-called Magnetically Arrested Disk (MAD) models, or not, 
as in Standard and Normal Evolution (SANE) models \citep[see][for definitions, reviews and references]{narqua23, eventh22c}. 
Most GRMHD simulations 
also predict, in certain cases, the emergence of collimated relativistic jets. 
Jet models, first proposed 
by \citet{falmar00} and \citet{markof01}, 
for which the base of the jet  
gives rise to the sub-mm bump and its extension to the low-frequency spectrum, are still 
compatible with the data. 
Indeed, at cm wavelengths the spectrum sticks out with flatter slope 
compared to the extension of the sub-mm bump (\Fig\ref{fig:SgrAstar-sed}). 
It seems dominated by an external non-thermal component,
possibly a dark jet that displays most of its power in kinetic, 
rather than radiative, energy. 

In addition the effective accretion rate reaching the BH horizon has been estimated to be only
$10^{-9}$--$10^{-7}\rm~M_\odot~yr^{-1}$ through measurements of linear polarization of \sgra millimeter emission, 
which implies limits on the Faraday depolarization and therefore on the plasma density around the source
\citep[][]{aitken00, marron07}. 
Such a low accretion rate implies somehow that the stellar-wind material captured at the Bondi 
radius does not reach the horizon and it is rather stored in the system
or expelled in outflows.

\subsubsection{\sgra black hole shadow and NIR emission}
Since the high-frequency radio/sub-mm emission must come from the very inner region close to the BH horizon,
VLBI measurements have been performed during the last decades (at increasingly higher radio frequencies 
where interstellar scattering is lower) to determine the size of \sgra.
These observations have recently led to the detection of the BH shadow with the Event Horizon Telescope (EHT). 
This world global interferometric array of radio-telescopes capable to attain resolutions 
of the order of tens of $\mu$as, 
performed, in 2017, 
accurate measurements of \sgra at $\lambda=1.3$~mm, 
and resolved the compact source emission in a 
thick ring of $51.8\pm2.3\rm~\mu as$ diameter (0.42~AU)
with a distinct central depression, providing, for the first time,
a reconstructed image of the SMBH shadow  
\citep[see][and ref. therein]{eventh22a, eventh22b}. 
These results are 
consistent with the GR predictions of the appearance of a Kerr BH at the GC,
with the estimated mass of \sgra and a prograde accretion disk viewed at low inclination ($i<50^\circ$).
The BH derived parameters are fully compatible with the averages of the NIR measurements of the S-star orbits
without being able to resolve the inconsistencies between the Keck and the VLT measurements.

\sgra has been detected as a flaring source of NIR light \citep{genzel03} soon after the discovery of \sgra X-ray flares (\Sec\ref{sec:SGRA}). 
The wealth of observations conducted since then 
show that \sgra NIR emission is continuously varying but likely presents two states: 
a quiescent one characterized by red-noise variability described by a log-normal process with a well-defined 
median flux density of $1.1 \pm 0.3$~mJy,
and a flaring state produced by sporadic flares that create the observed power-law extension of the flux distribution \citep{doddse11, gravit20b},
even though a single, but right-skewed or broken, distribution can fit the data \citep{witzel18, witzel21}. 
A larger flaring activity, in 2019 after the pericenter passage of both S2 and G2 was reported with Keck \citep{do19b}
and VLT/GRAVITY \citep{gravit20b} 
but without changes in persistent emission.  
The NIR flares are more frequent (3--4 per day) than X-ray ones, show variabilities on minute scales, 
present steep (red) power-law energy spectra
and are linearly polarized, pointing to thin synchrotron emission
of transiently accelerated electrons very close to the BH.
Spectral changes (flares redder when dim, e.g.\ during rise and fall) 
were detected with the VLT, confirmed recently with Keck and recently measured during a bright NIR flare
observed with VLT/GRAVITY and \spitzer in 2019. 
A quasi-periodicity  
in the light-curve of the first detected flare in 2003 was not confirmed by later flare observations,
but in 2018 VLTI/GRAVITY could measure the movement of a flare emission centroid,
showing that it follows a nearly circular clockwise orbit with a period of $\approx$ 40 min
and displaying polarimetric properties varing with a similar periodicity 
\citep{gravit18b,gravit23}. 
The data are interpreted as emission from a hot spot moving in an accretion disk at a distance of 3--5~R$_S$ from the black hole, viewed at low inclination angle (nearly face on), 
with a period of 33--65 min, i.e., close to the innermost stable circular orbit 
period of a non-rotating $4\times10^{6}\rm~M_\odot$ BH.
The derived flare orbit orientations are close to the one of the YNC clockwise stellar disk, 
suggesting the accretion flow is fueled by their winds, in agreement with recent simulations \citep{ressle20}.
A deepest analysis of the EHT ALMA observation of \sgra, performed after a prominent X-ray flare, 
led to the detection of mm polarization periodic variations, also consistent with 
the hot spot model at 5~R$_S$ viewed at low inclination angle as found with GRAVITY \citep{Wielgu22}.
Very recent EHT polarization measurements of the persistent mm emission on the event horizon scale \citep{eventh24} 
favour a clockwise rotating MAD around a very rapidly prograde spinning BH seen at very low inclinations. 
These recent data do not exclude the possibility of a jet, 
but, today, it is believed that the outflow from \sgra takes rather the form of a weak wind 
than the one of a focused jet \citep[][]{ciumor26}. 

Another recent result from ALMA is the detection of a sub-mm H30$\alpha$ hydrogen recombination line
displaying a broad (2200~km~s$^{-1}$) double-peaked profile arising within 0.3$''$ of the source
\citep{murchi19}, 
interpreted as emission from a cold ($\approx 10^{4}$~K) 
disk around \sgra.
However, the lack of detection of associated H Brackett-$\gamma$ line 
\citep{ciurlo21} 
implies that the H30$\alpha$ line must be amplified 
by a maser, which is problematic in a non-compact source like such an extended disk.
The HD models of the YNC stellar winds in the central cavity, 
some now including matter accretion from the CND \citep{solank23}, 
do allow the formation of such 
a disk, compatible with the sub-mm/IR measurements \citep{calder25},
but the results critically depend on the model parameters 
and the topic is still debated.\\

Some of the main features discussed in this section are reported 
in the summary finding chart (Fig.~\ref{fig:conclu}--\ref{fig:conclu2}) presented in the conclusion section (\Sec\ref{sec:Con}).


\section{Searches for the High-Energy Emission from the GC} \label{sec:SEAR}
The GC view presented above and based on low-frequency measurements
has been progressively enriched, starting from the 1960s, by high-energy observations of the region.
We summarize the historical developments of these measurements until 
the end of the last century in \Sec\ref{subsec:EarlyXG}, 
and then present the instruments
that have been used to observe it in X-rays (\Sec\ref{subsec:XTel}) and at higher energies 
(\Sec\ref{subsec:HX-Gam}) over the last 25 years. 
Most of them are still in operation and regularly monitor the region, 
while new telescopes have also started to explore the GC and are mentioned
along with planned future missions (\Sec\ref{subsec:HE-Recent}). 

\subsection{Early X/gamma-ray observations} \label{subsec:EarlyXG}

The detection of high-energy emission from the GC dates back to the 1960s, 
at the very beginning of X-ray astronomy, when,
using data of a 1964 Aerobee rocket flight, \citet{bowyer65} reported
the discovery of an X-ray source in Sagittarius.
Other measurements by sounding rocket experiments 
followed \cite{clark65,fisher66,gursky67,bradt68} 
showing that most of the newly discovered X-ray sources, later classified as X-ray binaries (XRB),
were clustering around the GC. 
In the 1970s, the region was monitored by the first X-ray satellites, 
\Uhuru, \Arielfive and \HEAOone. 
The central source (4U~1743--29) detected by \Uhuru appeared extended over $\approx$~2\deg,
either due to the combination of at least three point-like sources or to diffuse emission \citep{kellog71}, 
while \Arielfive \citep{eyles75}
and other experiments \citep[][]{crudda78} revealed, with large position errors, 
the presence of other bright, variable, or even transient objects (e.g.\ A~1742--294 and A~1742--289) 
and the occurrence of several X-ray bursts.
At the end of the 1970s, it was already clear that, although the GC appeared quite active, 
and dominated by bright and variable XRBs,
the X-ray luminosity of the nucleus itself was much lower that the one inferred for AGN \citep{procto78}.

With the launch of the \Einstein Observatory in 1978 it was possible, 
for the first time, to implement, at low energies, 
the grazing-incidence X-ray mirrors 
coupled to position sensitive detectors
to image the sky with resolutions of $\approx 1$\arcmin.
\citet{watson81} obtained the first X-ray images of the GC at these resolutions and 
showed that the central 20\arcmin of the Galaxy at $<4$~keV 
were dominated by diffuse emission and a dozen of point-like sources, 
one of which coincident with Sgr~A~West and 
including \sgra. This source was then resolved into 3 weak objects with \ROSAT 
more than 10 years later \citep{pretru94}, leading to a measurement of a soft X-ray luminosity of only
$10^{34}\rm~erg~s^{-1}$ for the one coincident within 10\arcsec with the nucleus. 

Important results were also obtained on the GC X-ray diffuse emission,
in particular by the \GINGA satellite, that discovered a prominent 6.7~keV iron line 
diffuse emission from a region encompassing the whole CMZ \citep{koyama89}. 
This component resembled the diffuse Galactic ridge emission (GRXE),
distributed along the Galactic plane and characterized by a thin hot thermal plasma spectrum, 
with temperatures of 7--10 keV,
which was discovered in the early 1980s with \HEAOone and \emph{EXOSAT} missions \citep{worral82, warwic85}.
The possibility that the GC hosted such a large hot plasma cloud raised the question of its origin and source
power, considering that, not being gravitationally confined, it needed to be refilled on rather short time scales
(\Sec\ref{subsec:DXHC}).
In the early 1990s the \Granat/ART-P instrument discovered a harder ($> 10$~keV) diffuse emission
associated with the molecular clouds of the region. 
This prompted speculation on whether this component could be due to Compton diffusion of 
X-rays emitted in the past by some very bright source \citep{sunyae93}, possibly \sgra\ itself,
leading to the prediction of Fe\Ka fluorescent line emission associated with the clouds.
In 1994, the \ASCA satellite was the first 
to separate the diffuse 6.7~keV line emission of ionized iron 
from the 6.4~keV one of neutral Fe atoms and showed that the distribution of the latter, 
unlike the one of the 6.7 keV line, was correlated with the molecular material \citep{koyama96}.
The hypothesis of a past outburst from \sgra illuminating the clouds
was then explicitly formulated (\Sec\ref{subsec:DXNT}).

While soft X-ray data revealed faint emissions from the GC, 
hard X-ray observations above 5--10 keV showed bright emissions in the direction of the GC. 
Early observations from 30 keV to few MeV detected a hard continuum and emission lines \citep{haymes69, haymes75}, 
in particular at 511 keV \citep{LinRam89} (\Sec\ref{subsec:511}), 
from the GC but with a localization uncertainty of a few degrees. 
These results were seen as evidence of the presence of a massive BH at the GC \citep{gentow87},
before coded mask imaging instruments, flown onboard satellites in the 1990s, 
increased the angular resolution of hard X-ray telescopes to better than 30$'$ \citep{golgro22}.
Observations in the 3--30 keV range with \emph{SpaceLab2}\xspace \citep{skinne87}
and ART-P \cite{pavlin94}, as well as in hard~X/soft $\gamma$-rays (30--1000 keV) 
with \Granat/SIGMA \cite{goldwu94}, then showed that the GC emission in these bands 
was rather due to the powerful hard XRBs in the area.
The brightest is 1E~1740.7--2942, an \Einstein source at 40$'$ from \sgra, rather trivial in soft X-rays but 
with such a large hardness ratio and high variability in hard X-rays
that it was quickly suspected to host a stellar-mass BH.
The SIGMA results, including the detection of variable emission around 500~keV from the source \cite{bouche91}, 
possibly explaining previous detections attributed to the GC nucleus 
(even if later measurements have not confirmed such a feature), 
prompted radio observations which led to the discovery of its extended radio jets 
and its classification as the first microquasar \citep{mirabe92}.
The SIGMA derived upper limits for \sgra  
implied a very low luminosity ($<$ 10$^{36-37}$~erg~s$^{-1}$) 
even at energies larger than 30 keV \cite{goldwu94}, 
difficult to reconcile with the BH accretion models 
and the expected accretion rates from the YNC winds \citep{melfal01},
questioning the very existence of the GC SMBH. 
This prompted the development of the ADAF models 
in order to explain the very low radiative efficiency of the central SMBH \citep{naraya95} (\Sec\ref{subsubsec:SgrAs}). 
In those years the OSSE instrument on \emph{CGRO}\xspace,
proved that the bulk of the 511~keV line emission was not variable, but rather diffuse 
and extended over the GB \citep{purcel97}, not confirming 
a direct link with \sgra (but see \Sec\ref{subsec:511}).

Meanwhile, high energy (HE) $\gamma$-rays (50 MeV to 100 GeV) 
from the Galactic plane, with an increasingly significant concentration in the spiral arms and in 
the inner Galaxy \citep{bignam75}, were progressively detected since the 1960s by the satellites 
\emph{EXPLORER~XI}\xspace, \emph{OSO-3}\xspace, \emph{SAS-2}\xspace and \emph{COS~B}\xspace,
that exploited the pair production process to detect photons at these energies. 
Interpreted as the result of cosmic-ray interaction with the ISM of the Galaxy, 
whose density increases towards the center, the measured gamma-ray emission by \emph{COS~B}\xspace 
seemed to imply an under-density of the CRs in the GC with respect to the local values \citep{blitz85}.
However, later, \citet{mayerh98}, using EGRET on \emph{CGRO}\xspace, 
detected at the GC a source sticking out from the expected emission,
also arguing that \emph{COS~B}\xspace data were compatible with their new results.
The source (3EG J1746--2851) was positioned slightly away from the nucleus ($\sim 0.2^{\circ}$)
but it could still be linked to \sgra given the large localization error \citep{hartma99}.

At very high energies (VHE, $>$ 100 GeV), the technique based on detection of atmospheric Cherenkov light
produced by particles generated by such gamma-rays was progressively 
implemented and refined in ground based observatories, until significant detection 
from the GC was claimed by \citet{tsuchi04} with Cangaroo-II and by \citet{kosack04} with Whipple
(see \Sec\ref{subsec:GRCS} for details on the HE/VHE gamma-ray GC source).

At the turn of the century, a new era of HE 
astronomy has been opened 
with the launch of the \chandra, \XMM (1999) and \suzaku (2005) X-ray observatories; 
with the launch of the hard X-ray/low-energy gamma-ray missions \INTEGRAL, \SWIFT 
and \nustar and of the HE gamma-ray one \FERMI (2002--2012),
and with the start of operations (2003--2007) of more sensitive VHE gamma-ray telescopes 
(\HESS, \Magic, \Veritas).

\begin{figure*}
	\centering
	\includegraphics[width=\textwidth]{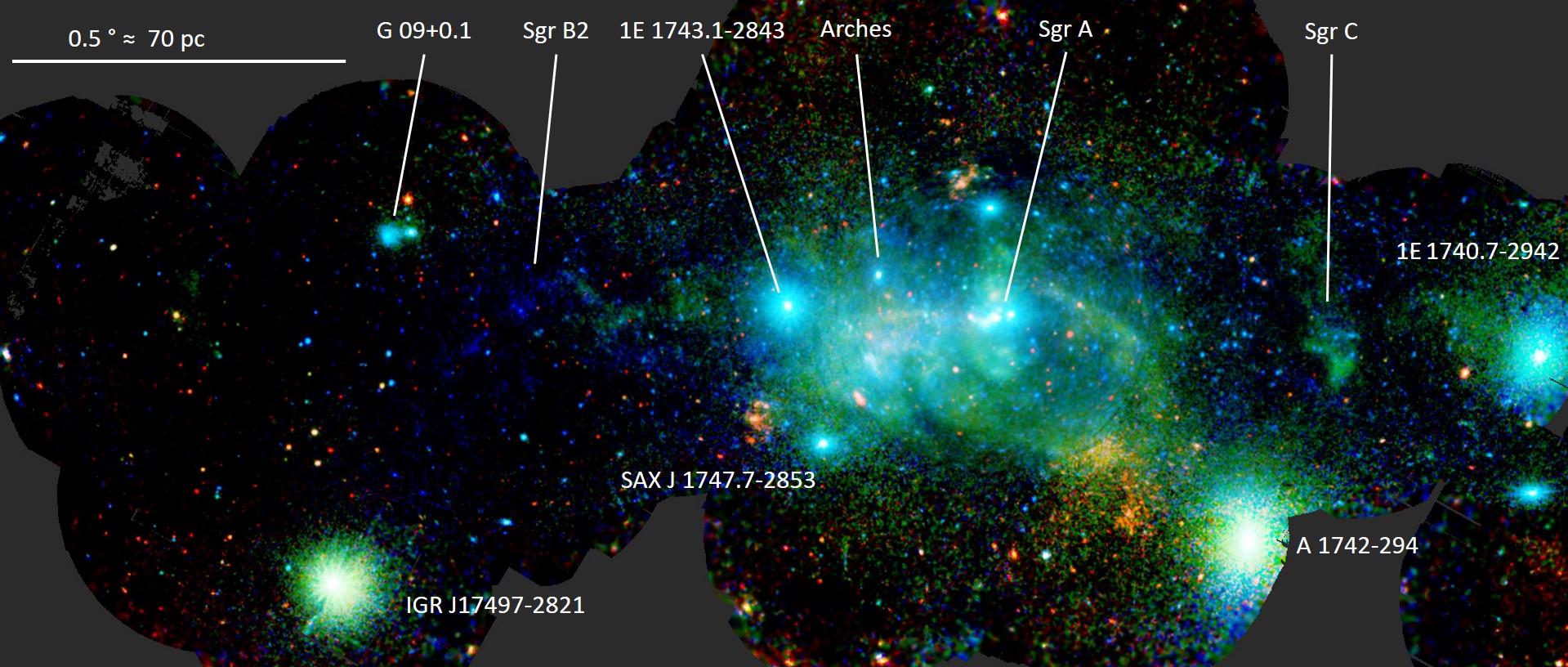}
	\caption{The \XMM GC Survey. Mosaic image of all \xmm observations performed until 2015 
	within one degree of \sgra
	where in red, green and blue are showing the emissions in the 0.5--2, 2--4.5 and 4.5--12 keV energy bands, which 
	include X-ray radiation from X-ray binaries, star clusters, supernova remnants, bubbles and superbubbles, \HII regions, 
	PWNe, non-thermal filaments, nearby X-ray active stars, Sgr~A complex and other features
	\AG{\citep[adapted from][original image credit ESA/XMM-Newton/G. Ponti]{ponti15b}}.
	}
	\label{fig:xmm-gc}
\end{figure*}

\subsection{X-ray telescope observations of the GC}\label{subsec:XTel}

NASA's \chandra X-ray Observatory \citep[CXO,][]{weissk02}, equipped with the most powerful 
grazing incidence X-ray mirrors ever,
reaches a still unequaled angular resolution of 0.5$''$ full width half maximum (FWHM) in a 17$'$ $\times$ 17$'$ field of view (FoV) over 0.1--10 keV, thanks to the set of four 10~m focal mirrors associated to the 
Advanced Charge-Coupled Device (CCD) Imaging Spectrometer camera disposed for imaging (ACIS-I).
While the spectral resolution of the CCD remains moderate (100--200 eV, FWHM), 
high spectral resolution (power 100--500) can be obtained 
using dispersive grating systems,
the one working at $>$ 2 keV being the High Energy Transmission Grating spectrometer (HETG)
coupled to a CCD set disposed for the use (ACIS-S).
Launched in July 1999 \chandra started observing the GC in the fall, 
which led to the first solid detection of \sgra in X-rays \citep[][]{bagano01}
and since then accumulated several millions of seconds (Ms) of exposure on the source, 
in particular with a dedicated X-ray Visionary Project (XVP) of 3~Ms in 2012 with HETG/ACIS-S,
and few large surveys of the CMZ and other regions \citep[e.g.][]{muno09} (\Fig\ref{fig:chimneys}, right).

\XMM \citep[][]{jansen01}, the European Space Agency (ESA) 
X-ray Observatory, was launched a few months after \chandra. 
Hosting three sets of 58 nested mirrors with 
three different European Photon Imaging Cameras (EPIC), 2 CCD MOS and 1 CCD pn, 
at their 7.5-m focal planes,
offers a lower angular resolution, $\approx$ 15$''$ half energy width (HEW), than \chandra
but with three times larger ($\approx$ 1000 cm$^2$) effective area
over a 30$'$ diameter FoV and better spectral resolution along a broader range (0.1--12 keV).
The numerous individual observations with the EPICs,
three major surveys of the CMZ in 2000--2001, 2012 and again in 2020,
and two large projects to explore the GC polar regions 
allowed the early detection of \sgra flares, the full mapping of the region (\Fig\ref{fig:xmm-gc},	\ref{fig:soft_plasma_map}, \ref{fig:chimneys} left) \citep[see e.g.][]{ponti15b, ponti19}
with several discoveries and studies, also of variable features \citep[e.g.][]{terrie18}
and the collection of a very large set of high-quality data. 

The JAXA observatory \suzaku \citep[][]{mitsud07},
carrying four X-ray mirrors coupled to CDD detectors,
was characterized by low background and good spectral resolution,
features that allowed several important studies in particular of the diffuse emission.
Several surveys of the GC were performed using
the XIS instrument working in the range 0.1--10 keV, with angular resolution of $\approx 2'$ (HEW) in a FoV of 17$'$
and effective area of 4 $\times$ 200 cm$^2$, 
particularly to the study the iron lines distribution in the region \citep[e.g.][]{uchiya11, koyama18}. 
\suzaku stopped operations in 2015.

NASA's Neil Gehrels \Swift Observatory \citep[][]{gehrel04} instead was used to monitor GC X-ray transients and flaring activity of \sgra using its X-Ray Telescope (XRT), composed by a mirror set 
coupled to an EPIC/MOS CCD detector, to image sky areas of 23$'$ size in 0.2–10 keV with effective area of $\approx$ 120 cm$^2$
and angular resolution of $\approx$ 20$''$ HEW. 
A program to regularly monitor the central 25$'$ of the Galaxy started in 2006, 
one year after the launch, with short (1 ks) snapshots every 1--4 days during the target visibility periods. 
It provided until 2014 some 1200 observations for a total exposure of 1.3~Ms \citep[][]{degena15} and it is still on-going. 

\subsection{Hard X-rays and gamma-rays GC surveys} \label{subsec:HX-Gam}
After the \Granat mission, the coded mask technique was refined and largely applied to image the sky in the hard X-ray range (20--300 keV) and in low-energy gamma-rays (300 keV to 10 MeV). 
The ESA International Gamma-ray Astrophysics Laboratory (\integral), was launched in 2002, 
carrying three coded mask instruments, JEM-X, IBIS and SPI to cover a broad energy range from 3 keV to 10 MeV
\citep[][and references therein]{winkle03} in FoVs that span between 13$^\circ$ and 45$^\circ$ sides. 
The large FoV of such instruments and the very regular \INTEGRAL surveys of the GC 
until the end of the mission early 2025 have provided, 
fine images of the CMZ and surrounding region in hard X-rays \citep[][]{belang06a},
thanks, in particular, to the Imager on Board the \integral Satellite (IBIS), 
and its unprecedented angular resolution of 12$'$ in 20--1000~keV over a 30$^\circ$ wide FoV. 
\INTEGRAL revealed a peculiar central excess \citep{belang04}, 
monitored the variable sources that dominate the GC hard X-ray emission, 
discovered the hard emission of Sgr~B2 \citep{revniv04}
and its monotonic decrease over years \citep[][]{terrie10}.
The Spectrometer on \INTEGRAL (SPI), 
with its high spectral resolution (power $\approx$~250),
provided by Germanium detectors cooled to low temperatures,
and moderate imaging capabilities ($\approx$~2.7$^\circ$ FWHM)
has been used to more precisely map the 511~keV line emission of the Galaxy and study its spectral shape \citep[][]{knodls05, sieger16}. 

A real revolution in imaging the low-energy part of the hard X-ray range has been 
accomplished 10 yr later with \nustar, a NASA mission launched in 2012, 
carrying a set of mirrors with a multi-layer coating able to reflect 3--80 keV X-rays 
to solid state detectors placed at 10~m focal length, thanks to an extendable mast \citep[][]{harris13}. 
This telescope features 1$'$ HEW resolution 
in a 12$'$ side FoV with an effective area comparable, at 6 keV, to the \XMM EPIC/pn one.  
Large GC surveys have been carried out with \nustar \citep[][]{mori15, hong16}, 
along with a number of observations of localized regions.
These have allowed to detect the \sgra flares at more than 20 keV, to resolve 
the complex hard X-ray emission of the central 20 pc and to study of a number of hard sources of the CMZ. 

Above 20 MeV, gamma-ray observatories rely on the pair creation process, directly observed in space in the high-energy range and through the atmospheric cascade it creates for ground-based instruments in the very high energy range (above 100 GeV). Launched in 2008, the \Fermi/LAT 
is able to operate from few tens of MeV up to few hundred GeV thanks to its precision tracker, its thick calorimeter and its segmented anti-coincidence shield which permits a very efficient background rejection \citep{atwood09}. With its very large field of view and its scanning observation mode, \Fermi/LAT 
observations cover the full sky in a few orbits and provide a very uniform sky survey. Besides a collection of more than 6000 sources \citep{abdoll20}, the \Fermi/LAT 
discovered large-scale structures connected to the GC: the \Fermi bubbles \citep{su10} and the Galactic center GeV excess \citep[see e.g.][]{ackerm17}. While the angular resolution is rather modest below the GeV range ($\geq 1^\circ$) making source confusion a major issue to study the inner degree, the good performances at higher energies ($\leq 0.1^\circ$ above 30 GeV) allow for more precise measurements \citep{cafard21}.   

Above $\sim$ 100 GeV,  Imaging Atmospheric Cherenkov Telescopes (IACT) observe the Cherenkov radiation produced in the atmosphere by the cascade of secondary charged particles induced by incident cosmic particles \citep[see e.g.][]{denmaz15}. This technique offers a very large collection area, essential in the VHE range where photons are very scarce. Modern instruments, such as \HESS, \MAGIC or \Veritas rely on stereoscopic observations of air showers which provide good angular resolution (around or below $0.1^\circ$) and a satisfactory background rejection.  \MAGIC is a system of two 17m diameter IACTs located on the island of La Palma, with a threshold of $\sim$ 50 GeV \citep{aleksi16}. VERITAS is an array of four 12 m IACTs located in southern Arizona \citep{adams21}. 
\HESS is an array of four 12m telescopes \citep{aharon06b}, plus a large 28m telescope added in 2012 at the center of the array \citep{hessco18d}.
It is located in Namibia, which is more suited to observe the inner Galaxy compared with northern hemisphere locations, where the region can only be observed at relatively low elevation, 
increasing the energy threshold of detectable gamma-rays. Deep surveys of the GC have been conducted by the three experiments and revealed several sources and diffuse emission in the inner degrees.

Finally, particle sampling arrays, such as HAWC \citep{abeyse23} and LHAASO WCDA or KM2A arrays \citep{Cao24}, directly measure secondary shower particles and can probe up to ultra-high energy (UHE) gamma-rays from several tens of TeV up to the PeV range with an angular resolution of about 0.3$^\circ$. Both observatories are located in the northern hemisphere and can only observe the GC at very low elevations. 

\subsection{Recent and future instruments} \label{subsec:HE-Recent}
Amongst the recent HE instruments that observed the GC,
\eROSITA \citep[][]{predeh21},
launched in 2019 on board the Russian-German \emph{SRG}\xspace mission, 
has a range, an effective area and an angular resolution comparable to \XMM, but 
larger FoV (1.3$^\circ$), and it was designed to carry out deep all-sky X-ray surveys.
One of its major results is the discovery of two gigantic polar bubbles extending to 80$^\circ$ above and below the GC,
enveloping the gamma-ray \Fermi bubbles \citep[][]{predeh20}
and it also provided results on the emission reflected by the CMZ clouds \citep[][]{khabib22}.
\erosita, despite being fully efficient, suspended operations in 2022 following geopolitical events\footnote{See {https://www.mpe.mpg.de/7856215/news20220303?c=450698}}. 

The other new facility used to explore the GC is \IXPE \citep[][]{weissk22}, 
the first observatory dedicated to imaging X-ray polarimetry, launched in 2021 and which carries 
three X-ray mirror telescopes with polarization sensitive detectors, for
an angular resolution of $\approx 30''$ and a FoV of 17$'$.
The main objective of the \IXPE GC observations was to search for polarization of the reflected emission 
from the molecular clouds \citep[][]{marin23}. 

Great expectations for the study of the GC are placed on the new generation of X-ray observatories 
equipped with non-dispersive high spectral resolution spectro-imagers based on X-ray calorimeters.
The recently launched (2024) \Xrism observatory \citep{tashir21} is the first of such missions to be operating.
It carries a micro-calorimeter 
that provides data with an unprecedented energy resolution of 5 eV (FWHM) over a 3\arcmin FoV 
with the angular resolution of $\approx$ 1\arcmin (HEW), and
is already providing interesting results on the GC \citep[][]{xrismc25}.
The future \Athena X-ray observatory, expected for the late 2030s, will carry the X-ray Integral Field Unit,
a spectral-imager 
with an exceptional resolution of 4 eV, angular resolutions of 8\arcsec--10\arcsec and an effective area of 1.4~m$^2$,
over a FoV of 4\arcmin diameter in the 0.2--12 keV range \citep{peille25},
greatly improving the performances with respect to \Xrism.

For the low-energy gamma-ray band, 300 keV--10 MeV,
called the MeV Gap because the least explored part of the electromagnetic spectrum, 
several experiments have been proposed in recent years, 
and at the moment the Compton telescope
\textit{COSI} \citep{tomsic24} is the one planned to fly soon. 
The expected sensitivity, over a very large FoV (1/4 of the sky) with an angular resolution of 4\deg--2\deg, 
is several times better than SPI at 0.5--2 MeV.
The design of future MeV missions of superior sensitivity is currently debated in the community for
operation in the next decade\AG{s} \citep{sieger22}. 

Important advances on the GC gamma-ray properties are also expected from the Cherenkov Telescope Array Observatory (CTAO). It consists of two hybrid arrays of large, medium and small size telescopes covering a broad energy range with sensitivity improved by a factor of up to ten compared with current IACTs. The southern part of the observatory will be deployed in Chile and will be fully operational at the end of the decade. A deep GC survey is foreseen among the observatory key science programs \citep{cheren19}. It will allow the exploration of the region in the 20 GeV to 100 TeV range with an angular resolution of a few arcmin and much larger sensitivity than the present IACTs. In addition, particle samplers in the southern 
hemisphere such as the Southern Wide-Field Gamma-ray Observatory (SWGO), project currently in the 
study phase, will open the window of the UHE gamma-rays at the GC.


\section{\sgra and its close environment at high energies}\label{sec:SGRA}
While the main goal of the early GC HE observations was to detect emission from the SMBH, only \Chandra has in fact the resolution
needed to disentangle the complex X-ray morphology of the central few parsecs.
\Chandra detected the persistent X-ray counterpart of \sgra (\Sec\ref{subsec:SGQE}) soon after its launch
and discovered one year later also its peculiar flares (\Sec\ref{subsec:SGFE}). 
We are certain that this source is associated with the SMBH because its flares are simultaneous to NIR ones, 
which can be positioned at $\mu$as precision on \sgra with GRAVITY.
The SMBH is surrounded and fed by hot gas produced by massive star winds (\Sec\ref{subsec:SGQE}) 
and by several peculiar X-ray sources in the inner 20~pc (\Sec\ref{subsec:SGCE}).

\subsection{\sgra quiescent emission and central hot plasma} \label{subsec:SGQE}

\begin{figure*}
	\centering
\includegraphics[width=0.45\textwidth]{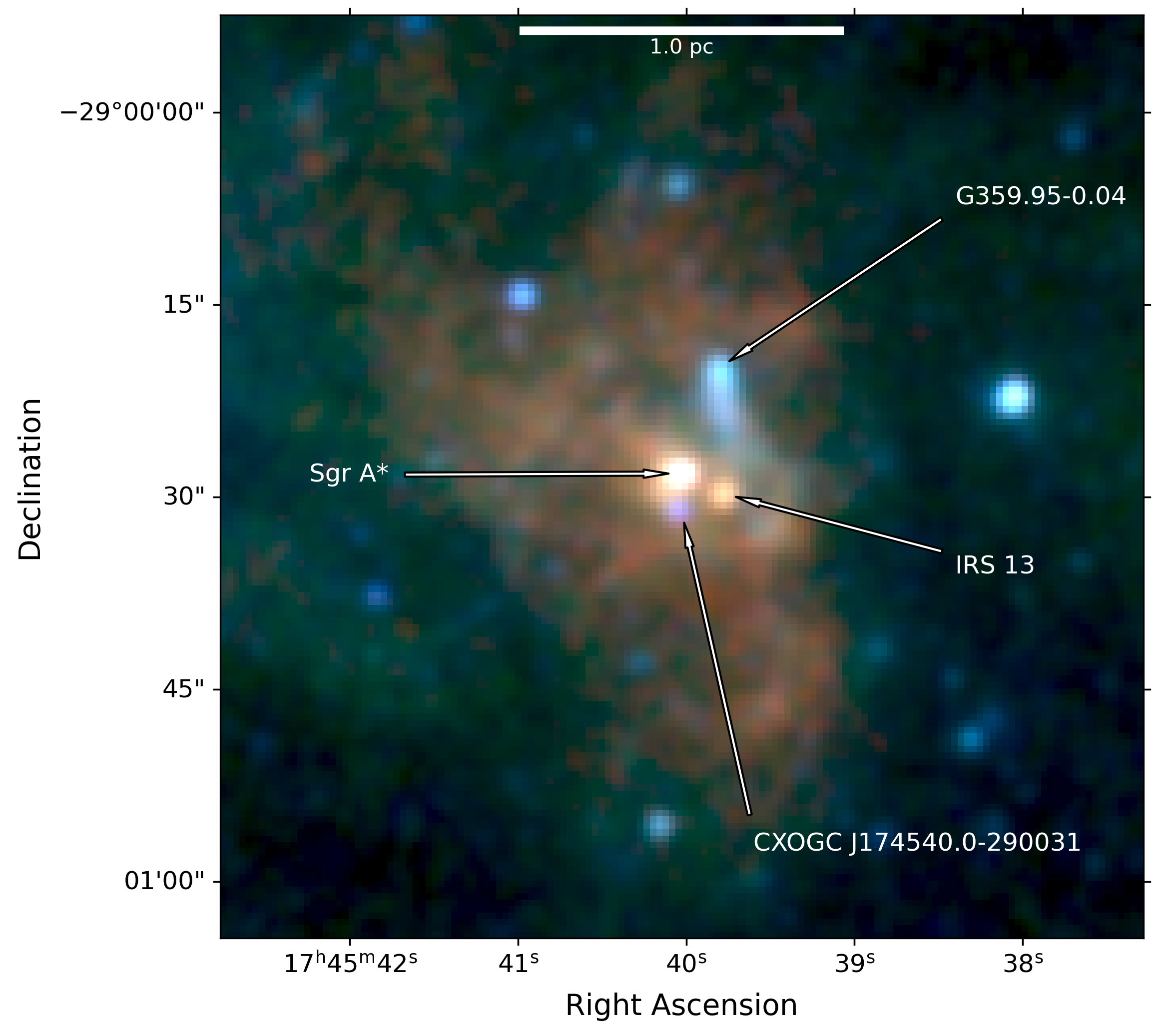}
 ~
 \includegraphics[width=0.53\textwidth]{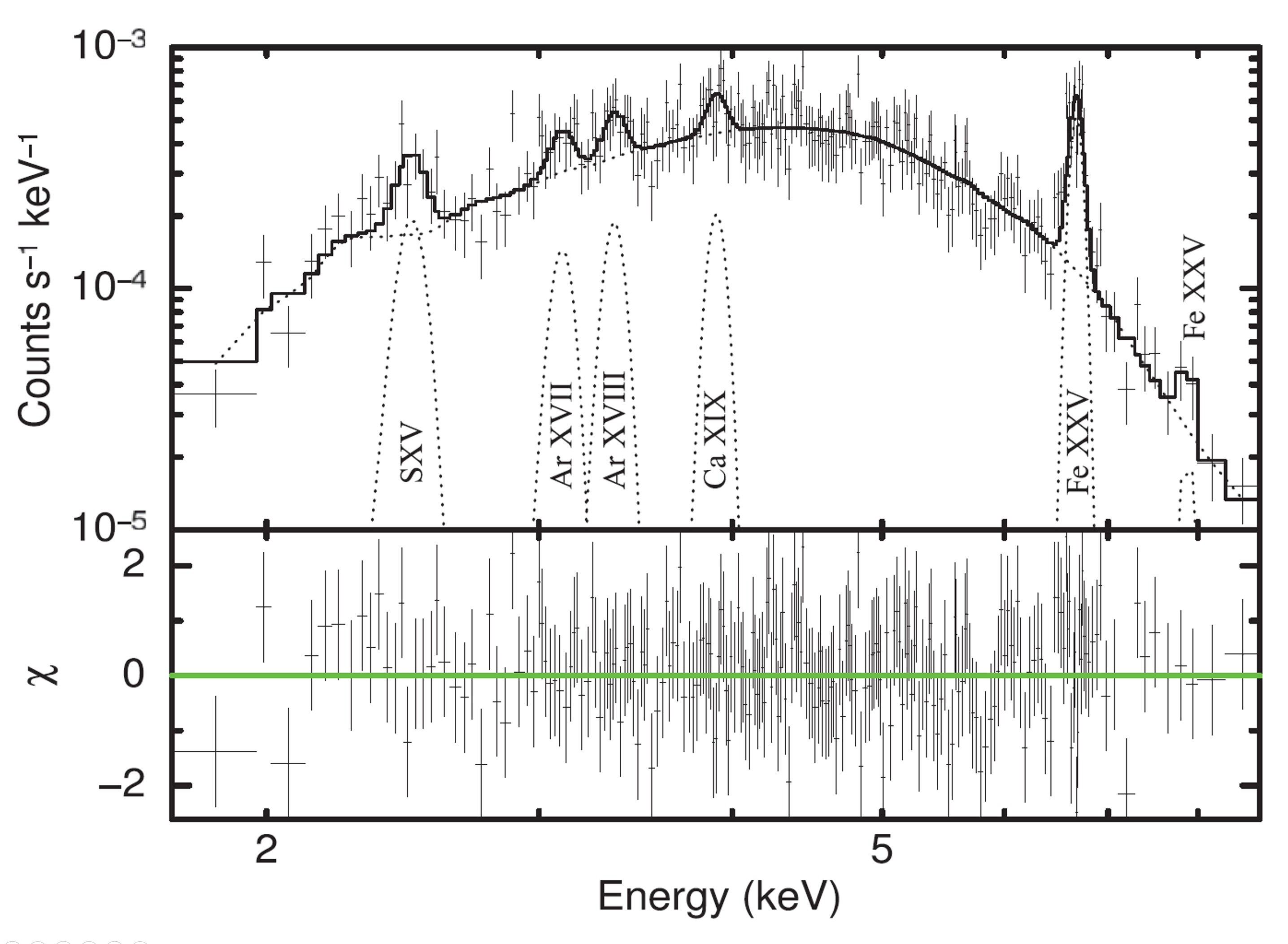}
	\caption{\sgra X-ray quiescent emission and surrounding sources. 
 Left: \Chandra 3-color X-ray image of the central 1.2$'$ (2.8 pc) (red: 1.5--2.6 keV, green: 2.6--4.5 keV and blue 4-5--7.5 keV). It shows the source associated with 
 \sgra at the center, the diffuse hot gas emission visible in red, the cometary-shaped PWN (G359.95--0.04), the massive stellar cluster IRS 13 as well as other point-like sources in the field. 
 Right: \Chandra HETG absorbed count spectrum of the \sgra quiescent emission, within 1.5$''$ and out of flare periods, 
 modeled with a thermal spectrum and Gaussian lines \citep[from][]{wang13}. 
	}
	\label{fig:SgrAstar}
\end{figure*}

\subsubsection{\sgra quiescent emission and accretion flow} \label{subsubsec:SGQEaf}
From the very first GC observations in 1999, \Chandra resolved the central \ROSAT source 
associated with Sgr~A into several components (\Fig\ref{fig:SgrAstar}, left)
and in particular detected, at $< 0.27''$ from \sgra, a very weak and steady source 
displaying a 2--10~keV luminosity of only $(2$--$3)\times10^{33}\rm~ers~s^{-1}$
at the GC and showing an extended morphology \citep{bagano03}.
The source intrinsic size (1.4$''$ FWHM) 
is comparable to the estimated Bondi radius of the accretion flow of \sgra, 
and later studies showed that the morphology appears elongated and oriented in a similar way than the clockwise disk of the YNC, even if the significance of this feature has not been quantified  \citep{wang13}. 
Between 4 and 20\% (depending on the analysis) 
of the quiescent emission
can be attributed to a central inner point-like component \citep{nowak12, neilse13, wang13}, 
possibly including undetected weak flares, 
and the overall emission appears steady, even after the pericenter passage of S2 and G objects
in their orbits around the SMBH \citep{yuawan16}. 

Building upon previous spectral studies \citep{bagano03,xu06}, 
detailed analyses of a very large \Chandra data set, including 
the high energy resolution ones from the XVP, 
and implementing proper treatment for 
contaminating sources, pileup and dust scattering, 
have provided high-quality spectra of the \sgra X-ray quiescent emission (\Fig\ref{fig:SgrAstar}, right),
allowing for the detection and characterization of several plasma emission lines \citep{wang13, corral20}.
The spectrum, within $R$ = 1.5$''$ after excluding flaring periods, 
could be modelled with a thin thermal plasma at 3.5~keV temperature in collisional ionization equilibrium
with the clear presence of a prominent, 700~eV equivalent width (EW), K$\alpha$ emission line of the He-like ionized iron (Fe~\textsc{xxv}) complex
at 6.67~keV 
and K$\alpha$ lines of several other species (He- and H-like ions of Ar, S and Ca), 
in particular the 3.1~keV line of Ar~\textsc{xvii} (200~eV EW), as well as of He-like Fe K$\beta$.

The \sgra steady X-ray emission is generally interpreted as thermal emission produced 
by the captured gas at the Bondi radius. 
Possibly after a thermalizing shock, this gas
is accreted onto the BH, heating up along its fall to virial temperatures.
The density, velocity and temperature profiles 
are driven by the physical characteristics of the accretion flow and by the initial conditions.
Precise line emission diagnostics can differentiate the possible type of accretion 
flow models \citep[][see also \Sec\ref{subsubsec:SgrAs}]{narray99, yuan03, xu06} 
and the \Chandra data appear, within uncertainties, compatible with most of 
the radiatively inefficient ones, except those without outflows, which predict too steep density profiles,
i.e., pure Bondi or pure ADAF models \citep[][]{corral20}.
The spectrum could be well fit by an absorbed thin thermal model that includes density and temperature distributions predicted 
by a RIAF with outflows, with an equivalent hydrogen absorption column density $N_{\rm H} = 1.38\times10^{23}\rm~cm^{-2}$
and abundances of about 1.5 solar, both compatible with expected values at the GC.
In a recent and detailed analysis of the XVP data, \citet{balakr24a} obtained similar results,
with a best-fit for the \sgra quiescent spectrum by a RIAF model with nearly as much outflow 
as inflow of matter, and with shallow variation of temperature, 
but with best-fit iron abundances below solar. 

Incidentally the derived upper limits on the 6.4~keV line of neutral iron in 
this emission \citep{wang13} are excluding an important contribution from coronal emission 
of low-mass stars in the NSC cusp which had been suggested by \citet{sazono12}.
The \Chandra \sgra quiescent emission slope is reported in the overall average SED represented in \Fig\ref{fig:SgrAstar-sed}, 
where it is compared to the model of \citet{yuan03}, a RIAF including outflows with a subdominant population
of non-thermal electrons explaining the low-frequency radio data and the seed from which the non-thermal flares can start, 
and where the bulk of the X-rays are attributed to thermal bremsstrahlung. 

\subsubsection{Surrounding hot plasma and influence on the accretion flow}

\Chandra also detected an X-ray diffuse emission in the central cavity, surrounding \sgra, 
within $\approx$ 10--15$''$ (0.4--0.6 pc) from it (\Fig\ref{fig:SgrAstar}, left). 
It is also characterized by a thermal spectrum but with an iron line centered at lower energies
than the Fe \textsc{xxv} complex, indicative of a non ionization equilibrium state.
This hot plasma with a patchy morphology is slightly peaked towards \sgra,
has a mean electron density of 26~cm$^{-3}$, a temperature kT = 1.3~keV and an estimated total mass of 0.1~M$_{\odot}$ \citep[][]{bagano03}.
It has been explained as the gas of the powerful stellar winds from massive stars in the nearby YNC,
heated by the shocks produced by their mutual collisions 
that efficiently converts the wind kinetic energy into gas internal energy \citep{rockef04}.
This is supported by detailed hydro-dynamical (HD) simulations of the YNC stellar winds interactions 
that take into account the spatial and velocity of the stars as well as the confining effect of the surrounding CND.
They reproduce the X-ray plasma properties observed with \chandra \citep{cuadra08, calder20}
as well as its morphology \citep{ressle18}.
The equilibrium state is reached in only a few thousand years from the onset of stellar-wind ejection, 
excluding a long-term impact by the Sgr~A East SNR shell passing-by or by putative ancient powerful BH outflows.
However, recent simulations have incorporated continuous and outburst BH outflows occurring 100--300 yr ago 
and indicate that they may still influence the plasma properties \citep[][see also \Sec\ref{subsec:DXNT}]{russel17}. 

This gas also provides the bulk of the material accreting onto the SMBH,
and from its X-ray measured properties, it is possible to establish the initial conditions of the flow.
The SMBH Bondi radius for this material is then estimated at 2.8$''$ (0.1~pc),
compatible with the \sgra quiescent source size, and the mass accretion rate at this radius 
at $10^{-6}\rm~M_\odot~yr^{-1}$, also compatible with the HD simulations.
The most recent 3D simulations, attempt to 
follow the gas fate all the way down to the inner regions of the accretion flow, 
incorporating radiation cooling and circularization of the flow in a disk, 
equatorial and polar outflows 
and GRMHD processes at play, establishing the conditions in the inner few tens of R$_S$.
They seem to show that the initial conditions are indeed crucial 
also to understand the very inner ($< 100\rm~R_S$) properties of the flow, 
in particular that the accretion disk, 
possibly generating 
the signal observed with the EHT (sub-mm quiescent emission)
or GRAVITY (non-thermal flares), may retain 
the initial orientation of the disk distribution of the stellar-wind-emitting stars \citep{ressle20}.
These results are still preliminary, in a very active area of investigation, where
large efforts are devoted to improve these simulations
and further constrain the models \citep[e.g.,][]{calder25}. 

\subsection{\sgra flaring activity} \label{subsec:SGFE}
\begin{figure}[ht]
	\centering
 \includegraphics[width=\columnwidth]{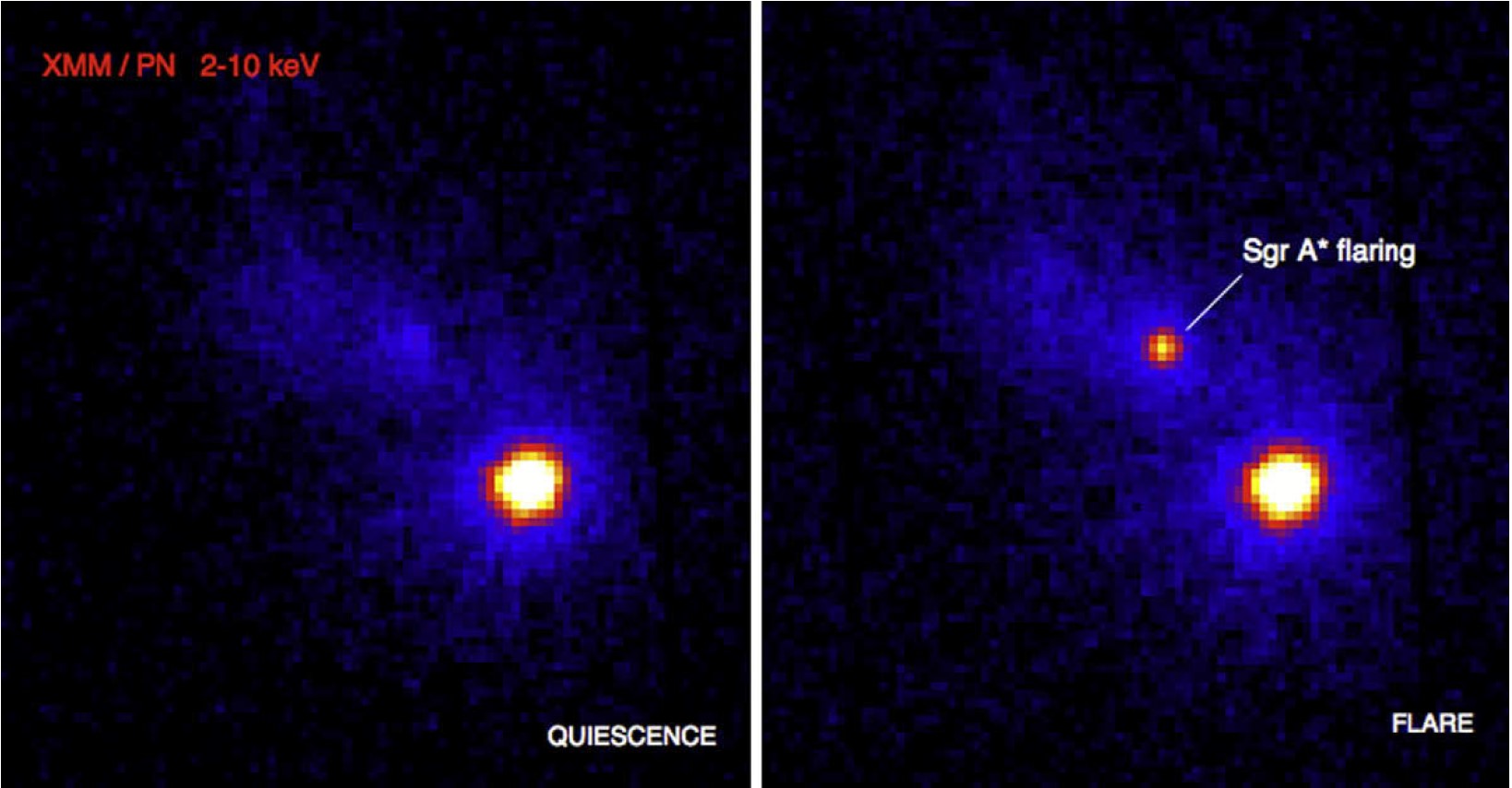}
	\caption{Bright \sgra X-ray flare observed with \XMM in 2007:
        the 2--10 keV image of the GC inner 15-pc region 
        before (left) and during (right) the flare \citep[from][]{trap10}.
        The bright source in the lower-right 
        side is AX~J1745.6--2901, a transient XRB  
        particularly active in 2007 at about 1.45\arcmin from the SMBH.
	}
	\label{fig:SgrAs-flares}
\end{figure}

One major result of the GC surveys with the new generation of X-ray telescopes was the discovery of the \sgra X-ray flares
(\Fig\ref{fig:SgrAs-flares}).
They were first observed with \chandra in 2000 \citep{bagano01}, confirmed soon after by \xmm \citep{goldwu03},
and then also detected in NIR in 2003 \citep[][see \Sec\ref{subsubsec:SgrAs}]{genzel03, ghez04}.
Given that these events allow us to explore the very inner region of the SMBH,
they have been searched for and studied with most HE observatories 
\citep{porque03b, belang05, trap10, nowak12}
including instruments in hard X-rays and gamma-rays \citep{trap11, aharon08b},
and finally detected also with \swift/XRT \citep{degena13} and \nustar \citep{barrie14}. 

\subsubsection{High-energy properties of \sgra flares}

During these flares, the source X-ray luminosity increases by up to 1--2 orders of magnitude, 
with the brightest events reaching peak values of L$_X \sim 5\times10^{35}\rm~erg~s^{-1}$ in the 2--10~keV band
\citep{nowak12, haggar19}. 
Spectra are featureless and can be described by absorbed power-laws with 
$N_{\rm H} \approx 10^{23}\rm~cm^{-2}$ 
(placing them at the GC) and photon indexes of 
$\Gamma \approx2.3$, harder
than the quiescent emission ($\Gamma \approx$ 3) and extending up to $>40$~keV \citep{barrie14}\footnote{First measurements found very hard spectra ($\Gamma \approx$ 1) that were later attributed to pile-up and scattering effects that had not been fully accounted for \citep[see][]{nowak12}.}.

\begin{figure}[ht]
	\centering
 \includegraphics[width=\columnwidth]{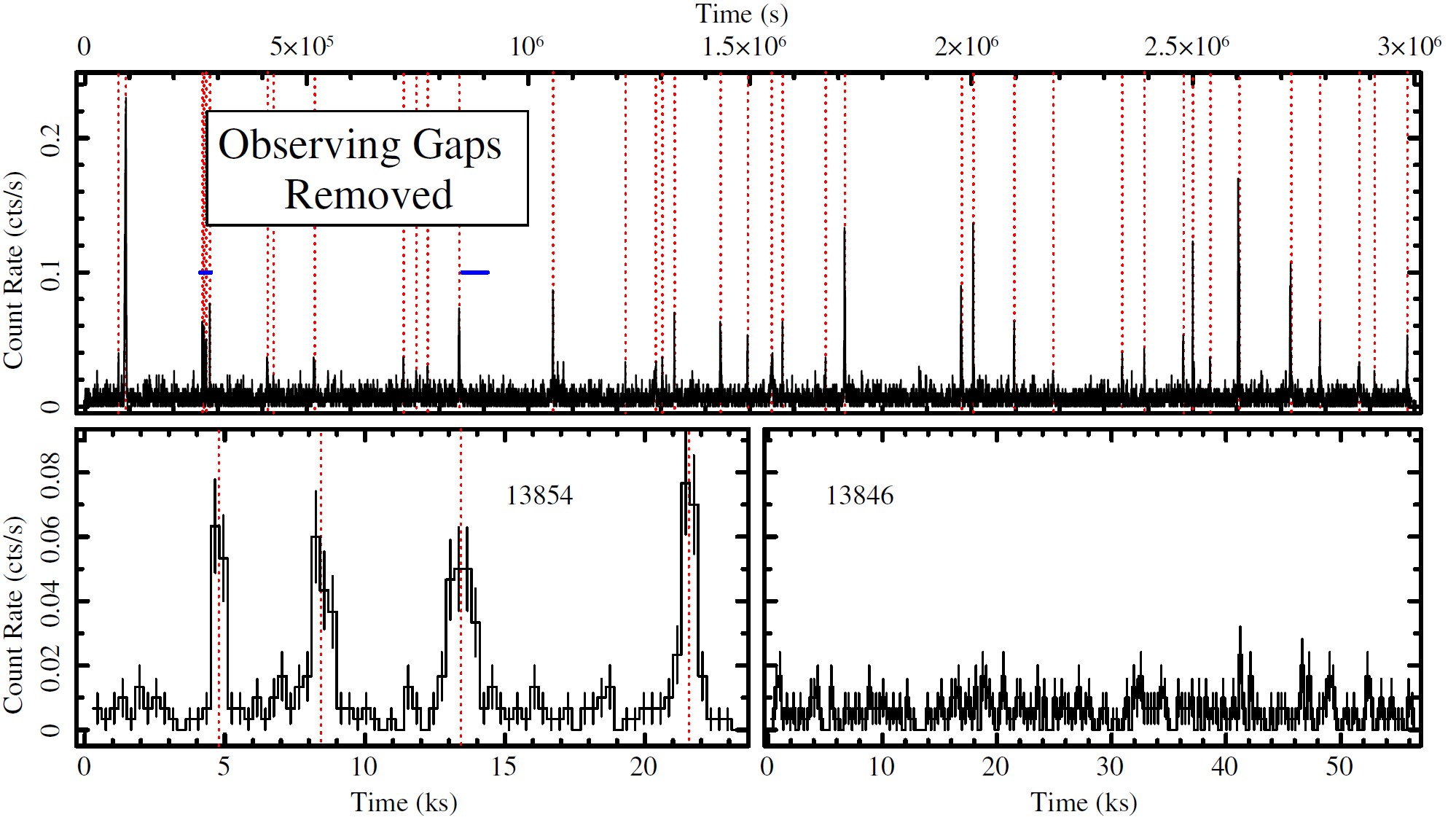}
	\caption{X-ray light curves of \sgra showing the flares 
	detected by \Chandra during the XVP observations (data gaps excluded) performed in 2012. 
	Lower panels show a zoom of the curve during observations with several flares (left) and without any (right). From \citet[][]{neilse13}.
	}
	\label{fig:SgrAs-flares-lc1}
\end{figure}

They occur randomly (Fig. \ref{fig:SgrAs-flares-lc1}), last from few tens of minutes to 3 hours, 
with light curves that appear roughly symmetrical around their centroid (\Fig\ref{fig:SgrAs-flares-LC}, bottom) 
but that show short time-scale variations down to 100~s,
which indicates that the emission is produced in regions as compact as 10~R$_S$. 

\citet{neilse13, neilse15},
using the 3 Ms XVP \chandra program,
established that the flaring rate is constant at $1.1^{+0.2}_{-0.3}$ per day 
and that the X-ray flux cumulative distribution function (CDF) is fitted,
above the quiescent emission described by a Poissonian process,
with an inverted power-law of slope 
$1.92^{+0.03}_{-0.02}$. 
This is reminiscent of the description of the NIR CDF 
with two laws respectively for the quiescent and flaring states \citep{doddse11, gravit20b}, 
which is still a controversial topic though. 
The power spectral densities measured during flares show the presence of red-noise, 
which makes the detection of significant periodicity quite tricky.
In fact no significant detections of quasi periodicities are found in the X-ray (and NIR) 
light curves of flares, 
in spite of early claims of a signal around 20--30 min \citep[][and references therein]{belang06b}, 
thus, as for NIR, if the orbital motion detected with VLT/GRAVITY is a characteristic of all flares, 
the process does not last enough to produce a significant coherent (quasi) periodicity signal in the emission. 
\citet{ponti15c},
from a re-analysis of all flares measured with \chandra and \xmm until 2015, found that the rate
of bright and very bright events significantly increased starting from summer 2014, 
and tentatively attributed the effect to the G2 2014 periastron passage.
While this result was confirmed by \citet{MosGro17}, 
\citet{bouffa19}, based on a longer  \chandra data set, 
claimed no variation of the X-ray rate,  while \citet{andres22}, using \swift, did measure significant
flaring activity increases in 2006--2007 and 2017--2019 periods, compared to the years 2008--2012.

The average flare X-ray spectra have been studied deeply for several bright flares.
Today the accepted best-fit X-ray spectral model 
is an absorbed inverted power law, 
with index $\Gamma = 2.20 \pm 0.15$ and $N_{\rm H} = (1.60 \pm 0.15)\times10^{23}\rm~cm^{-2}$,
where uncertainties are mainly due to the large absorption that has a strong effect on the slope determination
\citep[e.g.][and references therein]{ponti17}.
In spite of early claims, there is no significant change of slope from one flare to the other, so it does not seem to depend on their strength or duration. 
\nustar stacked spectra show that emission extends up to 79 keV with the same slope without significant spectral break \citep{zhang17}.
The absorption is similar to that observed in close-by objects (e.g.\ the central magnetar and other transients) and
is considered to be entirely due to intervening ISM along the LoS rather than local, and is therefore assumed to be constant.

\subsubsection{NIR flares and multi-wavelenth campaigns}

Since the detection in the NIR, several large multi-wavelength campaigns (MWCs) have been organized 
in order to study the simultaneous properties of the flares at all frequencies.
Given the random nature of the phenomenon, the presence of nearby transients, and the difficulty of coordinating ground and space observatories, not all MWCs were successful or complete. Their results were sometimes controversial, and some still are.
However they have clearly shown that all X-ray events have correlated NIR counterparts,
while the contrary is not true: IR flares show higher rates ($\approx$ 3--4 per day on average) than X-rays,
longer durations and some of them, even if bright, do not show correlated X-ray events.

\begin{figure}[ht]
	\centering
 \includegraphics[
 width=\columnwidth]{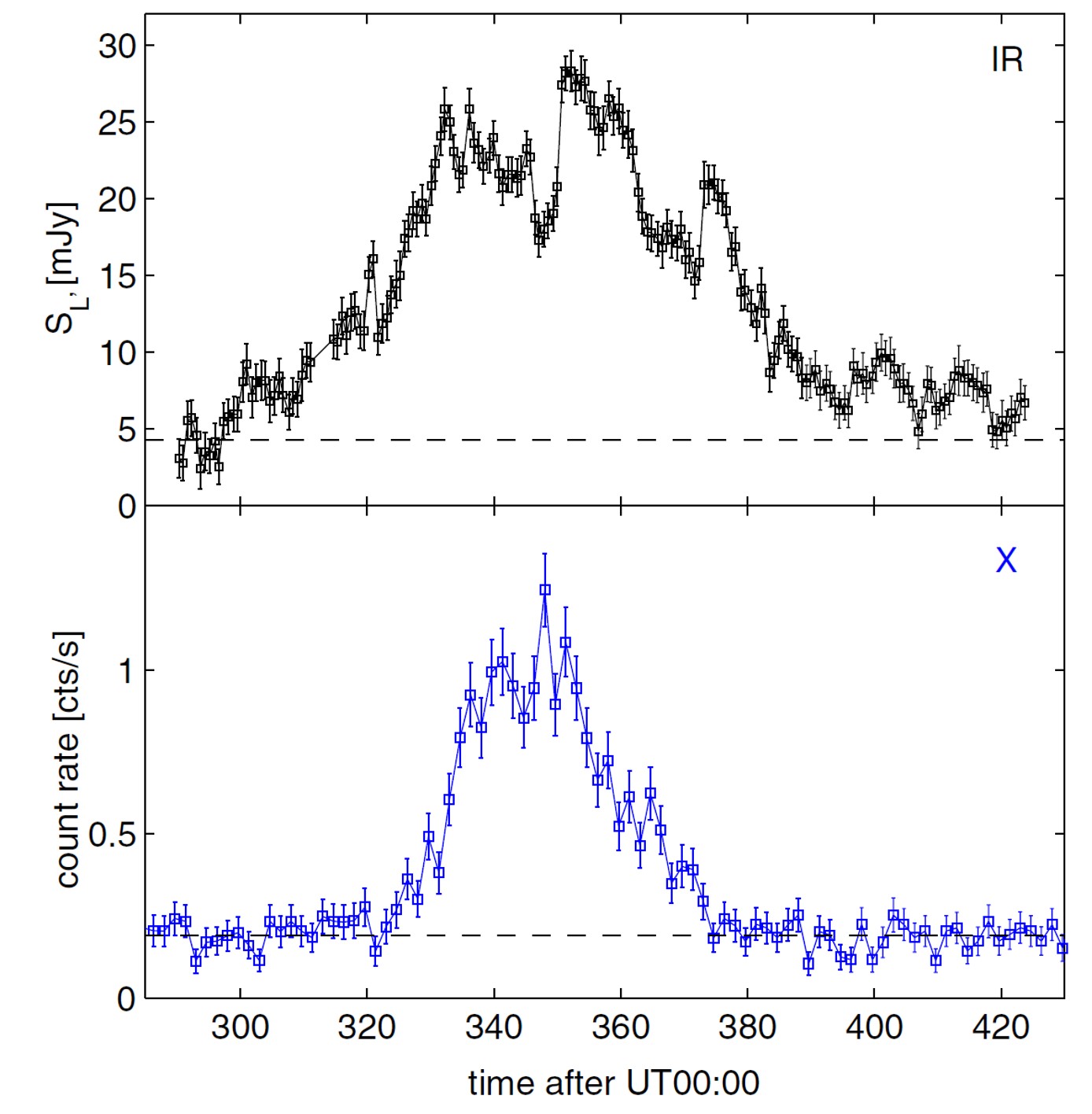}
	\caption{\sgra flare light curves obtained during a MWC:
	X-ray count rate in 2--10~keV (lower panel) and fluxes in the NIR L band (upper panel) 
	versus time (minutes) of the bright 2007 \sgra flare observed with \xmm and VLT/NACO 
	\citep[\AG{adapted} from][]{doddse09}.
	}
	\label{fig:SgrAs-flares-LC}
\end{figure}

Initial claims of X-rays delayed by up to 10 min with respect to NIR ones \citep{yusefz12}
that would suggest an external inverse Compton (EIC) mechanism for the origin of X-rays, 
were not confirmed (see also \Fig\ref{fig:SgrAs-flares-LC}). Instead, recent detailed works
based on \textit{Spitzer} and \chandra simultaneous observations, along with archival data,
concluded that no lags are significantly detected \citep{boyce19, boyce22}.
This rather supports synchrotron (SYN) or Synchrotron Self Compton (SSC) processes 
for the origin of X-rays
(as first proposed by \citet[][]{markof01, liumel02}) instead of EIC (see \Sec\ref{subsubsec:FlSpeMod}). 
Even if they peak at the same time, NIR flares generally start before and end after X-ray ones
and also appear more structured (\Fig\ref{fig:SgrAs-flares-LC}).

Search for significant time delays have been carried out also on sub-mm and radio variability
compared to IR or X-rays, but, up to now, no fully compelling detection is reported.
At these frequencies the variability is much lower, at level of 10--40$\%$, and
a solid identification of flares (and in particular their association to NIR or X-ray ones) is much more complicated.
Delays of peaks between low frequency bands 
were claimed in several occasions, 
and interpreted as due to a synchrotron plasmoid adiabatic expansion,
where the self-absorbed synchrotron peak moves to low energies as the plasma absorption decreases due to expansion.
In spite of the suggestive physical process, and the diagnostic power of the model \citep{vander66}, 
in several occasions it was shown that the predicted flux ratios, 
when considering delays with respect to X-ray peaks, 
do not fit expectations \citep{marron08, trap11}
and most of the claimed lags are not significant and/or the association of peaks at different bands not fully convincing.
In a recent work on \sgra variability in different frequency bands, \citet{witzel21}
claim an overall 30~min delay of the sub-mm variability with respect to the NIR one,
and discuss this in the frame of the plasma cloud adiabatic expansion process,
while additional results on the 2019 MWC \citep[][]{michai24} 
seem also to revive the interest in this process. 

\subsubsection{Constraints from multi-wavelength spectral analyses}\label{subsubsec:FlSpeMod}

Spectral properties over a broad range of frequencies are essential to establish the radiation mechanisms at work.
NIR and X-ray average spectra have different slopes.
In the $\nu$F$_\nu$ units, the average IR slopes are rising with frequency ($\beta \approx 0.3$)  
while the X-ray ones are decreasing ($\beta \approx -0.2$)\footnote{
NIR flux F$_{\nu}$ is often given in Jy units and the power-law slope by the frequency index $\alpha$  
(F$_{\nu} \propto \nu^\alpha$), while X-ray flux is in $\rm ph~cm^{-2}~s^{-1}~keV^{-1}$ units with inverted power-law photon index 
$\Gamma$ (F$_{X} \propto E^{-\Gamma}$),
and are often compared in $\nu$F$_\nu$ ($\rm erg~cm^{-2}~s^{-1}$) or luminosity $\nu$L$_{\nu}$ (erg~s$^{-1}$) units with $\beta$ index 
($\nu$F$_{\nu} \propto \nu^\beta$). Indices relations are then: $\alpha = 1 - \Gamma$, $\beta = \alpha + 1$. 
If photons are produced by an electron population with inverted power-law energy distribution of index p (n$_e \propto \gamma^{-p}$), by SYN or IC, then we also have $\beta = \frac{3 - p}{2}$.}.

However the average spectra are not enough to draw firm conclusions and  
the simultaneous measurements of spectral slope and variability in NIR and X-ray obtained in MWC 
are necessary to properly constrain the radiative models.
Little doubt 
exists for the mechanism responsible for the IR light. Linear polarization and non-thermal slope indicate that 
it consists of thin synchrotron emission produced by a transiently accelerated population of non-thermal (or hot thermal) relativistic ($\approx 1$~GeV i.e.\ $\gamma \approx 10^{2-3}$) electrons, 
at few R$_S$ from the BH, in the magnetic field (10--50 G) of the accretion flow or of the jet.
However, X-rays could be produced by several different mechanisms.
Either the transient electron energies extend up to high-energies ($\approx 100$~GeV, i.e.\ $\gamma \approx 10^{5-6}$), 
giving rise to X-rays also through synchrotron radiation \citep[SYN,][]{doddse09, ponti17}
or X-rays are produced by inverse Compton (IC) processes. 
The IC scattering is done by the same electrons that emit by synchrotron the NIR flare 
which either scatter the same IR photons to X-rays,
through the so called Synchrotron Self Compton process \citep[SSC,][]{eckart08}
or scatter the quiescent sub-mm thermal emission \citep[EIC,][]{yusefz06} 
of the accretion flow\footnote{The case in which the mildly relativistic (10--50 MeV) electrons that produce the quiescent
sub-mm emission scatter the transient synchrotron NIR radiation was also considered \citep{yusefz12} but \citet{doddse09} 
demonstrated that, at least in the optically thin case, 
it is always negligible compared to the primary EIC process (IR-electrons scatter sub-mm photons).}. 
In case of simple SYN, the slope shall extend to the X-rays unchanged, 
but since high-energy electrons cool faster (time scales of $\approx 30$~s) than low-energy ones ($\approx 30$~min), 
a spectral break ($\Delta\Gamma \approx -0.5$) is expected at a break frequency between IR and X-rays.
Because of the fast cooling, this model requires electrons to be continuously accelerated 
during flares, which last much longer than the cooling time of the electrons producing X-rays.
IC generally tends to produce boosted photons with the same spectral shape as the original seed photons, so IR and X-ray shall present the same slopes.

\begin{figure*}
	\centering
 \includegraphics[width=0.4\textwidth]{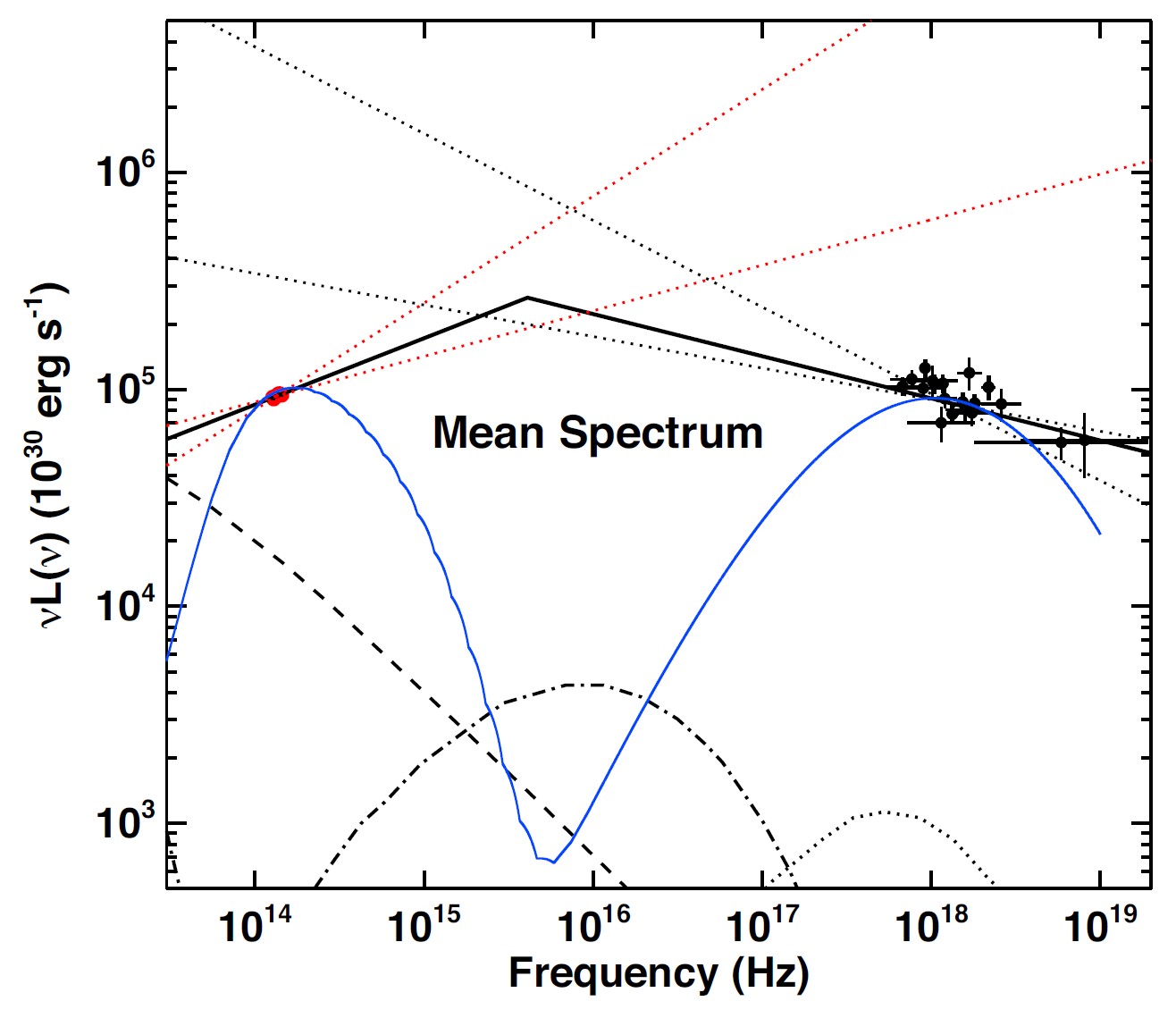}
 ~~~~~~~
 \includegraphics[width=0.54\textwidth]{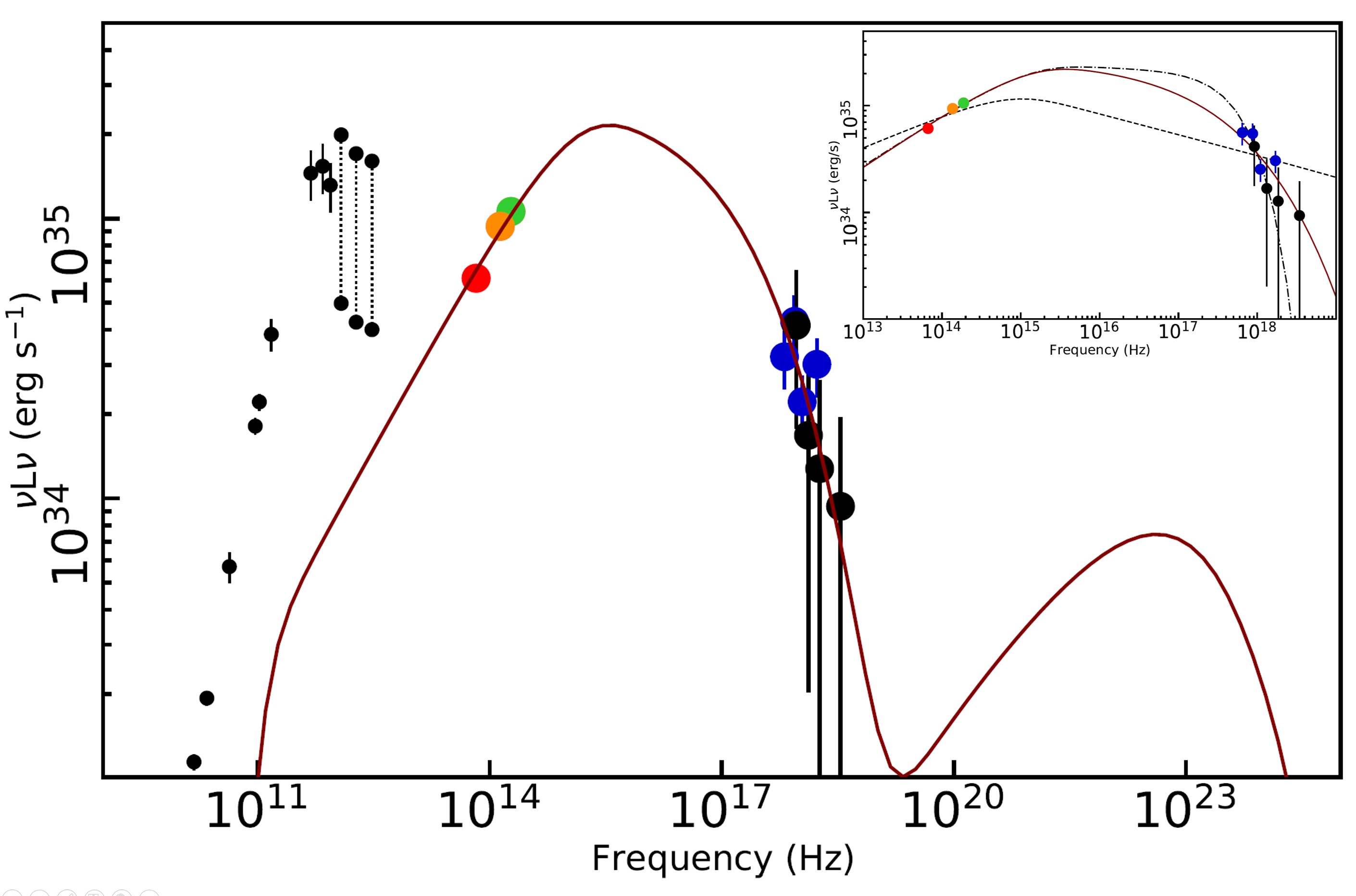}
	\caption{
	The average SEDs of \sgra's flares from successful multi-wavelength campaigns. 
 Left: NIR and X-ray spectra of the 2014 flare observed with \xmm, \nustar and VLT/SINFONI \citep[from][]{ponti17}.
The best fit SYN model with cooling break is shown (solid black line),
with uncertainties in IR (red dotted) and X-ray (black dotted) slopes, 
along with the best-fit SSC model (thin solid blue line)
and various IC components (others). 
Right: M/NIR and X-ray spectra of the 2019 flare observed with \spitzer (red point), VLT/GRAVITY (orange, green), 
 \chandra (blue) and \nustar (black) \citep[adapted from][]{gravit21}.
 Fluxes (in luminosity units) are corrected for extinction and absorption.
 Some not-simultaneous radio to sub-mm fluxes (and variability ranges) of the quiescent emission 
 are also indicated (small black points).
 The best fit SYN model (with cooling break and exponential cut-off) is
 plotted as solid brown line and includes the expected SSC component at gamma-ray energies (too low to be detected by \fermi).
 Inset panel shows a zoom on the IR/X-ray frequencies, with other tested models:
 SYN with break and no $\gamma_{max}$ (dashed line), SYN with break and $\gamma_{max}$ but sharp cutoff (dashed-dotted line). 
 }
	\label{fig:SgrAs-flares-sed}
\end{figure*}

While the first simultaneous X-ray/NIR detection of a weak flare was made in 2003 with VLT and \Chandra \citep{eckart04}, 
the first significant spectral constraints in both bands were obtained in a MWC for the very bright
2007 \sgra flare, 
observed simultaneously with \xmm and the VLT (Fig.~\ref{fig:SgrAs-flares-LC}). 
The VLT/VISIR measurement in the MIR \citep{doddse09, trap10},
simultaneous to a NACO one in the NIR, set an upper limit on the IR slope, 
showing that it had to be very hard compared to the X-ray one.
The authors interpreted the feature as supporting a broad-band synchrotron emission with a cooling break
and showed that IC and SSC could fit the data only if assuming very extreme physical parameters
(magnetic field, tiny emitting zone, huge electron densities).
The first real simultaneous spectra were however obtained in a MWC which involved the spectrometer SINFONI of the VLT
for another very bright flare,
observed simultaneously in 2014 also with \nustar and \XMM \citep[][and \Fig\ref{fig:SgrAs-flares-sed}, left]{ponti17}.
Again, while EIC is completely excluded by these data, the SYN model is preferred over the SSC, 
but it needs two additional features:
the maximum energy ($\gamma_m$) of the electron distribution must vary during the flare
as well as the cooling break frequency. 
These features can explain why IR flares last longer than X-ray ones:
at the beginning $\gamma_m$ is below the limit that gives X-rays, then rises to 10$^{5-6}$ producing the HE flare
and then decreases again below the X-ray threshold.
The varying break frequency can explain why NIR spectra are steeper (redder) at low intensities,
when it moves into the IR range. This variation
indicates a possible magnetic origin of the energy reservoir that powers the events, as
the cooling frequency is related to the magnetic field.

The most recent successful MWC involving GRAVITY, in coordination with \textit{Spitzer}, \nustar and \chandra,
caught a very bright NIR flare in 2019 associated with a moderate X-ray one \citep{gravit21}. 
The detailed time-resolved spectral analysis of this excellent data set again support the SYN model if it is assumed, 
in addition to the features identified previously (variable cooling break and varying $\gamma_m$ of the particle distribution), 
also an exponential decrease of the electron distribution above $\gamma_m$ rather than a sharp cut-off
(\Fig\ref{fig:SgrAs-flares-sed}, right).
Both EIC and SSC models are ruled out because they cannot fit the difference between X-ray and NIR slopes
and their relative fluxes, unless accepting really extreme parameters and notably an extreme magnetic field.
In addition to the poor fit, the authors argue that the extreme field implied by the IC models,  
would decrease the cooling time so much that variations shorter than tens of seconds shall be observed in both IR and X-ray
light curves, while they are not.
The only IC model that could fit the data is the one in which also IR are due to SSC  \citep[SSC-SSC,][]{witzel21} from
electrons producing sub-mm flare by synchrotron
(if IR are not due to SYN, B can stay low even though $\gamma$ is low)
but with the caveat that electron densities need to be very high ($\approx 10^{10}\rm~cm^{-3}$), 
which is at odds with the expected densities of the accretion flow \citep{gravit21}. 

\subsubsection{Summary and possible origins of the events}

In conclusion, at present, radiation mechanisms of the \sgra flares seem better explained 
if one uses simple single-zone models,
by those where both IR and X-rays are produced by synchrotron emission of transiently accelerated electrons, 
with varying maximal energy of the electron distribution and exponential cut-off above this energy,
possibly powered by the magnetic field, as in magnetic reconnection events.
In this picture \citep{ponti17, gravit21} the magnetic field 
decreases during the event and then goes back to its value of the steady accretion flow at the end of the flare.
This variation produces a shift of the cooling break frequency down to IR frequencies, 
producing the steeper IR slopes, detected at the beginnings and ends of the events, when the intensity is low.

In recent works on the 2019 MWC \citep{boyce22, michai24} 
the SYN (for sub-mm) - SSC (for NIR) - SSC (for X-rays) radiative model \citep{witzel21},
associated to a plasma cloud adiabatic expansion, has been re-proposed. 
However the data of the most recent MWC (2024),
when it was possible for the first time to detect a flare in MIR with the JWST, in absence of an X-ray flare 
but along with a 40 min delayed sub-mm event, 
can again be explained with a pure SYN model with gradual cooling of electrons accelerated by
magnetic reconnection and/or magnetized turbulence \citep{vonfel25}. 
As \citet{gravit21} conclude, certainly a fully coherent and complete model of \sgra flares
explaining both spectral and variability properties in all frequency ranges 
is still lacking.

The primary origin of the phenomenon is also not yet fully understood and
several ideas have been proposed to explain the process that powers the flares. 
While some consider just energy transfer from the plasma magnetic field to the particles, others 
evoke temporary larger accretion rates, 
that eventually power either MHD instabilities, shocks or magnetic re-connection in the inner flow \citep{yuan03}, 
or the activity of the jet \citep{markof01},
as for example accretion of stellar-wind higher density clumps, of asteroids, comets 
and planets \citep[][and references therein]{zubova12}, or of gas captured from transiently approaching objects.
As mentioned above, searches for varying X-ray flare rates have been carried out
in order to link the phenomenon to the known variations of the environment, 
like the periastron passages of the Gs gaseous objects (G1 in 2001, G2 in 2014) 
or of the closest approaches of the windy star S2 (2002, 2018).
No firm conclusions were reached from these X-ray studies and,
while no variations of the IR flaring rate were reported by \citet{witzel18},  
a burst of bright NIR flare rate was detected in 2019 with both Keck and VLT (\Sec\ref{subsec:SgrA}) 
and confirmed then by \citet{weldon23} who 
discussed the possible link to the G2 periastron passage.
Considering that a variation of the accretion rate at the Bondi radius takes a while to 
produce repercussions in the inner accretion flow, 
since the estimated viscous timescale is of 5--10 years, 
it is difficult to directly link these possible rate variations to the original accretion event, 
but it is certainly worth to continue the search for such connections.

\subsection{X-rays and hard X-rays from the central 20 pc} \label{subsec:SGCE}

\subsubsection{Discrete X-ray sources} \label{subsubsec:SGRCEds}

Several persistent X-ray sources are visible in the X-ray images of the central 20~pc \citep{bagano03, maeda02}.
Few weak objects are detected within 30$''$ from the BH (\Fig\ref{fig:SgrAstar}, left).
The most interesting feature is certainly the cometary-shaped PWN also detected in radio as G359.95--0.04 
only 10$''$ north from \sgra, which is a serious counterpart to the source of the TeV central emission (\Sec\ref{subsec:GRCS}).
Two other point-like sources have no stellar counterpart and were classified as weak XRBs.

A number of transient point-like sources have also been detected during the numerous observations of the GC
with \Chandra, \XMM and the other observatories, particularly \swift,
the most remarkable of which are certainly 
the 2004 transient (CXOGC J174540.0--290031, see \Fig\ref{fig:SgrAstar} left)
discovered at only 2$''$ from \sgra and seen to display microquasar-like radio jets
\citep[][and references therein]{muno05a} 
and the magnetar SGR J1745--2900, located 2.4$''$ from the SMBH, 
detected in outburst in 2013 \citep{kennea13} and which is now back to quiescence.
This cusp of X-ray transient sources will be discussed below along with the relevant source populations (\Sec\ref{subsec:POPS}).

On larger scales, more extended features include Sgr~A East (\Sec\ref{subsubsec:SgrAEast}), hard X-ray sources (\Sec\ref{subsubsec:hxrsrc}) and also another relevant X-ray jet feature detected with \Chandra. 
This is the linear filament named G359.944--0.052, 
located about 20$''$ from \sgra, and aligned towards the BH in the S-E direction,
which displays some peculiar spectral properties, no variability, and
which has been proposed as a possible X-ray counterpart of a parsec-scale jet from \sgra
\citep[see][and references therein]{zhu19}. 

\begin{figure*}
	\centering
 \includegraphics[width=0.45\textwidth,trim=0mm 0mm 0mm 0mm, clip]{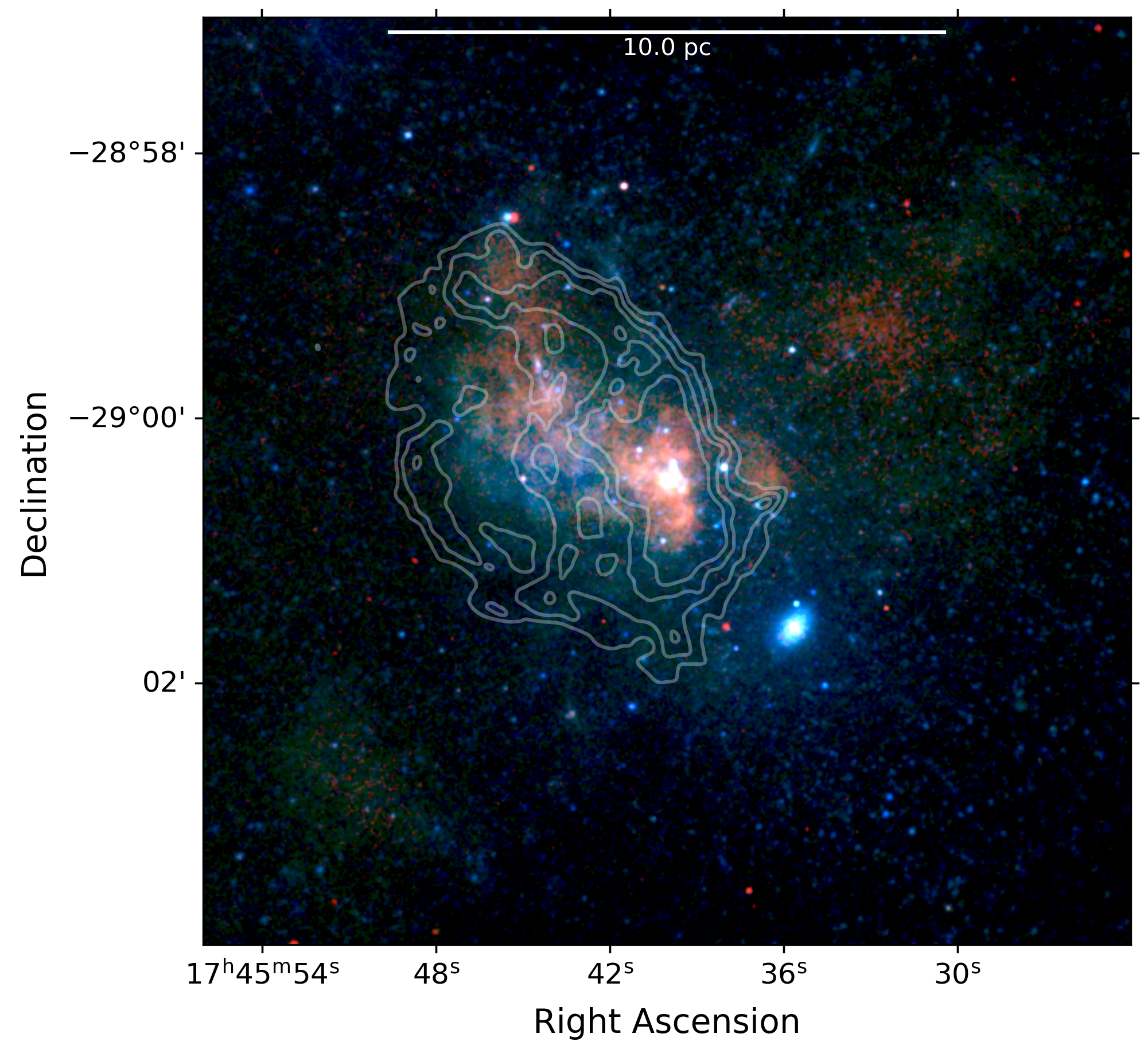}
 ~
  \includegraphics[width=0.53\textwidth]{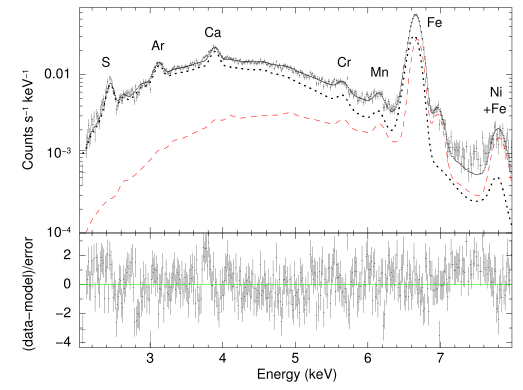}
	\caption{Left: \Chandra 3-color X-ray image of Sgr~A East (red: 1.5--2.6 keV, green: 2.6--4.5 keV and blue 4-5--7.5 keV) with the 6~cm MeerKAT radio contours of its radio shell overlayed. The hot thermal emission fills the interior of the radio shell. The Cannonball is visible on the northern part of the shell. Right: The X-ray spectrum of Sgr~A East central part with the best-fit thermal models \citep[from][]{zhou21}.
	}
	\label{fig:SgrAEast}
\end{figure*}

\subsubsection{Sgr A East X-ray emission}\label{subsubsec:SgrAEast}
A distinct bright thermal emission from a hot plasma with prominent lines of S, Ar, Ca, and Fe ions,
appears coincident with the inner region of the non-thermal radio shell of the Sgr~A East SNR 
(\Fig\ref{fig:SgrAEast}).
First detected at low resolution with \ROSAT, \ASCA and \BeppoSAX 
\citep{koyama96, sidmer99} Sgr~A East was clearly resolved from the GC diffuse X-ray components 
with \Chandra \citep{bagano03, maeda02} and then studied deeply \citep{park05}, also with \XMM \citep{sakano04}
and \Suzaku \citep{koyama07a}.
With a total 2--10 keV luminosity of $8 \times 10^{34}\rm~erg~s^{-1}$, the SNR emission is rather soft
and characterized by a two-temperature (1~keV and 5~keV) plasma with high metal abundance 
at the center of the radio shell and by one temperature (1.3~keV) plasma with solar abundances 
in the external northern part. This has been interpreted as emission
dominated by stellar ejecta heated by a reverse shock in the center 
and by hot swept up ISM close to the shell.
Sgr~A East was then classified as a single metal-rich mixed-morphology rest of 
a core collapse SN of a $< 20\rm~M_{\odot}$ star 2--4~pc behind \sgra. 
The SN was thought to have occurred about 10~kyr ago,
releasing about $5\times10^{50}$~erg and possibly leaving behind 
the high-velocity neutron star (NS) remnant, now seen as the X-ray 
\citep{park05}  and radio 
PWN known as the Cannonball, just outside the north side of the radio shell.

However, since the recent detection of Mn, Ni and Cr lines with \Suzaku \citep{ono19},
a younger (1--2~kyr) event, possibly even due to a thermonuclear white dwarf (WD) deflagration 
of a type Iax SN, has been considered as origin of the remnant \citep{zhou21},
in order to explain the \Chandra X-ray feature the ridge, at the edge of the central cavity,  
as due to the interaction of the SNR ejecta and the stellar-wind outflow from the YNC \citep{zhang23}. 
In this model the Cannonball is not the NS remnant of the SN, 
the central plasma is not due to ejecta heated  
by conventional collisional ionization of the SNR reverse shock 
(that needs more than 2~kyr to develop) but 
rather a recombination plasma, photo-ionized by a large outburst of \sgra in the past \citep{ono19}. 

Whatever the type and age of the original SN, 
Sgr~A East is a very important object in the HE phenomena context 
given its unique position close to the SMBH.
Note that recent modelling of the SNR and its interaction with the GC environment \citep{zhang23} 
does not support the initial hypothesis that it may have caused 
a large rise in the BH mass accretion several centuries ago as the swept-up dense material passed through \sgra,
and does not explain either the present BH starving 
due to the ISM having been cleared out from its vicinity.
Its interaction with the 50~km~s$^{-1}$ cloud and, towards the center, with the CND 
and even with the YNC stellar-wind outflow
can lead to prominent CR acceleration and production of gamma-ray emission 
by their subsequent interaction with the molecular material.
Very recently \Xrism has confirmed the presence of over-ionized plasma within Sgr~A East, 
interpreted as due to a past outburst of \sgra \citep{xrismc25}.
A conclusive interpretation of the Sgr~A East SNR is therefore still to be drawn, and may benefit from future high-spectral resolution X-ray telescopes with higher angular resolution (\Sec\ref{subsec:HE-Recent}).

\subsubsection{The central hard X-ray sources}\label{subsubsec:hxrsrc}
In hard X-rays, the \INTEGRAL surveys revealed a significant central ($R < 20$~pc) emission, 
extending up to $> 100$~keV,
compatible with the 12$'$ FWHM PSF of the IBIS instrument, centered at only 1$'$ from \sgra \citep{belang04,belang06a}, 
but clearly too bright to be compatible with the extension of the \sgra spectrum at high energies. 
The nature of this source (IGR~J17456--2901) 
was clarified by \nustar \citep{mori15}, which 
could unambiguously identify its harder emission with the central PWN
G359.95--0.04 detected by \Chandra (discussed in \Sec\ref{subsec:GRCS}) 
and the softer one (predominant in the 20--40 keV band) 
with a diffuse emission that the authors attributed to an unresolved population of magnetic 
cataclysmic variables (mCV) \citep[][see also \Sec\ref{subsec:DXHC}]{perez15},
as initially suggested by \citet{krivon07}.
Above 10 keV, \nustar did not find non-thermal X-rays from Sgr~A East but could detect,
in addition to the \sgra flares mentioned above,
few other hard X-ray features in the central 10~pc, none attributed to PWNe.
G359.97--0.038 detected up to 40~keV, is located just outside the Sgr~A East shell, close to the Plume
and it is probably illuminated by non-thermal bremsstrahlung or inverse Compton processes of electrons
originated by the interaction between the shell and the 50~km~s$^{-1}$ cloud \citep{nynka15}.
Another filament appears associated to a radio feature (G359.89--0.08)
and was attributed to synchrotron emission from $\approx100$~TeV electrons trapped in magnetic flux tubes, 
which could be generated 
by the interaction of relativistic protons accelerated by \sgra or by the Sgr A East SNR 
with the nearby 20~km~s$^{-1}$ cloud \citep{zhang14}.
Other non-thermal features discovered with the \nustar GC survey outside the Sgr~A complex are discussed below
(\Sec\ref{subsec:PONT}).


\section{X-ray source populations and discrete structures in the CMZ} \label{sec:XRPO}

Typical X-ray point-like sources detectable at a distance of 8~kpc and under strong absorption are accretion-powered systems, including white dwarfs, neutron stars or black holes, 
which are tracers of the old stellar population but with open questions on the specificity of the compact remnant population at the GC (\Sec\ref{subsec:POPS}). 
Part of these compact sources and other energetic phenomena also accelerate particles that are responsible for more extended features detected in X-rays (\Sec\ref{subsec:PONT}). 
In addition, massive stars with strong winds, such as WR or O-stars, tracing recent star formation processes, 
can also be detected as X-ray point sources and be associated with extended high-energy thermal emission (\Sec\ref{subsec:POTH}).

\subsection{The CMZ populations of point-like X-ray sources} \label{subsec:POPS}

\begin{figure}[ht]
		\centering
 \includegraphics[width=\columnwidth]{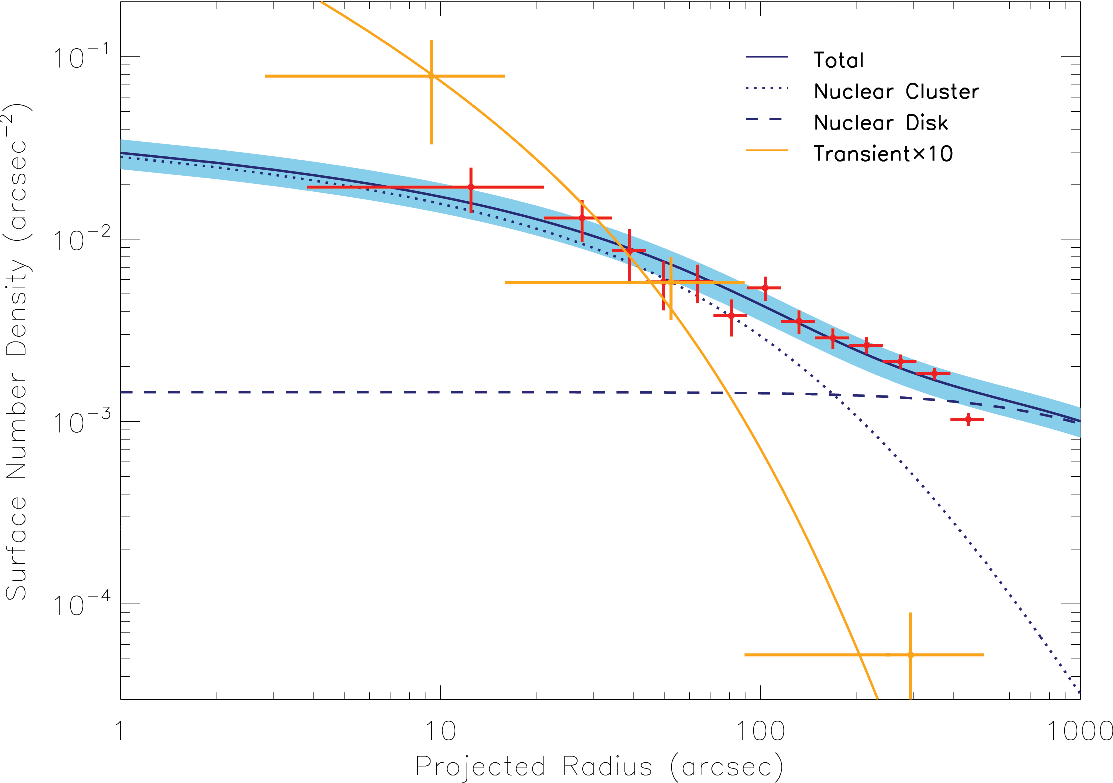}
	\caption{Surface density profile of X-ray sources with a luminosity  $L_{\rm2-10keV} \gtrsim 6\times10^{31}\rm\,erg\,s^{-1}$ in the inner 40~pc of our Galaxy (in red), tracing a population of mCVs. The best fit model derived from the stellar mass distribution (solid blue line, and uncertainties in shaded blue) includes the contributions from the nuclear star cluster (dotted line) and from the nuclear disk (dashed line). The surface density profile of known transients (multiplied by 10, in orange), tracing a population of quiescent LMXB, appears much steeper than the stellar distribution. Figure adapted from \citet{zhu18}.
	}
	\label{fig:pop-profile}
\end{figure}

In the inner region of our Galaxy, the X-ray point source distribution follows the stellar population seen in the near-infrared. The highest density is measured towards the central few parsecs, where the contribution from the nuclear star cluster dominates, and smoothly decreases outwards, following the more-extended nuclear disk component \citep[see][and \Fig\ref{fig:pop-profile}]{muno06,hong09,zhu18}.
The majority of these sources are persistent and most likely associated with a population of magnetic cataclysmic variables (mCVs, \Sec\ref{subsubsec:mCVs}). Known transients reveal a sparser population of X-ray binaries (\Sec\ref{subsubsec:XBs}), and rarer classes of point sources may be part of additional populations that remain to be constrained (\Sec\ref{subsubsec:otherpop}).

\subsubsection{Magnetic cataclysmic variables} \label{subsubsec:mCVs} 

Existing deep \chandra observations covering the CMZ revealed close to 10,000 point sources within this region \citep[after excluding foreground sources,][]{muno09,zhu18}. The X-ray information available for the vast majority of these objects is insufficient to classify them individually, but their statistical and cumulative properties support the identification of a large fraction of these sources as intermediate polars (IPs), a subclass of magnetic CVs, i.e.\ white dwarfs accreting from a low-mass companion \citep[for a review on the X-ray properties of these objects, see][]{mukai17}. Indeed, most of the X-ray point sources detected in the CMZ have a 2--10~keV luminosity between $10^{31}$ and $10^{34}\rm\,erg\, s^{-1}$, and do not exhibit strong variability on either short or long timescales \citep[less than 15\% of the sources were detected as significantly variable,][]{muno09,zhu18}. Among them, few sources do have periodic variability detected, with time periods compatible with typical IP spin periods \citep[e.g.][]{muno03}. The surface density of persistent sources follows the stellar mass traced by the near-infrared starlight (see \Fig\ref{fig:pop-profile}) with a ratio compatible with the abundance of mCVs in the solar neighborhood, supporting the identification of a large fraction of these sources as mCVs. 

Their cumulative spectrum reveals a hard continuum and prominent iron emission lines at 6.4 (Fe\Ka), 6.7 (Fe\,\textsc{xxv}~He$\alpha$) and 7.0~keV (Fe\,\textsc{xxvi}~Ly$\alpha$), which are also typical of mCVs \citep[e.g.][]{xu16}.
Further investigation for the X-ray sources overlapping the NSC 
in the central 20~pc demonstrates that the equivalent width of the lines and their flux ratio are different for sources above and below $L_{\rm 2-10keV} = 6\times10^{31}\rm\,erg\,s^{-1}$. Above this threshold, the properties of the cumulative spectrum is fully compatible with what is expected for IPs while a larger fraction of non-magnetic CVs, namely dwarf novae, may be present among fainter sources. 

Contribution from populations of sources with no Fe\,\textsc{xxv} emission line (i.e.\ other than CVs and coronally active binaries, ABs) 
should affect the value of the equivalent widths measured in the cumulative spectra, and therefore cannot exceed few percents of the total amount of persistent sources detected in the 2--10~keV range (see also \Sec\ref{subsubsec:XBs}). Above 10~keV, only the brightest part of these populations have been detected, with few dozens of sources resolved by \nustar \citep{hong16}. IPs likely contribute to at least 40--60\% of this bright hard subsample, along with a possibly higher fraction of X-ray binaries (see \Sec\ref{subsubsec:XBs}). The faintest part of the CV population is yet to be resolved and likely contributes to several components of diffuse hard X-ray emission detected at various spatial scales around the Galactic center (see \Sec\ref{subsec:DXHC}).

\subsubsection{Low mass X-ray binaries} \label{subsubsec:XBs}

A population of X-ray binaries at the Galactic center is expected from stellar evolution and stellar dynamics, and should be strongly linked to the total number of NS and BH present in these inner regions. The XRB population 
has primarily been detected through a few bright persistent sources and dozens of X-ray outbursts. Across the CMZ, about twenty five transients reaching luminosities $\rm L_{\rm 2-10keV}\gtrsim 10^{34}\rm\,erg\,s^{-1}$ have been identified. Most of them are located in the central 50~pc and have been closely monitored by \swift/XRT starting from 2006 \citep[][see also \Sec\ref{subsec:XTel}]{muno09,degena15}, which now provides the complete sample of all transients from this region having recurrence times below fifteen years. Their radial distribution is different from the bulk of mCVs described previously (see \Fig\ref{fig:pop-profile}). It is steeper 
and consistent with an overabundance in the central parsec, suggesting an important role of dynamics (through mass segregation and stellar capture) in the formation of, at least, part of these systems \citep{muno05a,zhu18}. 

Sources reaching luminosities above $10^{36}\rm\,erg\,s^{-1}$ with outbursts lasting for several months have been classified as low mass X-ray binaries (LMXBs), i.e.\ NS or BH accreting from a low-mass stellar companion \citep[for a review on the properties and demographic of these systems, see][]{sazono20}. 
The GC \swift/XRT monitoring also revealed very faint X-ray binaries (VFXB) that have outbursts lasting less than about a month with luminosities in the $10^{\rm34-36}\rm\,erg\,s^{-1}$ range, and recurrence times similar to the ones of brighter LMXBs \citep{bahdeg23}. 
They likely constitute an inhomogeneous subclass of LMXBs in which the accretion is limited, due to e.g.\ a small accretion disk, a high magnetic field or wind-feeding \citep[][and references therein]{degena15}. However, the high extinction towards the CMZ makes it hard to further classify these faint GC transients. Among all bright or faint and persistent or transient sources, NS-LMXBs have been securely identified through the detection of type I X-ray bursts or of coherent pulsations \citep[see e.g.][]{degena12}. The identification of BH-LMXBs is not as straightforward but few BH candidates have also been proposed \citep[see e.g.][]{muno05b,mori19}.

In quiescence, LMXBs may be detected as very faint non-thermal point sources. Investigation of the spectral properties of \chandra sources close to the supermassive black hole revealed twelve point sources with power-law like spectra ($\Gamma\sim2$, and no iron emission lines) in the central parsecs, contrasting with both the spatial distribution and the spectral properties of mCVs \citep{hailey18,mori21}. None of these sources has ever been detected in outburst but several of them show moderate long term variability. Their exact classification has been debated. They are presented as most likely BH-LMXBs \citep{mori21} but NS-LMXB with recurrence time longer than the ones that have been observed at the GC so far, or millisecond pulsars (MSP), cannot be ruled out for at least part of these systems.

Mass segregation around SMBHs predicts the high density of massive compact remnants in the inner regions traced by detected LMXB. However, the number of NS- and BH-XRBs could be altered with respect to isolated ones, by e.g.\ SN kicks \citep{bortol17} or 3-body encounters \citep{genero18}, leaving important uncertainties on the extrapolations that can be done from the observations. Current predictions state that more than a hundred of BH-LMXB could be present in the central parsec, together with thousands of isolated BHs \citep[][and references therein]{mori21}, while the population of NSs should be also closely related to the population of pulsars (see \Sec\ref{subsubsec:otherpop}).

\subsubsection{More elusive pulsar populations} \label{subsubsec:otherpop}

One X-ray transient, SGR J1745--2900, detected within 0.1~pc from \sgra was not classified as a LMXB but as a magnetar \citep{mori13,rea13}, i.e.\ an isolated neutron star powered by a very strong magnetic field \citep[see][for a review on magnetars]{meregh15}. This source went into outburst in 2013, reaching an X-ray luminosity of few 10$^{35}\rm\,erg\,s^{-1}$, followed by a slow exponential decay over several years \citep{rea20}. The NS has a spin period of $\sim3.76\rm\,s$ detected both in X-ray and in radio \citep{mori13,shajoh13}. Its dispersion measure is the highest ever measured towards a pulsar, $DM = 1778\pm3\rm~cm^{-3}~pc$, and together with its high rotation measure is indicative of a strong magnetic field in its surrounding medium, $B \gtrsim 50~\rm\mu G$ or higher \citep[see discussion by][]{eatoug13}. This makes it the first X-ray pulsar unambiguously identified close to \sgra\footnote{There may be additional pulsars within identified NS-LMXB (see \Sec\ref{subsubsec:XBs}). GRO~J1744--28 is currently the only known pulsating NS-LMXB towards the GC, but its LoS 
distance has been debated \citep{degena12,sanna17,dorosh20}.}. 
Since magnetars are relatively rare, this finding could be consistent with former predictions of a large population of pulsars around \sgra \citep{pfaloe04} and, on larger scale, across the CMZ \citep[][see also \Sec\ref{subsec:GRDM}]{wharto12,spitle14}. The objects with large spin-down powers are expected to emit in X-rays and could account for a small fraction of unidentified X-ray point sources \citep{possen02,bertea21}. The detection of their spin would be the definitive proof of their existence. However, due to the scattering of radio waves off the dense interstellar medium, extensive radio pulsation searches have only revealed a handful of long period pulsars \citep{johnst06,deneva09,wongph24} and one MSP \citep{lower24} in the  CMZ (distance estimates based on dispersion measures).
Only one of them, PSR~J1747--2809, has a counterpart identified in X-ray (see \Sec\ref{subsubsec:G09+01}). 
Across the CMZ, additional extended X-ray features, such as PWNe, 
are further hints of the presence of X-ray emitting pulsars within the GC environment (see \Sec\ref{subsec:PONT}).

\subsection{Filaments and pulsar wind nebulae} \label{subsec:PONT}

High spatial resolution X-ray observations of the CMZ revealed a total of about thirty elongated features, with lengths of the order of 1~pc or more, referred to as X-ray filaments \citep[see the review by][for the most complete list of known filaments in the central $1^\circ \times 0.5^\circ$]{koyama18}\footnote{Additional filaments, characterized by a strong 6.4~keV emission line, are also detected. They are ignored through this section and will be discussed in \Sec\ref{subsec:DXNT}.}. Most of the known X-ray filaments are located within 12~pc from \sgra, possibly due to the highest exposure available towards this inner region \citep[][see also \Sec\ref{subsec:SGCE}]{johnso09,koyama18}.

Their exact morphology varies from one filament to the other, in terms of length, width, curvature and flux gradients. Several works have highlighted the cometary shape of a significant fraction of these filaments, part of them even having a point source (often called the head) associated with their extended emission \citep[their tail, see e.g.][]{lu08,johnso09}. The majority of known X-ray filaments has spectral properties consistent with a non-thermal emission best fit by an absorbed power-law with $\Gamma\sim1$--$3$ and a total 2--8~keV luminosity within $10^{32}$--$10^{34}~\rm erg~s^{-1}$ \citep{muno08,lu08,johnso09}. Four of them have also been detected above 10~keV by \nustar \citep{mori15}, including two that also coincide with gamma-ray sources (see \Sec\ref{subsec:GRDS}). 
Despite their resemblance to the numerous non-thermal radio filaments, which are explained by the synchrotron radiation of GeV electrons trapped by enhanced magnetic features close to the Galactic center \citep[e.g.][see also \Fig\ref{fig:MeerKat-CMZ-index}]{heywoo22}, only a handful of X-ray filaments have radio counterparts \citep[e.g.][]{lu03,muno08,ponti15b}. 

The origins of the non-thermal X-ray filaments are diverse. The properties listed above are fully consistent with the ones of PWNe 
\citep[see][for a review]{gaesla06,kargal17} 
having filamentary-like structures due either to the motion of the young pulsars or to strong magnetic field in their environment. In agreement with the massive star formation rate, it has been proposed that a large fraction ($\sim20$ out of the 34 X-ray filament detected within 20~pc of \sgra) of the non-thermal X-ray filaments could be PWNe \citep[][]{muno08,lu08,johnso09}.
However, no pulsation detection ever proved the presence of a pulsar in any of these sources (see also \Sec\ref{subsubsec:otherpop})\footnote{PSR J1747--2809 (see \Sec\ref{subsubsec:G09+01}) is the only exception but it is outside of the central $1^\circ \times 0.5^\circ$ considered by filament studies discussed in this Section.}. 
Among the most secured PWNe identifications are G0.13--0.11 \citep[][]{zhang20,churaz24} and G359.95--0.04 \citep{wang06a,mori15}, which will be discussed along with their possible gamma-ray emission in \Sec\ref{subsec:GRDS}, 
and the Cannonball \citep[][]{park05,nynka13} already mentioned in \Sec\ref{subsec:SGCE}.
Other mechanisms can provide similar X-ray signatures. One example is the interaction of SNRs with surrounding molecular clouds, which has been proposed for one filament close to Sgr A East \citep[][see also \Sec\ref{subsec:SGCE}]{nynka15}. A second example is the interaction of TeV electrons with magnetic structures (similar to radio NTF), which has been proposed for few filaments, such as Sgr A-E \citep[e.g.][]{zhang14}. 
In addition, several elongated features close to \sgra could also be linked to the supermassive black hole activity (see \Sec\ref{subsec:SGCE}).

Finally, thermal spectra including a 6.7~keV emission line have also been reported for at least one X-ray filament. Ram-pressure or magnetic field confined stellar winds from massive stars could explain such thermal structures \citep[see e.g.][]{johnso09}.

\subsection{Massive stars and star clusters} \label{subsec:POTH}

Star formation occurring within the CMZ (see \Sec\ref{subsubsec:SFR}) sustains the presence of numerous massive stars, namely Wolf Rayet and O supergiants, characterized by fast stellar winds. 
When colliding, these winds can produce a harder thermal emission than their parent star and be detected in X-rays despite the Galactic center strong LoS absorption \citep[see][for a review on the X-ray emission from interacting wind massive binaries]{raunaz16}.

At the Galactic center, the most obvious signatures of these high-energy processes are the three young and massive stellar clusters: the Arches, the Quintuplet and the YNC clusters 
(their main characteristics are given in \Sec\ref{subsubsec:SFR}). The Arches and Quintuplet clusters have been resolved by \chandra into a collection of point sources (4 and 5, respectively, if ignoring one foreground source for the Quintuplet cluster) together with an extended diffuse emission
\citep{yusefz02b,lawyus04,wang06b,tsujim07}.
The spectral shapes of the point sources are consistent with emanating from a hot diffuse gas having a twice-solar abundance and temperatures ranging from kT~$\sim2$~keV to $\sim9$~keV, together with large 
foreground absorption.  The brightest of these sources is within the core of the Arches cluster and has a luminosity $L_{\rm0.3-8keV}\sim10^{34}\rm~erg~s^{-1}$ \citep[e.g.][]{wang06b}. 
As several other X-ray point sources across the CMZ ($\sim 30$ in total), the NIR counterpart of the cluster point sources has been identified as WR or O supergiants \citep[][and references therein]{mauerh10}. Because isolated such massive stars cannot produce the X-ray emission detected, these sources have instead been classified as likely colliding wind binaries \citep[these X-ray sources would represent less than about 5\% of the massive star content of these clusters, see e.g.][]{wang06b,mauerh10}. This type of source could also explain the 70\% X-ray flux enhancement of at least 4 days detected towards the Arches cluster in 2007 \citep{capell11}.

In the Arches and Quintuplet clusters, the diffuse emissions has a thermal\footnote{The large-scale non-thermal emission reported in these works for the Arches cluster is now associated with the past activity of \sgra, see \Sec\ref{subsec:DXNT}.} 
origin with a strong 6.7~keV emission line and temperatures consistent with their point source contents  \citep[e.g.][]{yusefz02b,lawyus04,wang06b,tsujim07}. The winds from all massive stars present in these two massive young clusters can also be responsible for, at least, a significant part of the diffuse thermal emission detected \citep[][]{wang06b}. However, collective emission of point sources below the detection threshold should also contribute. This likely includes fainter colliding wind binaries, as well as young stellar objects.
X-ray emitting young stellar objects are tracing the young low-mass stars \citep[0.3--3~M$_\odot$,][]{feigel05} and the clusters' diffuse emission can be used to put constraints on these lighter populations and thereby on their stellar initial mass function (IMF). 
These studies highlight a deficit of X-ray diffuse emission compared with what would be expected from a standard IMF, in all GC clusters, including the YNC \citep{naysun05,wang06b}. In agreement with other works focusing on the stellar populations detected at lower wavelengths, this has been interpreted as a top-heavy IMF, i.e.\ an overabundance of massive stars in the main clusters of the CMZ \citep[see also][]{paumar06,hosek19,hussma12},
which could be due to physical and dynamical properties of the CMZ \citep{chadum24}.

Across the CMZ, close to $100$ massive stars are known, including $\sim30$ with an X-ray counterpart \citep[][]{mauerh10,clark21}. Most of them are in apparent isolation. However, stellar clusters could be responsible for part of these isolated massive stars over a significant fraction of the CMZ, due to their proper motion, Supernovae kicks and tidal stripping from the CMZ dense environment \citep[e.g.][]{habibi14,clark21}.
While colliding wind binaries have been the favored scenario to explain the X-ray counterpart from these massive stars, quiescent high mass X-ray binaries 
could also account for few of these sources and possibly many more below the detection threshold of current X-ray surveys.

Through their winds and violent death as Supernovae and the plausible subsequent CR accelerations, these massive stars contribute to the energetics of the central regions.
Constraints on the SN rate and the high-energy properties of the SNR detected within the CMZ will be discussed in \Sec\ref{subsec:DXSC}, while their link with gamma-ray emission is presented in \Sec\ref{sec:GRCR}.

\begin{figure*}
		\centering
		\includegraphics[width=\textwidth,trim =0mm 0mm 16mm 0mm, clip]{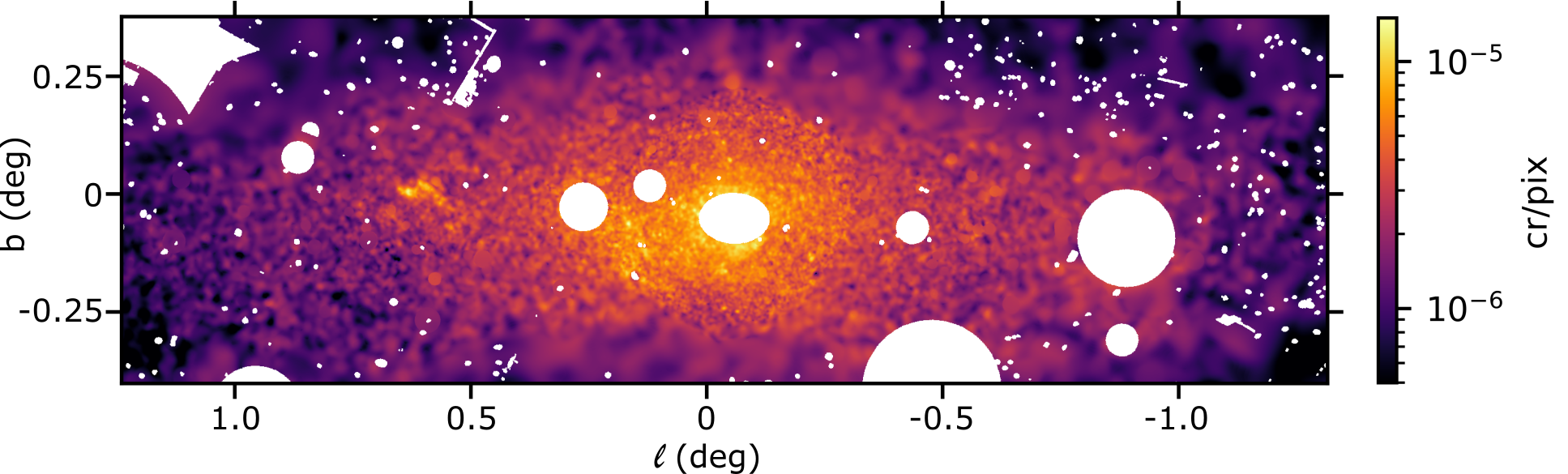}
		\caption{Hot thermal emission detected by \xmm within the CMZ. The color highlights the surface brightness of the signal  
		detected in the 6.62--6.80~keV band in logarithmic scale, after correction for contamination from non-thermal emission estimated from the 6.3--6.5~keV band. An adaptive smoothing has been applied to the mosaic and regions around bright sources have been excluded to avoid contamination (in white). Figure slightly adapted from \citet{anasta23}.}
		\label{fig:hotplasma}
\end{figure*}

\section{Diffuse X-ray emission from the CMZ and beyond}\label{sec:DXRE}

The Galactic ridge X-ray emission (GRXE, see also \Sec\ref{subsec:EarlyXG}) is characterized by a two-temperature plasma  \citep[see e.g.][]{yamauc96,kaneda97}, a softer component with a temperature of about 0.8 keV and a hotter component of 7 keV. The softer component is understood as being caused by interstellar gas shocked by supernova remnants and stellar winds \citep[][\Sec\ref{subsec:DXSC}]{kaneda97}. The hotter component is understood as being caused by the cumulative emission of faint unresolved sources. In the GC, the X-ray diffuse emission has different characteristics. An excess of the hot component peaking at the center is observed and it has been proposed to be caused by a genuine gravitationally unbound plasma (\Sec\ref{subsec:DXHC}). Additionally, a non-thermal component extending in the hard X-ray range with a prominent neutral iron K$\alpha$ line is observed from ISM gas clouds in the CMZ. This emission is time variable and attributed to illumination by a bright past outburst of \sgra (\Sec\ref{subsec:DXNT}). The contribution of low-energy CRs to this emission has been debated. 
While, other tracers such as $\rm H_3^+$ indicate a larger density in the CMZ than in the rest of the Galactic disk, it does not seem sufficient to represent a large fraction of the apparently steady emission (\Sec\ref{subsec:CRNT}). 

\subsection{The hard diffuse component: very hot plasma or unresolved point sources?} \label{subsec:DXHC}

Among the diffuse emissions detected towards the CMZ, there is a hot thermal component, having a temperature of $kT\sim 7$~keV, traced by a strong Fe~\textsc{xxv} He$\alpha$ and Fe~\textsc{xxvi} Ly$\alpha$ emission lines at 6.7 and 6.97~keV (see \Fig\ref{fig:CMZ_soft_plasma_spectra}). 
This hot component peaks at the GC and smoothly decreases outwards (see \Fig\ref{fig:hotplasma}), connecting to the bulge diffuse emission at higher latitudes and to the ridge one at larger longitudes. Its emission is largely accounted for by 
the populations of unresolved X-ray sources and can therefore be used to study the cumulative properties of these sources (\Sec\ref{subsubsec:DXHC-ps}). The origin of the excess emission relative to the estimation of the unresolved point source contribution has been debated and could originate from an extended hot plasma (\Sec\ref{subsubsec:DXHC-plasma}).

\subsubsection{Populations of unresolved point sources}
\label{subsubsec:DXHC-ps}

The large-scale morphology of the hot thermal component 
follows the stellar mass traced by NIR light, with a flux ratio compatible with the local one \citep{revniv06b,anasta23}. Towards the bulge, more than 80\% of the 6.7~keV emission 
has been resolved into point sources and the remaining signal is likely attributed to the X-ray point sources that are below the sensitivity limit of \chandra deepest observations \citep{revniv09}. Resolved sources include mCVs, which were initially thought to be the dominant contributor to the diffuse emission \citep[see e.g.][]{krivon07,hong12,yuasa12}. However, further investigation of the bulge component by \suzaku and \nustar allowed to better constrain the cumulative spectrum of these sources, including the EW of its iron emission lines \citep{nobuka16} and its average temperature \citep[$kT\sim8\rm~keV$,][]{perez19}, challenging previous interpretations. These results are instead fully consistent with a population dominated by dwarf novae \citep[see also][]{reis13,xu16}. 

Towards the Galactic center, harder emission and stronger EWs are detected. In the inner 10~pc, \nustar revealed an unresolved hard X-ray emission possibly associated with the NSC, strongly peaked towards the center, with a hard X-ray luminosity $L_{\rm 20-40keV} \sim 2 \times 10^{34}\rm~erg~s^{-1}$ and a Bremsstrahlung temperature above 35~keV \citep{perez15}. This hard emission has been attributed to a population of IPs with WD masses $M_{\rm WD} \gtrsim 0.8$~\Msol \citep{hailey16, xu19}. These unresolved IPs would be similar to the ones detected as individual sources by \nustar across the CMZ \citep[][see also \Sec\ref{subsubsec:mCVs}]{hong16} and consistent with the average WD mass measured by the Sloan Digital Sky Survey 
\citep{zoroto11}. 
 The fraction of IPs likely decreases outwards, together with a relative increase of non-mCVs, which could explain why large-scale spectral analyses including the CMZ or adjacent regions detected significantly harder emission than in the bulge, and initially hinted towards a single population of less massive IPs rather than dwarf novae \citep[][]{krivon07,yuasa12}. 
However, combining populations of mCVs, non-mCVs and ABs still fails to reproduce the EWs detected towards the CMZ, which may indicate different source properties or contamination by truly diffuse components (see \Sec\ref{subsubsec:DXHC-plasma}).

\subsubsection{Constraints on the contribution from a hot plasma}
\label{subsubsec:DXHC-plasma}

The normalization factors derived for the bulge/ridge to explain the level of 6.7~keV emission from the stellar mass distribution (SMD) 
models have failed to reproduce the strong 6.7~keV emission from the Galactic center, providing evidence for a significant excess within $\lvert l \lvert \,\lesssim 1^\circ$ and $\lvert b \lvert \,\lesssim 0.25^\circ$ \citep[e.g.][]{uchiya11,anasta23}. 
The precise description of this excess depends on the analyses performed and the SMD model used. The bulk of this excess has the same profile as the SMD 
but with normalisation ratios ranging from 1.3--1.5 for the most recent estimations \citep[][]{anasta23} up to values $\gtrsim 2$ for previous works \citep[][and references therein]{nakash13,heawar13a,yasui15} and a significant increase towards the NSC, 
as well as local enhancements towards known point and extended sources such as XRBs, SNRs and stellar clusters (see also \Fig\ref{fig:hotplasma}). 

There is no consensus on the nature of this emission. The precise description of the 6.7 and 6.97~keV emission lines (i.e.\ their exact energy, flux ratios, and EWs) has been interpreted by several studies as due to a truly diffuse optically thin thermal plasma close to collisional ionization equilibrium, with a temperature $\sim7$~keV \citep{koyama07b,koyama09,yamauc09,uchiya13,yamauc16}. Such a hot plasma would not be gravitationally bound and would escape the central region within few $10^4$~yr, requiring either a huge energy input to sustain it or a confinement mechanism to maintain it. A large-scale toroidal magnetic field within the CMZ, as suggested by NIR polarimetric results, could provide sufficient pressure support to relax the energetic problem by maintaining the hot plasma over its $10^7$--$10^8$~yr radiative-cooling timescale \citep[see][and references therein]{nishiy13}. 
In terms of heating, possible contributions from SN explosions, 
violent past activity from \sgra or magnetic reconnection have all been discussed, along with their respective limitations to reach a thermal energy of $\sim10^{53}$~erg and a temperature of $\sim 7$~keV \citep[see e.g.][]{uchiya13, nishiy13, koyama18, anasta23}. 

Other studies have pointed to the close morphological correspondence between the region covered by the excess and the NSD, 
and argue that it could be entirely due to unresolved point sources. In this case, the X-ray enhancement could be explained by a modification in the underlying source populations, higher metallicities or an improper modeling of the stellar mass in these inner regions \citep{heawar13a,anasta23}. Unlike the large-scale hot plasma scenario, this point source interpretation would also be compatible with the  warm diffuse gas with a large filling-factor reported by \citet[][see also \Sec\ref{subsec:CRNT}]{oka19}. 
However, this would not rule out the possibility of having smaller scale plasma components contributing to local 6.7~keV enhancements, including softer components with $kT\sim1.5~\rm keV$, which are also detected towards the CMZ 
\citep[][see also \Sec\ref{subsec:DXSC}]{heawar13a}.

\subsection{The soft plasma component: hot plasma bubbles, loops and chimneys} \label{subsec:DXSC}

\begin{figure}[ht]
	\centering
 \includegraphics[trim=25mm 15mm 60mm 40mm, clip, width=\columnwidth]{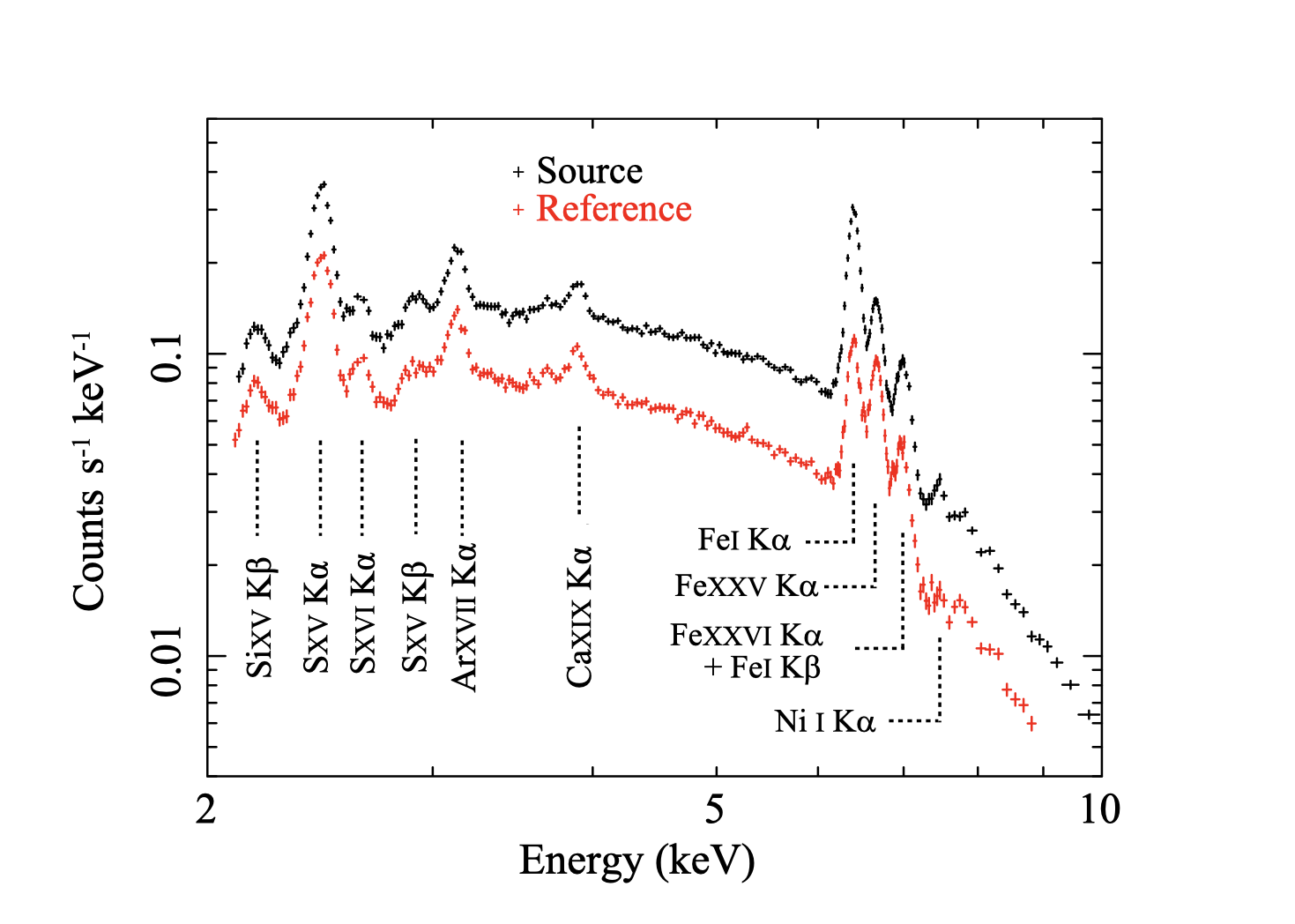}
	\caption{\Suzaku spectra of various soft plasma emitting regions in the CMZ \citep[see][for more details]{nobuka10}. 
 The more prominent lines from the soft plasma component are S~\textsc{xv} at 2.46~keV, 
 Si~\textsc{xiii} at 1.86~keV (not shown), Ar~\textsc{xvii} at $\sim3.12$~keV and Ca~\textsc{xix} at $\sim 3.9$~keV. The neutral Fe K$\alpha$ line at 6.4~keV and the helium-like Fe~\textsc{xxv} line at 6.7~keV, associated with other diffuse components, are also visible. Figure adapted from \citet{nobuka10}. 
	}
	\label{fig:CMZ_soft_plasma_spectra}
\end{figure}

Contrary to the hot plasma, the soft plasma component pervading the inner 300 pc is unambiguously a truly diffuse but structured and patchy emission \citep{wang02a, muno04}. It is characterized by a number of strong helium-like 
and hydrogenoid lines of Sulfur, Silicium and Calcium. The brightest one is the S~\textsc{xv}~He$\alpha$ at 2.46 keV (see \Fig\ref{fig:CMZ_soft_plasma_spectra}).  
They trace a plasma distribution with typical temperatures of about 0.7--1.0 keV \citep{kaneda97, muno04} and densities of about $0.7$--$0.1\ \rm cm^{-3}$.

\begin{figure*}
	\centering
 \includegraphics[width=\textwidth]{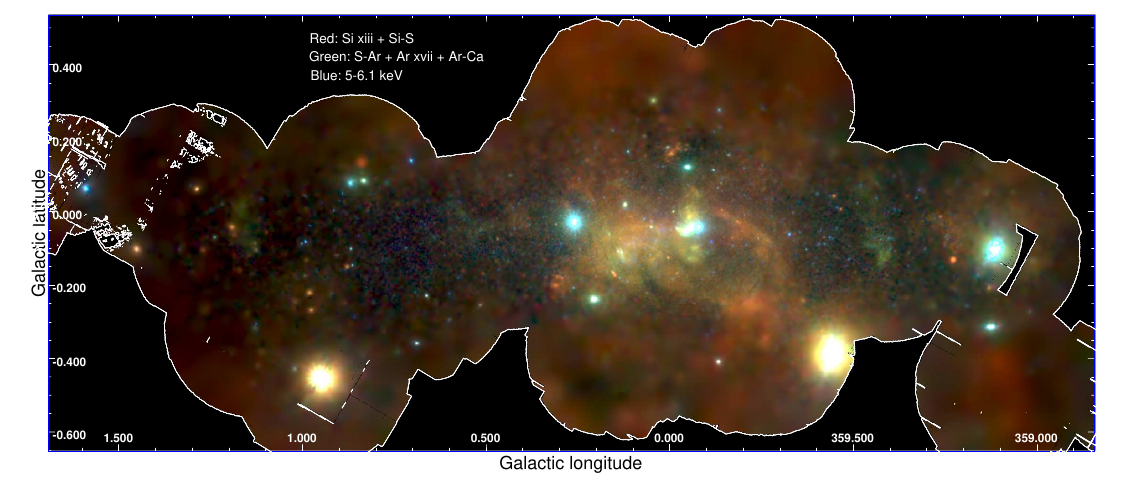}
	\caption{\xmm 3-color image of the soft plasma X-ray emission in the CMZ. Red band is within 1.8--2.3 keV and contains the Si line, green band is within 2.7--3.7 keV and contains the Ar line, blue band is 5--6.1 keV. The large ring like feature west of \sgra is clearly visible, as well as the Radio Arc bubble. The bipolar outflows are visible about 0.1$^\circ$ north and south of \sgra.  Figure extracted from \citet{ponti15b}. }
	\label{fig:soft_plasma_map}
\end{figure*}

While the soft plasma was initially identified with \asca, subsequent surveys of the GC region by \chandra \citep{wang02a, wang21}, \xmm \citep{sakano03, ponti15b} and \suzaku \citep{koyama07b, yamauc18} have resolved various localized features.  
The spatial distribution of the soft plasma emission is indeed highly non-uniform as visible from \Fig\ref{fig:soft_plasma_map}. In the inner 20~pc, \chandra observations reveal brightness variations between $(0.2 - 1.8) \times 10^{-13} \rm ~erg~ cm^{-2}~ s^{-1}$. Correcting for absorption, \citet{muno04} finds bolometric luminosities of the soft plasma ranging from 0.6 to $9 \times 10^{34} \rm~ erg~s^{-1}$ within the inner 20 pc.

Similarly to the soft X-ray plasma component observed in the GRXE, the GC soft plasma is likely heated by supernovae shocks and by winds of massive stars. Energetically, the radiative cooling rate of the soft plasma within the inner 20~pc of the Galaxy could be balanced by 1\% of the kinetic energy of one supernova every $3 \times 10^5$~yr. \xmm X-ray survey reveals about 10--12 supernova remnants candidates through their thermal emission, implying a supernova rate as large as $0.3$--$1.5 \times 10^{-3} \rm~yr^{-1}$ in the CMZ \citep{ponti15b}. 

In the following, we discuss a few specific features. For in depth reviews of these features and overall emission, we refer the reader to \citet{ponti15b} and \citet{koyama18}. 

\subsubsection{A ring-like superbubble southwest of \sgra}

\citet{mori09} discovered two soft plasma features, G359.77--0.09 and G359.79--026 with \suzaku, with absorption column densities consistent with the GC distance and ionization timescales of $\sim 30 \rm~kyr$. \xmm observations by \citet{heawar13b}  confirmed the characteristics of these features. \citet{mori09} and \citet{heawar13b} have both proposed that they are physically connected and part of an apparent elliptical structure centered on $l \sim 359.9^\circ$ and $b \sim -0.125^\circ$ with minor and major axes of 18 and 28 pc at the GC distance. \citet{ponti15b} observed that S~\textsc{xv} emission fills the elliptical structure showing it is the projection of a large shell of hot gas. The total thermal energy of the structure is estimated to be $\sim 10^{51} \rm~erg$, which is too large for a single regular supernova remnant \citep{mori09, heawar13b}. It could have been energized by supernova explosions from a collection of massive stars. Several massive stars escaped from the Quintuplet cluster are indeed found in the region \citep{habibi14} and could be plausible progenitors \citep{ponti15b}. Another alternative origin is a single very energetic event. \sgra being in projection inside the ellipse, the shell could have been produced by an unbound remnant of a stellar capture by the SMBH (a tidal disruption event or TDE) occuring in the last 30~kyr \citep{ponti15b}. The expected recurrence time of TDEs in the GC is in the range of 10--100 kyr \citep{alexan05, merrit10}, 
making the existence of such remnant today plausible. This mechanism could release energies up to $10^{52} \rm~erg$ \citep{khomel96}, large enough to power the superbubble, although more recent studies suggest unbound debris might have lower kinetic energies of about $10^{50} \rm~erg$ \citep{guillo16}.  

\subsubsection{The Radio Arc bubble}

A very bright region of soft X-ray emission is visible around $l \sim 0.1$\deg\ and $b \sim -0.1$\deg\ (see \Fig\ref{fig:soft_plasma_map}). This region is located between the Sgr A complex and the Radio Arc, 
overlapping the Radio Arc bubble, a large mid-IR shell \citep{levine99}. The bubble has a 23~pc (10$'$) diameter and encompasses the Quintuplet cluster \citep{simpso07}. \citet{figer99} has proposed the IR shell to be caused by winds from the cluster massive stars. \citet{rodrig01} noted that the arc bubble is filled with continuum X-ray emission measured with \asca which they assumed could be due to a population of sources. With deep \xmm observations, \citet{ponti15b} observed that the bright soft plasma X-ray emission in the region extends over 7$'$ and completely fills the Radio Arc bubble. The total thermal energy of the X-ray plasma inside the arc bubble could be as large as $1.5 \times 10^{51} \rm~erg$ requiring multiple supernova events to energize it. The emission is not uniform with several cavities possibly due to different explosion events. 

With its IR shell morphology filled with thermal X-rays, the Radio Arc bubble is therefore a likely superbubble associated to the young and massive Quintuplet cluster. 

\subsubsection{G359.41--0.12 and the Sgr~C-chimney} 

In the Sgr C region, \citet{tsuru09} have observed a peculiar pair of soft X-ray plasma emission regions, named G359.41--0.12 and the chimney. The former is an 5$'\times$3$'$ elliptical region, while the latter is a vertical structure emanating from the edge of G359.41--0.12 and extending more than 0.15\deg\ towards the Galactic halo. 
Both structures have comparable spectral properties, consistent with being located at the GC and being physically connected. \citet{tsuru09} have measured thermal energies of $5.9 \times 10^{49} \rm~erg$ and $7.6 \times 10^{49} \rm~erg$  and dynamical timescales of $2.4 \times 10^4 \rm~yr$ and $4 \times 10^4 \rm~yr$ for each structure, respectively. The morphology, energetics and timescales of G359.41--0.12 are suggesting this is a thermal SNR. The origin of the chimney is less obvious. \citet{tsuru09} proposed that it is an outflow from the SNR, possibly caused by ambient gas distribution and extending up to 30~pc. Conversely, \citet{ponti15b} discuss the morphological similarities of the Sgr~C-chimney with the Radio Arc and proposed it is part of the GC lobe detected in radio \citep[][see also \Sec\ref{subsec:chimneys}]{law10}.

\subsubsection{The GC bipolar outflows}

The bipolar \sgra lobes are symmetric structures extending 15 pc above and below the Galactic plane. They were initially observed in the deep \chandra image of the GC \citep{bagano03, morris03, markof10} 
and were interpreted as a possible outflow from \sgra.  Further studies with \xmm demonstrated the thermal nature of the emission \citep{heawar13b}. The temperatures range between 0.7 and 1.1 keV, electrons densities between 1--10~$\rm cm^{-3}$, somewhat larger than in the rest of the CMZ. The lobes surface brightness and gas pressure decrease rapidly with increasing latitude indicating outflowing gas \citep{ponti19}. Edges are relatively sharp in the X-ray images.
Radio observations with the VLA show that the north lobe has a clear radio counterpart characterized by free-free emission. The presence of a radio counterpart to the southern X-ray lobe is less clear \citep{zhao16}. The thermal energy in the lobe has been estimated to be a few $10^{50} \rm ~ erg$ and the sound crossing time of the lobes is a few tens of kyr \citep{ponti19}.

Several scenarios have been invoked to explain the lobes \citep[see e.g.][]{morris03, markof10, heawar13b, ponti15b, ponti19, clavel19}. 
Most of them involve collimation of the plasma injected close to the GC by the CND. They can typically be split in two categories: quasi-stationary outflows of hot plasma, or burst-like injections connected to transient events occurring near the SMBH. Quasi-stationary outflows can be caused by the stellar winds from the massive stars of the YNC. The latter process is unlikely due to the age difference between the lobe (tens of kyr) and the massive stars, see \Sec\ref{subsubsec:YNC}.
The accretion flow onto \sgra being radiatively inefficient, it has been proposed that a large fraction of the power and mass accreted at the Bondi radius is dissipated in the form of outflows rather than accreted by the black hole \citep{wang13}. The power of these outflows could be sufficient to energize the bipolar lobes. Since the clumpy morphology of the lobes and their sharp edges cannot be easily explained by steady sources, transient or episodic phenomena have been discussed.

The X-ray lobes could have been produced by a supernova occurring in the YNC.  
With an estimated supernova rate of $0.3$--$1.2 \times 10^{-4} \rm \ yr^{-1}$ \citep{jouvin17}, it is likely that an explosion has occurred sufficiently recently to have produced the lobes. The energy released should be sufficient to account for the total thermal energy. \citet{ponti15b} have proposed that the lobes could be the remnant of the explosion at the origin of the magnetar SGR~J1745--290. Hydrodynamical simulations show that a SN can inflate the lobes on a few kyr \citep[see e.g.][]{yaline17}. 
Finally, episodic outbursts from the SMBH could be powerful and frequent enough to create the lobes. The energy released by a stellar tidal disruption is sufficient to power the structure and the recurrence rate at the GC is high enough \citep{alexan05} 
so that one such event could be at the origin of the lobes.  \citet{markof10} have also argued that a collection of prominent outbursts of the SMBH has caused the emission and is responsible for the observed knots.  

\begin{figure*}[ht]
 \includegraphics[width=1\textwidth]{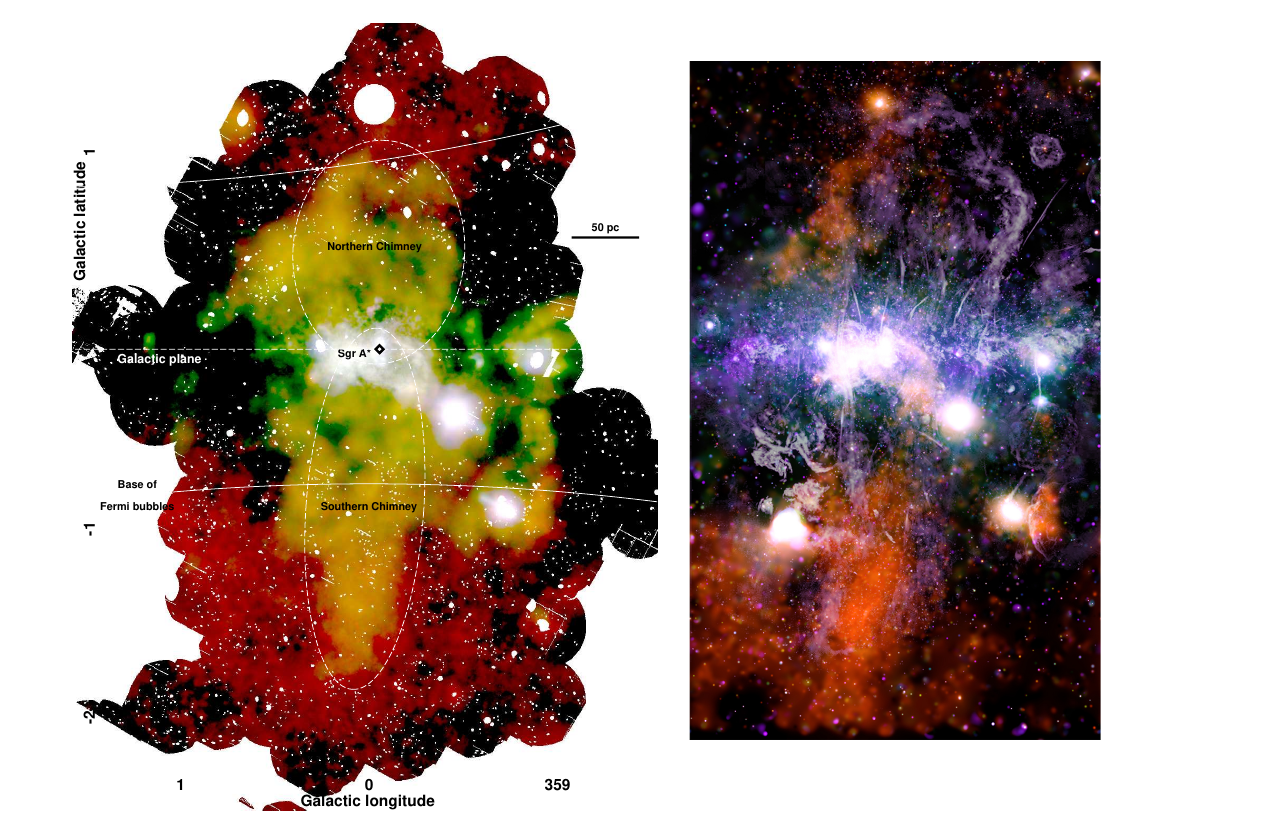}
	\caption{Left: Three-color image of the X-ray emission from the inner 1$^\circ \times 3^\circ$ (red: continuum emission in the 1.5--2.6 keV band, green: 2.35--2.56 keV band (S \textsc{xv} line), blue: continuum emission in the 2.7--2.97 keV band.
    Two symmetric features are clearly visible extending $\sim$ 150--200 pc north and south of the GC. These features
    are reported and called the chimneys by \citet{ponti19}. The northern part is co-spatial with the GC lobe visible in radio, while the southern lobe has been discovered by \citet{nakash13} and interpreted as 100 kyr old recombining plasma. Figure extracted from \citet{ponti19}.
    Right: Composite image of the inner 2.2$^\circ$ by 4.2$^\circ$, from \chandra data \citep[orange, green and purple,][]{wang21} and from MeerKAT \citep[gray,][]{heywoo19}. The spatial correlation of the X-ray chimney or plume with the radio bubble, especially in the southern part is clearly visible. Figure extracted from the 2021 \chandra photo album (https://chandra.harvard.edu/photo/2021/gcenter/)\AG{, credits X-ray: NASA/CXC/UMass/Q.D. Wang; Radio: NRF/SARAO/MeerKAT}. 
	}
	\label{fig:chimneys}
\end{figure*}

\subsubsection{The X-ray chimneys}
\label{subsec:chimneys}

There are additional evidences of an outflow from the GC on scales of $\sim$ 150 pc. 
The Galactic Centre lobe is a loop-like structure detected in radio and extending north of the Galactic plane by more than one degree \citep{sofhan84}. 
A large X-ray thermal emission region has been observed with \suzaku one degree south \citep{nakash13} of the GC. The plasma is optically thin and is over-ionized with a recombining time scale of $\sim 10^5 \rm\ yr$ which had been interpreted as the result of a possible past outflow of the SMBH. A similar degree-scale plume was found in the north of the GC with \suzaku supporting such a scenario \citep{nakash19}.

The high-latitude \xmm survey of the GC has shown that these plumes are part of a larger structure extending 160 pc to the north and south of the GC, called the X-ray chimneys \citep[][see \Fig\ref{fig:chimneys}]{ponti19}. They are well confined in longitude and present sharp edges. Their base is a relatively large region of 50 pc surrounding \sgra. Their thermal energy content is estimated to be $\sim 4 \times 10^{52}~ \rm erg$. The chimneys are not symmetric which might be due to ISM asymmetries. 
Contrary to the bipolar lobes the X-ray surface brightness and plasma electron density are relatively flat over the north and south chimneys. The latter varies from $0.2$ to $0.1\ \rm cm^{-3}$ from the base to the top. Based on their different plasma physical characteristics, \citet{ponti19} argued that the bipolar lobes and the chimneys are distinct features and that the lobes are embedded in the larger volume of the chimneys.

MeerKAT radio continuum observations have observed a pair of 
radio bubbles roughly symmetric around the GC with a total span of $140 \times 430\ \rm pc$ \citep{heywoo19}. The bubble emission is edge-brightened which suggests it has been caused by a time bounded event, rather than steady state outflows. The cosmic-ray energy in the bubbles is estimated to be $7 \times 10^{51}\ \rm erg$ and the total energy content could reach  $7 \times 10^{52}\ \rm erg$.

The X-ray and radio structures match well in particular in the southern part where the X-rays are enclosed within the the radio emission \citep[see][]{heywoo19, wang21, ponti21}. The association is less clear in the northern part, since the two emissions do not exactly overlap \citep{wang21}. Some authors even propose that part of the northern radio bubble, coincident with the thermal radio emission in the west of the GC lobe, is in fact a foreground \HII region \citep{tsuboi20}.  Yet, \citet{ponti21} suggest that the chimneys could be tilted towards us so that the northern structure is in front of the GC. Finally, \citet{ponti21} found that the bubble correlate well with spurs of warm dust emission surrounding it (visible in WISE IR images) which they interpret as dense gas heated by the shock traced by the radio emission. 

The X-ray chimneys and radio bubbles are interpreted as signatures of an outflow from the GC towards the halo and as such are likely connected to the Galaxy scale \Fermi and \erosita bubbles (see \Sec\ref{subsec:Con-LinkG}).
Several type of processes might have created them: past activity from the SMBH or nuclear star formation processes. 
\citet{zhang21} have performed numerical simulations showing that supernovae occurring in the nuclear stellar disk naturally form a collimated outflow, and by adjusting the SN rate and vertical magnetic field strength they could reproduce the morphology of the X-ray chimneys. Another possible interpretation is that the chimneys act as a channel connecting the intermittent activity of the GC with the larger scale features like the \Fermi bubbles \citep{ponti19, ponti21}. This is supported by the observation of linear plasma structures  in the southern chimney and interpreted as an `exhaust vent' by \citet{mackey24}. Measurements of proper motions or Doppler shifts which might be accessible with micro-calorimeters would give credit to this hypothesis.  

\subsection{Non-thermal emission by molecular clouds: past activity of \sgra} \label{subsec:DXNT}

 A bright and highly variable non-thermal emission correlated with the molecular clouds has been first detected 
through the CMZ in the 1990s (see \Sec\ref{subsec:EarlyXG}), traced by both a strong Fe\Ka fluorescent emission line and a hard X-ray continuum. Its detailed properties (\Sec\ref{subsubsec:DXNT-prop}) 
are characteristic of an illumination by photons produced by the intense past activity of a hard X-ray source. These photons, most likely originating from \sgra, are
currently propagating in the inner region of our Galaxy (\Sec\ref{subsubsec:DXNT-origin}). When these light echoes leave the clouds, a fainter level of non-thermal emission remains, whose origin has also been investigated (\Sec\ref{subsubsec:DXNT-faint}).

\subsubsection{Properties of the bright non-thermal emission}
\label{subsubsec:DXNT-prop}

\begin{figure*}
		\centering
 \includegraphics[width=\textwidth]{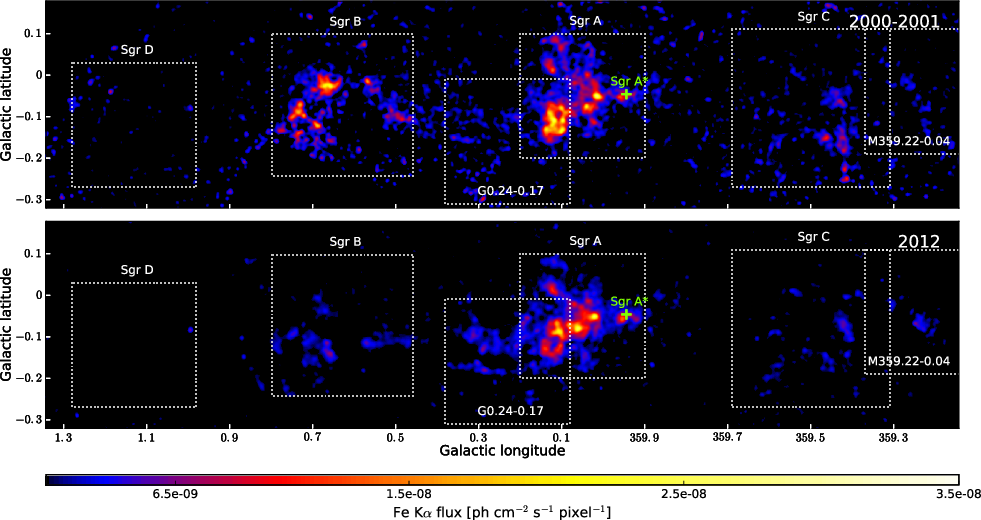}
	\caption{Morphology and variability of the non-thermal emission at the Galactic center, traced by the Fe\Ka fluorescent line detected by \xmm\ in 2000-2001 (top) and in 2012 (bottom). Figure from \citet{terrie18}\AG{, reproduced with permission © ESO}. 
	The white dotted squares highlight all molecular complexes with 6.4~keV emission detected (see also \Fig\ref{fig:cmz-compIR}). 
	}
	\label{fig:xmm-echoes}
\end{figure*}

The morphology of the non-thermal X-ray emission  
detected at the Galactic center closely follows the distribution of matter in the CMZ, with the brightest regions being Sgr~B, 
Sgr~A and Sgr~C molecular complexes (see \Fig\ref{fig:xmm-echoes}, \citealt{koyama18,terrie18}, and references therein). 
Within these molecular complexes, the non-thermal emission traced by a prominent 6.4~keV fluorescent line is highly structured, with bright clumps or filaments and more diffuse and extended features matching molecular material also seen in submillimeter line surveys \citep[][]{ponti10,clavel13,clavel14,terrie18,brunke25,albosl25}. 
This correlation indicates that these clouds play a key role in the emission process. It is also important to note that while the bright X-ray non-thermal emission matches with molecular structures, within each molecular complex not all clouds are emitting X-rays and the ones that are emitting are not necessarily the densest ones, indicating that the signal is restricted to a fraction of the CMZ only \citep[e.g.][]{ponti10,terrie18}. 

Moreover, the non-thermal emission of the CMZ is highly variable both in flux and in morphology. Clear signatures of flux variability have been first reported towards Sgr~B \citep{inui09,terrie10} and are now detected across the whole CMZ \citep[][and references therein]{terrie18}. The collection of X-ray bright clouds has also changed over the past decades, indicative of a signal propagation through these regions \citep[e.g.][]{clavel13,churaz17a,terrie18}. 
Such propagation has been characterized to be superluminal across the Bridge cloud inside the Sgr~A region \citep{ponti10}, and such motion can only be reproduced by projection effects of light echoes from a source that is distant from the reflecting clouds.
Thereby, these results ruled out previously competing scenarios involving LECR interacting with MC \citep[e.g.][]{tatisc12}. 

\begin{figure}[ht]
		\centering
    \includegraphics[width=\columnwidth]{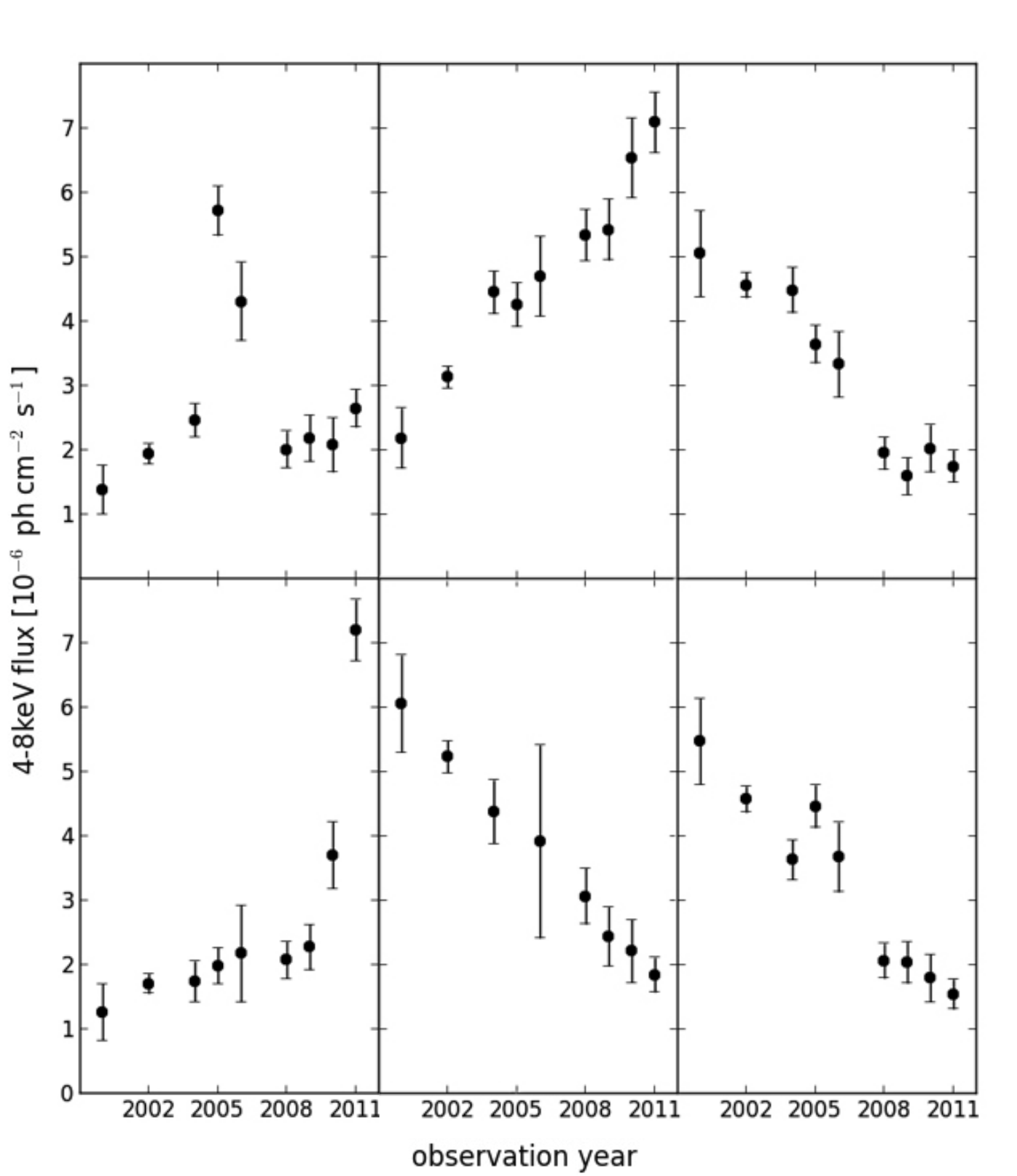}
	\caption{The two different time behaviors of the non-thermal emission observed within the Sgr A molecular 
	complex. The 4--8~keV light curves are extracted from $15''\times15''$ sub-regions of the Bridge cloud (left) and of three other molecular clouds (middle and right). Figure adapted from \citet{clavel13}\AG{, reproduced with permission © ESO}.
	}
	\label{fig:xmm-echoes-lc}
\end{figure}

Dedicated variability studies of the CMZ have shown that, for a given cloud, the amplitude and the time-scale of the variability depend on the size of the region considered. For instance, larger regions tend to have lower amplitude and slower variations, which can be explained by the dilution of the variable signal and/or a sum of substructures varying differently \citep[][]{clavel13, rogers22}. 
The variability properties can also change from one cloud to another, which can be explained by either different density distribution of the clouds along the LoS 
and/or differences in the incident light front itself \citep[][see also \citealp{sunchu98,crasun02,odaka11}, for theoretical estimations of MC emission light curves]{clavel13}. 
In particular, the variability study performed on the smallest possible scale with \chandra\ 
in the Sgr~A region has revealed two different time behaviors: a very short peaked signal that lasts no longer than two years, and slower linear variations, either increasing or decreasing over a decade or so \citep[\Fig\ref{fig:xmm-echoes-lc},][]{clavel13}. The variations detected across the CMZ are all compatible with these two time behaviors, and all bright features have been significantly variable over the past decades \citep[][]{chuard18,terrie18}. 
The amplitudes between the lowest and the highest non-thermal fluxes detected for these individual clouds are generally a factor of a few. The strongest variation reported so far is a flux increase of at least a factor of 10 over less than a year of a 1-pc scale filament within the Sgr A molecular complex \citep{clavel13}. 
As of 2025, the latest X-ray observations of the CMZ show a decreasing trend of the once brightest molecular structures, with only a few brightening features. The current brightest region is coincident with the eastern part of the Bridge molecular cloud \citep[][]{khabib22,marin23}.

X-ray observations with \asca, \suzaku, \xmm and \chandra have shown that the spectrum 
of the bright non-thermal emission is characterized by a strong emission line created by the fluorescence of neutral iron, Fe\Ka at 6.4~keV, with an equivalent width EW~$\sim 1 \rm\, keV$, and the associated Fe\Kb line at 7.06~keV with an EW which is about seven times lower. These lines are associated with a power-law continuum which is absorbed at low energies, with column densities ranging from few $10^{22}$ to $\sim10^{24}\,\rm cm^{-2}$, in agreement with the $N_{\rm H}$ integrated along the line of sight in the direction of the CMZ. The current spectral measurements are also compatible with the presence of an absorbing Fe K edge at 7.1~keV 
\citep[see][for examples of detailed spectral analyses]{muraka00,muraka01b,inui09,ryu09,ponti10,nobuka11,ryu13}. Observations at higher energies with \nustar\ and \integral\ have detected this non-thermal component up to 40~keV and even to 100~keV, in the brightest regions, allowing to better constrain the power-law  photon index of the signal to $\Gamma \sim 2$  \citep{revniv04,krivon14,mori15,terrie10,zhang15,krivon17}. 
Spectral analyses performed on observations from different years show that the 6.4~keV emission line and the underlying power-law continuum vary synchronously \citep[][]{terrie10,nobuka11,clavel14} 
and the typical surface brightness reported for the 6.4~keV line at the peak ranges from few $10^{-5}\,\rm cm^{-2}\, s^{-1}\, arcmin^{-2}$ for the brightest region in Sgr~A down to few $10^{-7}\,\rm cm^{-2}\, s^{-1}\, arcmin^{-2}$ in Sgr~D \citep{terrie18}.  

Finally, \ixpe has detected polarization towards the Sgr A molecular complex, with a polarization degree of the non-thermal continuum emission of $(31\pm11)$\% and a polarization angle of $(-48\pm11)^\circ$ \citep{marin23}.
These values further prove the reflection nature of the signal and constrain the scattering geometry. 
In particular, the direction between the cloud and the illuminating source is likely perpendicular to the direction of the polarization detected\footnote{This statement assumes that the incident light is not polarized \citep[see][for more details]{khabib20}.}, 
making the \ixpe measurement compatible with a past illumination from \sgra \citep{marin23}.

\subsubsection{X-ray echoes and past activity of \sgra}
\label{subsubsec:DXNT-origin}

The variability, the spectral shape and the polarization of the bright non-thermal emission, along with its correlation with molecular clouds, are all consistent with it being caused by the reflection of X-ray light onto molecular material present in this region \citep[][]{sunchu98,churaz02}. 
This incident X-ray light is scattered by the cloud, creating the power-law continuum with the same photon index than the incident spectrum with significant photoelectric absorption within the cloud (and possibly also all along the light path), inducing both a spectral absorption at low energy and fluorescence. Several elements have their K-shell fluorescence lines in the X-ray energy range. The strength of these emission lines depends on both the abundance of the element and the probability of fluorescence after photoelectric absorption by atoms of this element, which explains why the Fe\Ka line dominates the spectrum of the reflected signal.

The exact shape of the spectrum, including the intensity of the 6.4~keV emission line, depends on the cloud properties (size, density, metallicity), on the incident signal properties (luminosity, spectrum and duration), as well as on the geometry of the reflection (LoS position of the cloud and its distance to the source) and possible absorption along the light path \citep[e.g.][]{sunchu98,odaka11,walls16}. 
Thus, the 6.4~keV flux of a spherical cloud at a distance $D$ from the observer is a function of the luminosity $L_X$ of the incident event, and can be written as:
\begin{equation}
    F_{\rm 6.4keV} \propto f_{Fe} \, \frac{\Omega}{4\pi D^2} \, \delta_{Fe} \, N_{H} \, \sigma_K^{Fe} \, L_{X}
\end{equation}
where $f_{Fe}$ is the iron fluorescent yield, $\delta_{Fe}$ the iron abundance with respect to hydrogen, $\sigma_K^{Fe}$ the iron K shell photoabsorption cross section at 7.1~keV, $\Omega$ the solid angle of the cloud from \sgra, and $N_H$ the illuminated column density within the cloud.
Using reasonable values for all these parameters, the observed Fe\Ka flux can therefore be used to estimate the luminosity of the incident event \citep[][]{sunchu98}. If we assume the source responsible for the X-ray echoes is at the Galactic center, the corresponding events need to have an X-ray luminosity in the 2--10~keV range of at least a few $10^{39} \rm \,erg\,s^{-1}$ for several years to explain the brightest molecular clouds \citep[e.g.][]{koyama96,muraka01a,ponti10}. 
The actual value of \sgra's past luminosity depends on the line-of-sight position of the cloud as well as the duration of the illuminating event. Clouds that are further away from \sgra than their projected distance will intercept a smaller fraction of the incident signal, while shorter events will illuminate only a fraction of the clouds at a given time, resulting in a possibly large underestimation of the past luminosity of \sgra \citep[][]{churaz17a}.

\begin{figure}[ht]
		\centering
 \includegraphics[width=\columnwidth]{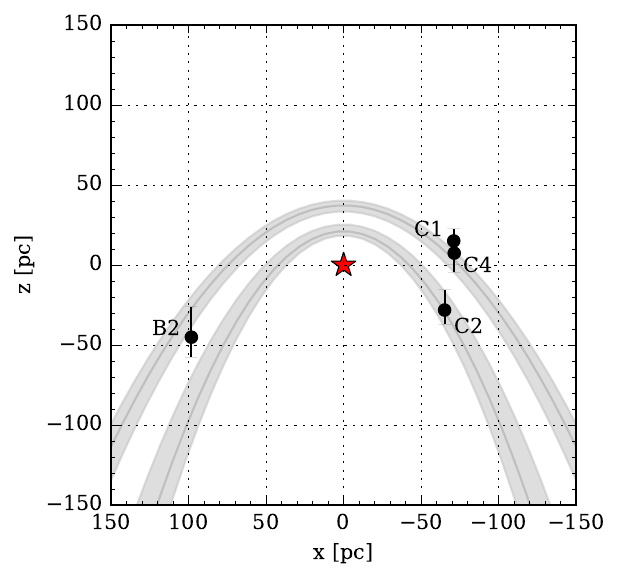}
	\caption{Echoes in the Galactic center viewed from the Galactic north pole. The line-of-sight distance ($z$) of clumps within Sgr~B (B2) and Sgr~C (C1, C2 and C4) derived from their spectral shape are shown in black. The red star marks the position of \sgra. The grey parabolas correspond to the two past events illuminating the CMZ derived by \citet{chuard18}, together with their age uncertainty (parabola widths). They corresponds to a relatively short event occurring $\sim140$ years ago and a longer one occurring $\sim240$ years ago. Figure from \citet{chuard18}\AG{, reproduced with permission © ESO}. 
	}
	\label{fig:echo-parabolas}
\end{figure}

The energetic involved and the spectral constraints are such that among the variety of X-ray transients in the inner region of our Galaxy, only the supermassive black hole \sgra can be responsible for the X-ray echoes detected across the CMZ \citep[][]{clavel13}. 
The time delay $t$ between the direct detection of an event from \sgra and of its reflection onto a given molecular cloud, depends on the projected distance $x$ and the line-of-sight distance $z$ between the cloud and the black hole. Thereby, at a given time, all clouds illuminated by the same event are distributed along the paraboloid defined by
\begin{equation}
	z(t) = \frac{1}{2}\left( ct - \frac{x^2}{ct}\right),
	\label{eq:parabola}
\end{equation}
where $c$ is the speed of light \citep[\Fig\ref{fig:echo-parabolas};][]{sunchu98}. Therefore, the extension of the CMZ along $z$ and $x$ allows, in principle, to probe \sgra's past activity from the past 1000 years or so. 

However, reconstructing the past lightcurve of \sgra from the X-ray echoes is not trivial, since the line-of-sight position of the clouds are poorly known (see \Sec\ref{subsubsec:MC} and \Fig\ref{fig:cmz-mcdist-models}). 
Therefore, different techniques have been developed through the years to constrain the past activity of \sgra from the properties of these bright X-ray echoes, using their variability pattern, their spectral shape and, more recently, also their polarization properties.

The light curves of the clouds are the result of a convolution between the past light curve of \sgra and the density profile of the clouds. Therefore, studying the variability of the 6.4~keV emission on the smallest possible scale, to avoid averaging substructures that vary differently, can provide strong upper limits on the duration of the illumination. Such studies have shown that for all bright clouds the illumination cannot be longer than about a decade \citep{terrie18}, and even shorter than a few years for several clouds in Sgr~A \citep{churaz17a, clavel13}. Furthermore, under the assumption that clouds that are experiencing the same time behavior are illuminated by the same past event from \sgra, it is possible to constrain their relative position following equation~(\ref{eq:parabola}). However, the age of the event~$t$ remains unknown, unless there are constraints about the line-of-sight position of the reflecting clouds. Methods have been developed to infer it from the spatial and temporal variation themselves \citep{churaz17a}, or from the 6.4~keV flux and the relative density of the clouds obtained from molecular tracers \citep{ponti10, clavel13}. However, the results obtained strongly depend on the number of illuminating events assumed by the authors.

A precise modeling of the X-ray emission detected towards the clouds can provide independent information on their line-of-sight position. The cloud will absorb a higher fraction of the X-ray emission from the ambient plasma if located closer to the front edge of the CMZ, providing information on its position within the CMZ \citep{ryu09,ryu13}. The spectrum of the echo itself also depends on the line-of-sight position of the cloud: the properties of the continuum emission change with respect to the scattering angle, while the characteristics of the fluorescent lines are isotropic. 
Hence, the equivalent width of the emission lines depend on the scattering angle \citep{capell12}, and dedicated spectral models accounting for the geometry of the reflection process have been developed to retrieve the line-of-sight position of the reflecting clouds \citep[e.g.][see also \Fig\ref{fig:echo-parabolas}]{walls16, chuard18}. Given the possible degeneracies between the spectrum of the X-ray echo, the astrophysical background emissions and the foreground absorption, absolute positions obtained from these techniques can suffer from poorly controlled uncertainties.

Lastly, the polarization degree of the X-ray echo also depends on the scattering angle, with decreasing values for increasing absolute values of the line-of-sight position, $|z|$ \citep{churaz02}. The current brightness of the reflection signal across the CMZ, restrict polarization measurement with \ixpe to a single subregion within the Sgr~A molecular complex (see \Sec\ref{subsubsec:DXNT-prop}). The polarization degree reported is compatible with a two-century old event from \sgra \citep{marin23}. However, any unpolarized signal contributing to the non-thermal emission or the possible contribution from different clouds having different polarization properties would lower this value. 

Based on the variability, spectral and polarization constraints obtained, there is a consensus on the fact that \sgra was at least a million times brighter than today for at least a short period of time within the past three centuries or so. 
The number of past outbursts currently propagating in the CMZ, their exact age, duration and luminosity, are however still debated \citep[see e.g.][and \Fig\ref{fig:echo-parabolas}]{clavel13,churaz17a,chuard18,stel25}. 
The nature of the accretion events that likely caused these past outbursts is also unknown, and several scenarios have been proposed, including the tidal disruptions of low-mass objects (cold clumps from stellar winds, gas clouds, planets), the partial capture of a star or a `failed' TDE \citep[see e.g.][]{cuadra08,yu11,zubova12,czerny13,saclod19}.

\subsubsection{Fading signals and constraints on what remains}
\label{subsubsec:DXNT-faint}

The light front created by \sgra's past outbursts are currently propagating within the CMZ, creating the bright X-ray echoes described in \Sec\ref{subsubsec:DXNT-prop}. These light fronts are expected to progressively leave the clouds that have been illuminated in the past decades, possibly reaching other features along the line of sight, and will eventually leave the CMZ. The monitoring of the variations over the last decades has shown a global decrease of the emission in most regions \citep{terrie18,khabib22}. The once brighter molecular cloud Sgr~B2 follows this trend, while on small scales \xmm still detects limited variability surrounding the densest core, Sgr~B2 itself is at its record low in terms of 6.4~keV flux \citep{rogers22}. Starting from 2011, on larger scales, the \integral  light curve of Sgr~B2 has entered a regime consistent with a constant flux in the hard X-rays \citep[30--80~keV,][]{kuznet22}, questioning the nature of the remaining stationary signal detected from this molecular cloud and from others \citep[see e.g.][]{kuznet19}.

A stationary and fainter non-thermal emission could be created by the same process as the bright X-ray echoes, for instance by the reflection of a lower-luminosity plateau in the light curve of \sgra, or by sparser material with a larger extension along the line of sight. However,  following the passage of the light front across the densest molecular clouds of the CMZ, an emission due to multiple scattering could persist over several years, creating a stable non-thermal signal with a harder spectrum and a 6.4~keV emission line including a large Compton shoulder \citep{odaka11}. A possible alternative to a reflection-related origin of this stationary emission is the interactions between LECR and the molecular clouds (see \Sec\ref{subsec:CRNT}).
The spectral constraints currently available do not allow to conclude on the respective contribution of these different processes to the faint level of stationary non-thermal emission \citep[][]{cherny18,kuznet22,rogers22,zhang15}. However, LECR are likely not responsible for the major part of Sgr~B2's current level of non-thermal emission, either due to unphysical spectral parameters \citep[for electrons,][]{zhang15,rogers22} or to incompatibilities with constraints based on other tracers \citep[for protons,][see also \Sec\ref{subsubsec:ioni-6.4}]{kuznet22}.

\subsection{Ionization and non-thermal emission from cosmic rays} \label{subsec:CRNT}

As we will see in \Sec\ref{sec:GRCR}, gamma-ray observations of the GC region reveal the presence of ongoing particle acceleration there, not associated to the ubiquitous background of cosmic rays that pervades the entire Galactic disk (for a general overview on Galactic cosmic rays see \citealt{gabici22}).
It is therefore natural to explore whether such additional component in the CR intensity plays a role in the generation of the non-thermal X-ray emission in the CMZ discussed above.
In particular, the interactions of LECRs (sub-GeV) with the gas in the CMZ might result in a stationary X-ray emission.

A way to constrain/predict the contribution from LECRs to the stationary hard X-ray emission from the CMZ 
is to search for other manifestations of the presence of such particles in the region.
With this respect, one possibility is to search for absorption or emission lines  associated to molecular species which are produced in clouds as a result of the cosmic-ray ionisation of interstellar matter \citep[e.g.][]{dalgar06,tielen13}.
Particles in the sub-GeV domain are most effective in ionising interstellar matter.
The amount of LECRs present in the region can be then derived from the intensity of these lines, provided that the chemistry operating in the cloud is well known.

\subsubsection{A very large cosmic ray ionisation rate in the CMZ revealed by molecular lines}

LECRs play a crucial role in molecular clouds, as they are the only ionising agents able to penetrate large column densities of gas \citep[see][for reviews]{gabici22,padova20}.
This is because the ionising UV radiation is absorbed in the external layers of the cloud.
The classification of molecular clouds is a complex matter \citep[see e.g.][]{snomcc06}, but for simplicity here we name them {\it diffuse} or {\it dense}, depending on whether photons or cosmic rays dominate the ionisation of the gas in the cloud, respectively.
With this definition, the transition from photoionisation to cosmic-ray ionisation takes place at a hydrogen column density of $N_H \approx 8 \times 10^{21}$~cm$^{-2}$ \citep{mckee89}.

A quantity which is generally adopted to quantify the ionisation of interstellar matter due to LECRs is the CR ionisation rate of molecular hydrogen, $\zeta_{CR}$, which has units of s$^{-1}$ and represents the number of CR-induced ionisations per second and per hydrogen molecule.
For reference, the value of this quantity in the local interstellar medium could be measured with unprecedented accuracy after the Voyager probes crossed the heliopause and entered interstellar space \citep[see][and references therein]{cummin16}.
For CR nuclei, the local ionisation rate slightly exceeds $\zeta_{CR}^N \sim 3 \times 10^{-17}$~s$^{-1}$, while for CR electrons, due to the steepness of their energy spectrum, this quantity cannot be constrained but certainly exceeds $\zeta_{CR}^e \sim 0.5 \times 10^{-17}$~s$^{-1}$ \citep[e.g.][]{cummin16,gabici22}.

In diffuse molecular clouds, UV radiation mainly ionises carbon and is rapidly attenuated.
On the other hand molecular hydrogen, due to its larger ionisation potential, is mostly ionised by cosmic rays.
Once produced, H$_2^+$ interacts very rapidly with H$_2$ to produce protonated molecular hydrogen, H$_3^+$.
This molecule plays a pivotal role in molecular clouds, as it triggers most interstellar chemistry \cite[e.g.][]{dalgar06}. 
Rovibrational transitions of H$_3^+$ can be observed in the infrared absorption spectra of stars embedded in (or located behind) a molecular cloud \citep{linmcc01,oka06,miller20}.

The typical ionisation fraction in diffuse molecular clouds is $\approx 10^{-4}$, mostly determined by photoionisation of carbon.
The abundance of free electrons is quite small, but still suffices to determine the rate at which H$_3^+$ is destroyed (through dissociative recombination).
The rate of production of H$_3^+$ per unit volume is then proportional to the CR ionisation rate, $\zeta_{CR}$, while its destruction rate is proportional to the dissociative recombination Langevin rate constant, which has been measured in laboratory experiments \citep{mccall03}.
This explains why the CR ionisation rate can be estimated from the measured intensity of H$_3^+$ lines.
In fact, as astronomical observations provide the column density of H$_3^+$, rather than its volumetric one, the actual quantity that can be constrained from observations is the product $\zeta_{CR} \times L$, where $L$ is the LoS path length inside the cloud.
This introduces an important uncertainty on the estimate of the CR ionisation rate, as it is not trivial to measure $L$.

\citet{oka05} reported on the detection of H$_3^+$ absorption lines in stellar spectra along lines-of-sight crossing the CMZ.
From the measured column densities of H$_3^+$, they obtained values of the product $\zeta_{CR} \times L$ exceeding ${\cal O}(10^4)$~cm~s$^{-1}$.
This implies that values of the CR ionisation rate similar to those measured in the local ISM would require unreasonably large values of the path length, namely, $L$ exceeding a kiloparsec, which is larger than the size of the CMZ!
They were forced to conclude, then, that the ionisation rate close to the GC has to be much larger than the local one.

New infrared observations were performed by the same team in order to increase the number of sightlines explored \citep[e.g.][]{goto14} and 
\citet{oka19}, based on spectral observations of about 30 stars located through the entire CMZ (from 140 pc west to 120 pc east of \sgra), estimated an astonishingly large CR ionisation rate of $\zeta_{CR} \sim 2 \times 10^{-14}$~s$^{-1}$.
An independent study based on a similar set of H$_3^+$ data and on a more detailed chemical modelling also obtained extremely large values of the ionisation rate in the range $\zeta_{CR} \sim 1$--$11 \times 10^{-14}$~s$^{-1}$ \citep{lepeti16}.
No significant spatial variations of the ionisation rate were observed across the CMZ.
These measurements probe the warm ($\sim$~200~K) and diffuse ($<$~100~$cm^{-3}$) gas phase, which dominates the volume of the CMZ \citep{oka19}.

The detection, in stellar spectra, of absorption lines from the oxygen-bearing ions OH$^+$, H$_2$O$^+$, and H$_3$O$^+$ can also be used to constrain $\zeta_{CR}$ in diffuse atomic clouds.
This is because the chain of reactions leading to the formation of such ions is initiated by the CR ionisation of atomic hydrogen, followed by charge exchange with oxygen, and then by a chain of reactions with H$_2$ to produce the sequence of hydrides O$^+ \rightarrow$ OH$^+ \rightarrow$ H$_2$O$^+ \rightarrow$ H$_3$O$^+$. 
The analysis of {\it Herschel} spectral data performed by \citet{indrio15} showed that, for lines-of-sight crossing the CMZ, the abundance of such molecules implies very large ionisation rates in the range $2 \times 10^{-15} {\rm~s}^{-1} \lesssim \zeta_{CR} \lesssim 2 \times 10^{-14} {\rm~s}^{-1}$.

Finally, the high gas temperatures (from $\sim$ 60 K to 100 K and beyond) measured in the dense clouds within the CMZ are most likely due to turbulent heating, and this allows to infer an upper limit on CR heating which would correspond to a quite large upper limit on the CR ionisation rate, $< 10^{-14}$~s$^{-1}$.
However, this applies to the dense gas, and CRs may play a more prominent role in the diffuse phase (where H$_3^+$ measurements are relevant) \citep{ginsbu16}.

The results summarised above show that very large values of the CR ionisation rate are consistently found in the diffuse gas phase of the CMZ, based on observations of several chemical tracers, and this suggests that the intensity of low-energy (sub-GeV) CRs has to be much larger (by several orders of magnitude!) there than in the local ISM. 
This is indeed very puzzling because, as discussed in \Sec\ref{subsec:GRDE}, the interpretation of gamma-ray observations of the CMZ do not require any large excess in the intensity of CRs in the $\approx$~GeV energy domain.
A simultaneous explanation of gamma-ray and astrochemical data would then require a very fine-tuned shape of the spectrum of CRs, exhibiting an extremely large enhancement in the sub-GeV energy domain  \citep{raviku25}.

Sightlines crossing the CMZ and characterised by large gas column densities (dense clouds) can also be probed, even though the large gas column densities might prevent the observation of background/embedded stars and therefore the search for absorption lines.
As an alternative, one can search for molecular emission lines at submillimeter/millimeter wavelengths, and also in this case enhanced values of the ionisation rate have been reported. 
\citet{vander06} combined APEX and IRAM 30-m telescope data to map the distribution of H$_3$O$^+$ and H$_2$O in the Sgr B2 region.
Based on astrochemical modeling, they found that the observed H$_3$O$^+$/H$_2$O line intensity ratio requires a CR ionisation rate of the order of $\zeta_{CR} \sim 4 \times 10^{-16}$~s$^{-1}$.
This value is about one order of magnitude larger than the ionisation rate in the local ISM, as estimated based on data from the Voyager probes, but much smaller than that estimated in the diffuse molecular and atomic gas phases within the CMZ (absorption lines of H$_3^+$ and hydrides in stellar spectra, see above).
These differences might reflect the fact that the penetration of LECRs into dense clouds is hindered by the severe ionisation energy losses suffered by energetic particles when the gas density is large.
This would naturally reproduce a gradient in the CR intensity, which would be suppressed inside dense clouds such as Sgr~B2 \citep{dogiel15}.

An accurate interpretation of these spatial variations of the CR ionisation rate would require a modeling of the penetration of LECRs into dense clouds.
Unfortunately, this is a very complex problem, due to the intrinsic non-linearity of CR transport: CRs themselves excite the magnetic turbulence that regulates their penetration into the cloud. 
An extensive literature exists on the subject, from the pioneer studies by \citet{skistr76} and \citet{cesvol78}, to recent ones (e.g. \citealt{morgab15}, \citealt{phan18}, \citealt{ivlev18}, or \citealt{dogiel18} for an application to the CMZ).

\subsubsection{Problems with the interpretation of the ionisation rate measurements}

Even though modeling the transport of LECRs in the CMZ is a challenging task, some conclusions which are almost model-independent can be drawn from the data.
In this respect, a solid argument can be put forward to illustrate the main difficulty encountered in explaining the large values of the ionisation rates.
Particles of energy below the GeV domain are those which contribute most to ionisation, and as a consequence they suffer severe energy losses.
The ionisation energy-loss time decreases for lower particle energies, and is inversely proportional to the ambient gas density.
In dense environments such as the CMZ, the rate at which freshly accelerated particles have to be injected to compensate for those lost due to ionisation has to be extremely fast, requiring an exceedingly large amount of energy \citep{recchi19}.

This issue has been investigated in detail by \citet{raviku25}, who modeled the injection and transport of CR from very low to very high energy in the CMZ.
They found that the power needed to maintain a population of sub-GeV particles at the level needed to explain ionisation rates at a level $\zeta_{CR} \sim 2\times 10^{-14}$~s$^{-1}$ is exceedingly large.
For CR nuclei, the injection power has to be $\lesssim 10^{41}$~erg~s$^{-1}$, which is comparable to the estimated power in CR nuclei of the entire Galaxy \citep{strong10}.
The situation is even worse for CR electrons, that would require an injection power $\gtrsim 10^{40}$~erg~s$^{-1}$, which is about an order of magnitude larger than the entire CR electron power of the Milky Way \citep{strong10}.
To further stress the enormity of the required CR powers, one can notice that they are comparable to the average power needed to inflate the huge \eROSITA bubbles over the past tens of millions of years \citep{predeh20}.
Similarly large CR power requirements were also found by \citet{liuyan24}.

Besides the prohibitive values of the CR powers, other observations are difficult to reconcile with the large values of the ionisation rates. 
As we will see below (see \Sec\ref{subsec:GRDE}), gamma-ray observations of the CMZ suggest the existence of a source of CRs of high energy (GeV and TeV domain) located very close to the GC.
If such source is also assumed to produce the sub-GeV ionising particles in a continuous way (to compensate for those lost due to energy losses), one would expect the intensity of CRs to decline away from the GC, at odds with observations.
To fix this problem, one might consider a scenario where LECRs are injected everywhere within the CMZ by a population of (unspecified) sources.
In this case, the flat spatial profile of ionisation rates could be reproduced, but the energy reservoir problem would be even more severe, as the source of the power would not be connected with the SMBH located at the GC, which is the most powerful source in the region.
Finally, if LECR electrons are invoked to explain the ionisation rates, the associated non-thermal Bremsstrahlung emission from these particles would have been detected in the sub-MeV domain by the SPI instrument on \INTEGRAL as a steady source coincident with the CMZ, in disagreement with what has been reported by \citet{bouche08}.

All these issues show that it is extremely difficult to build a self-consistent scenario where LECRs are responsible for large ionisation rates observed within the CMZ \citep{raviku25}.
The only other plausible ionising agent is UV/X-ray radiation, but also this scenario does not seem to be very promising.
UV photons are absorbed very quickly after crossing moderate column densities of gas \citep{mckee89}, and even X-rays coming from past flares of \sgra would not suffice \citep{dogiel13}.
The origin of the large ionisation rates remains a true puzzle.

\subsubsection{Implications for high-energy observations: nuclear gamma-ray lines and 6.4 keV Fe K$\alpha$ line}\label{subsubsec:ioni-6.4}

An enhancement in the LECR intensity in the CMZ has at least two significant implications for high-energy observations.
First, excitation of nuclear levels follows from the interactions of LECR nuclei with ambient matter, and the resulting de-excitation lines may be observed in the MeV domain \citep{ramaty79}.
Second, the CR ionisation of iron atoms in a cold gas is accompanied by emission at 6.4 keV \citep[e.g.][and references therein]{tatisc12}.
The 6.4 keV line emission from LECRs would be stationary, and therefore distinguishable from the X-ray reflection component discussed in \Sec\ref{subsubsec:DXNT-prop}.

\citet{liuyan24} computed the expected intensity of nuclear line emission under the assumption that an ionisation rate of $2 \times 10^{-14}$~s$^{-1}$ (the largest reported value, \citealt{oka19}) is produced by LECR nuclei in the CMZ.
They found that the line intensities are within the capabilities of future proposed MeV missions (see \Sec\ref{subsec:HE-Recent}).
However, these estimates are based on the most optimistic value for the ionisation rate in the region.
A more conservative choice of this parameter (but still consistent with other values reported in the literature, for example those based on {\it Herschel} data which give $2 \times 10^{-15} {\rm~s}^{-1} \lesssim \zeta_{CR} \lesssim 2 \times 10^{-14} {\rm~s}^{-1}$) would make the perspective of a detection of nuclear lines much less appealing.

Similar conclusions can be drawn for the intensity of the 6.4 keV Fe K$\alpha$ line resulting from LECRs in the region.
Several estimates from the line intensity can be found in the literature, for both LECR electrons and nuclei \citep{dogiel09,dogiel15,yusefz07,yusefz13,raviku25}.
The most recent estimates by \citet{raviku25} indicate that the LECR-related line intensity is much smaller (few orders of magnitude) than the currently observed one, which is due to X-ray reflection \citep{terrie18}.
The value reported by \citet{raviku25} is a spatially averaged one over the entire CMZ, and therefore more optimistic conclusions can be certainly drawn for dense concentrations of mass such as the Sgr~B2 cloud \citep[e.g.][]{dogiel15}.
Thus, while the detection in the near future of the stationary signal of the Fe K$\alpha$ line associated with LECRs seem to be unlikely for the entire CMZ, pointed observations of most massive clouds might provide interesting constraints.


\section{Gamma-ray emission and cosmic rays in CMZ and beyond} \label{sec:GRCR}

Gamma-ray observations from the HE to VHE domains reveal a bright source positionaly coincident with \sgra. We review its measured characteristics and discuss some of the interpretations that have been proposed in \Sec\ref{subsec:GRCS}. In addition to the GC source, several gamma-ray sources are detected in the CMZ, in particular PWNe, which we discuss in \Sec\ref{subsec:GRDS}.  Diffuse emission coincident with massive clouds in the inner 200 pc has also been detected. It reveals an excess of CRs at the GC, likely injected by an accelerator located in the inner few pc (see \ref{subsec:GRDE}). On larger sizes, excess gamma-ray emission from the Galactic bulge is visible at the GeV and in soft gamma-rays at 511 keV. We briefly review  these emissions and discuss their possible interpretations in \Sec\ref{subsec:GRDM}.

\subsection{The central gamma-ray source} \label{subsec:GRCS}

\subsubsection{Observational results}

The GC is known to harbor a gamma-ray source since EGRET detected a bright emission at a position compatible with that of \sgra \citep{mayerh98}. The source, named 3EG~J1746--2851 \citep{hartma99}, was found to be located within 0.2\deg\ of the GC. Because of the poor angular resolution of EGRET and the large confusion in the region, the exact localization and possible confusion has lead to questioning its association with the GC \citep{pohl05}. 

\begin{figure}[ht]
	\centering
    \includegraphics[width=\columnwidth]{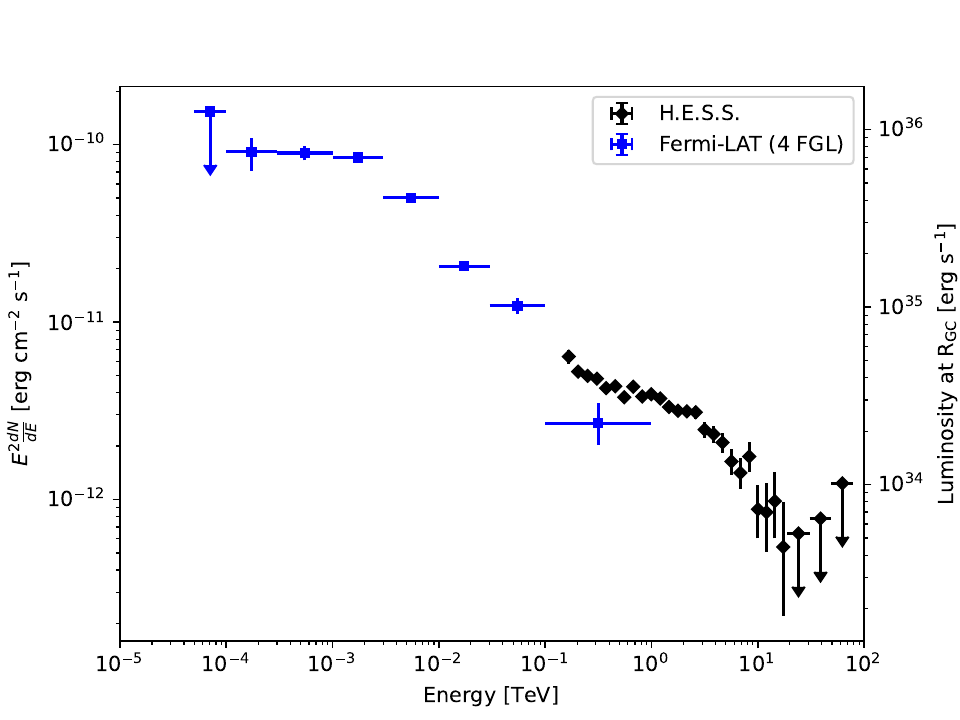}
	\caption{Broadband SED 
	of the gamma-ray source at the GC as reported by the 4th \Fermi catalog 4FGL~J1745.6--2859 \citep{abdoll20}, and by \citet{hessco16} for \hessGC. The equivalent luminosity at the distance of 8.3 kpc is indicated on the right side axis. 
	}
	\label{fig:hess-fermi-GC-source-sed} 
\end{figure}

The Galactic center has been a major target of interest for gamma-ray instruments in the Very High Energy 
domain for years before it was finally detected in 2004 by several instruments, namely Whipple \citep{kosack04},  Cangaroo-II \citep{tsuchi04} and \hess \citep{aharon04b}. All reported the observations of a relatively bright and steady point-like source coincident with the SMBH position, which we name \hessGC in the following. Its luminosity, assuming it is located at the GC, is found to be of the order of $5 \times 10^{34}$ \ergs \citep{aharon04b, albert06, archer16}. First results showed a rather hard power-law spectrum of index 2.2--2.3 without any significant curvature \citep{aharon04b, albert06}. With deeper observing time, a  curvature clearly appeared in the measured spectrum of \hessGC. \citet{aharon09} found it to be well reproduced by an exponential cutoff power-law spectrum and report an index of $\sim 2.10$ and a cutoff energy of $\sim 15$ TeV (see \Fig\ref{fig:hess-fermi-GC-source-sed}). 
Later measurements confirmed this result and found indices $\sim 2.1$ and energy cutoff values $\sim 10$ TeV \citep{hessco16, ahnen17, adams21}. It should be noted that the previous 
measurements do not fully correct for the contamination of the GC diffuse gamma-ray emission and of the Galactic foreground emission \citep[see e.g.][]{viamou13, hessco18b}.

The origin of the cutoff is an important element to understand the nature of \hessGC. It can be the result of the maximum acceleration energy of the particles producing the gamma-ray emission or, equivalently, result from a change in the particle transport properties or even from pair creation photo-absorption in the strong ambient photon field. Yet, we note that the absorption by the ambient Galactic radiation field on the line of sight does not predict a pronounced suppression of the gamma-ray source flux in this energy range \citep{moskal06, celli17}. 
If the cutoff is caused by absorption, it must take place in the immediate surrounding of the acceleration and gamma-ray production site. 

\begin{figure}[ht]
	\centering
 \includegraphics[width=\columnwidth,trim={2.5cm 1.5cm 2.5cm 1.5cm}, clip]{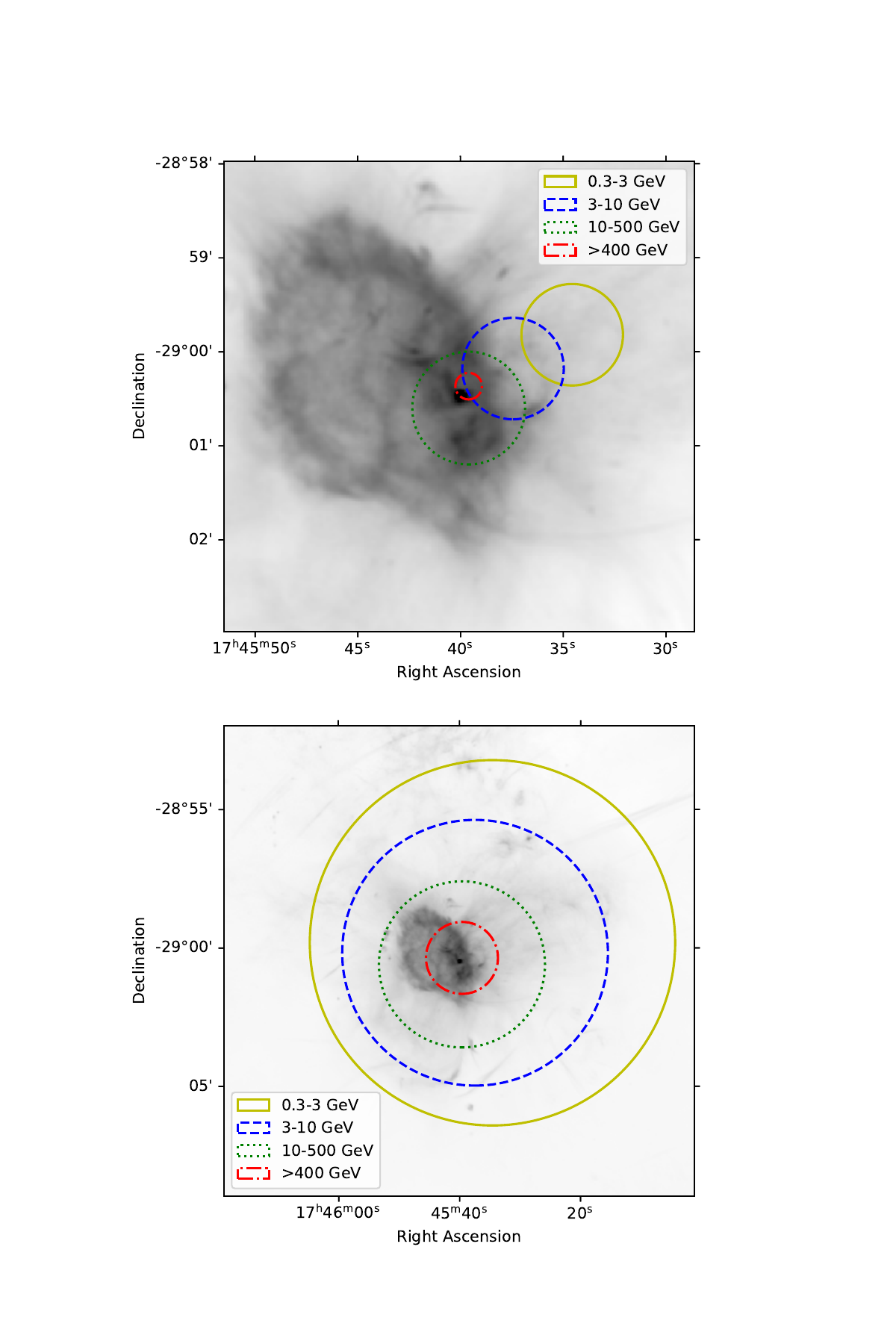}
	\caption{GC gamma-ray source position errors (upper) and extension upper limits (lower panel) at various energies overlaid on the 1.28 GHz brightness of Sgr A East seen with MeerKAT. Measurements in the 0.3--3, 3--10 and 10--500 GeV were obtained with the \Fermi/LAT \citep{cafard21}, while the measurement $>400$ GeV is obtained with \HESS \citep{acero10}.}  
	\label{fig:gamma-gc-position}
\end{figure}

Early detections had a position uncertainty of the order of 1\arcmin. After several years of observations and thanks to a very detailed work on the instrument pointing and the associated systematic errors, the \hess collaboration was able to reduce the position uncertainty down to 20\arcsec \citep[6\arcsec of statistical uncertainty and 13\arcsec of systematic uncertainty,][]{acero10}. The angular separation between \hessGC and \sgra is found to be of only 8\arcsec which is consistent with the SMBH position but also with the PWN \GCPWN (see upper panel of \Fig\ref{fig:gamma-gc-position}).

\citet{acero10} found the \hessGC source to be point-like, with an upper limit on its radius of 1.3\arcmin (95\% C.L.) assuming a gaussian morphology. This size limit is significantly smaller than that of the radio shell of the SNR Sgr A East which shows that the latter can be excluded as the main source of VHE photons (see lower panel of \Fig\ref{fig:gamma-gc-position}). 

To this day, these are the most precise constraint on the position and morphology of the GC gamma-ray source. Improving on this result will require  increased statistics as well as a much better constrained pointing and PSF systematic uncertainties. This should be possible with CTAO which should be able to resolve the ambiguity between the SMBH and PWN positions \citep{cheren19}. An intrinsic difficulty of such a measurement also lies in the underlying diffuse emission which is expected to be structured on small scales in this region due to the distribution of interstellar medium towards \sgra (see next sections).

If \hessGC is connected to the SMBH, its flux might display time variations on various timescales, from the frequent hour-long flares that \sgra experiences to longer timescales accretion rate variations. Variability of the GC gamma-ray source is therefore a crucial observable to interpret its nature. So far, observations of \hessGC have not shown any evidence of variability. Based on observations spanning over three years, light curves obtained by \citet{aharon09} did  not reveal any variability on timescales longer than 28 minutes (the duration of an observing run). A factor of two increase compared to the steady flux 
would be required to detect a flare with timescales of an hour. Similarly, \citet{ahnen17} found no significant variation with MAGIC observations taken from 2013 and 2015 (in particular during the G2 periastron passage). Finally, \citet{adams21} did not find any variability on day or months timescales in the \Veritas data set extending from 2009 to 2017.

To search for a possible flux variation in coincidence with an IR or X-ray flare, VHE observations have been conducted in the context of multi-wavelength campaigns aiming at \sgra. Such campaigns are crucial to test the existence of a signal in VHE gamma-rays associated with the regular flares observed in NIR and X-rays (see Sect. \ref{subsubsec:SgrAs} and \ref{subsec:SGFE}). 
During one such campaign, \Chandra detected a bright flare during a \hess observation. The latter did not reveal any variation of the gamma-ray signal; a factor of 2 variation of the flux could be excluded at 99\% confidence level \citep{aharon08b}. 
CTAO will be sensitive to 20\% flux variations of the GC source over 30 minutes \citep{viana19}. Current models of \sgra flares predict that some of the brightest flares might have IC emission levels close to the CTAO sensitivity at low energy \citep[see e.g. ][]{petrop24}.

In the energy range from 100 MeV to tens of GeV, the excellent performances of \fermi/LAT revealed a point source coinciding with the position of \sgra. 
This source was analyzed by \citet{cherny11} and \citet{malysh15} who found it to be brighter than \hessGC with a luminosity reaching $\sim 8\times 10^{35}$~\ergs. Its spectrum was found to be flat in the range of few hundred MeV to few GeV and steeper in the energy range above a few GeV. Yet, the region is complex and subsequent \fermi catalogs have described the GC GeV emission with several sources. \citet{abdoll20} finds the source 4FGL~J1745.6$-$2859 within 0.01\deg\ of the GC. \citet{cafard21} have reanalyzed more than 10 years of \fermi data to precisely characterize the source morphology in several energy bands (the constraints they obtain on the source position and size are shown in  \Fig\ref{fig:gamma-gc-position}).

It is unclear whether the GeV and TeV sources are caused by the same phenomenon. The measured position of 4FGL~J1745.6--2859 is only marginally consistent with that of \hessGC below 3 GeV and is in good agreement above it \citep{cafard21}. The spectrum of 4FGL J1745.6-2859 is also steep in the range of 5 GeV to 300 GeV while the VHE spectrum is flat, suggesting that another component is appearing in this energy range.

So far, no counterpart to the GC gamma-ray source has been detected by neutrino telescopes. Upper limits on the neutrino flux obtained by IceCube \citep{abbasi25} or Antares \citep{adrian14} are still larger than the measured gamma-ray flux. Improved limits would provide important constraints on the hadronic or leptonic nature of the gamma radiation.

\subsubsection{Interpretation}

High-energy gamma-ray emission is usually produced through the interaction of energetic particles with their environment. Two types of scenarios are usually invoked: leptonic ones involving energetic electrons and positrons, and hadronic ones involving ions. Energetic electrons and positrons interact with the ambient radiation fields via the inverse Compton process. They can also interact with neutral ISM and produce gamma-rays via bremsstrahlung radiation. On the other hand, energetic ions propagating in 
the ISM will undergo inelastic collisions with nuclei and produce gamma-rays \citep[for a general overview, see e.g.][]{aharon04a}. 

We can define different categories of models. The first one interprets the gamma-ray emission by the interaction of particles accelerated in the SMBH accretion flow in the neighboring environment (\Sec\ref{para:GRCS_interpretation_sgra}). The predictions regarding the spectrum, the variability or the size of the source depend on the distance of the acceleration site and the particle transport in the flow as well as the interaction with the ambient radiation fields or target matter. A second category connects the emission to specific objects and accelerators distinct from the SMBH (\Sec\ref{para:GRCS_interpretation_PWN} and \ref{para:GRCS_interpretation_sgraeast}). Finally, a third class of models interprets the signal in terms of collective effects from populations of objects in the vicinity of the GC (\Sec\ref{para:GRCS_interpretation_collective}).
A number of models, in particular those designed after the \Fermi results, try to account for the  broad band spectrum. The rest is more focused on the VHE source interpretation specifically.

\paragraph{\sgra.}\label{para:GRCS_interpretation_sgra}

Particles can be accelerated in the SMBH accretion flow or in a putative jet. The IR and X-ray flares prove that non-thermal particle acceleration occurs down to the inner few Schwartzschild radii. Because of the very low luminosity of \sgra, the photon field density in the inner accretion flow is not large enough to cause significant photo-absorption of TeV photons. 
The region is expected to be transparent to VHE $\gamma$-rays, particle acceleration and gamma-ray production can therefore occur in the inner accretion flow and be detectable as suggested by \citet{ahaner05}. They  
propose several radiative mechanisms that could account for the measured TeV luminosity.  It is nevertheless likely that particles accelerated in the inner tens of $\rm R_S$ can escape the region and interact further away in the accretion flow or beyond in the nearby ISM. This idea was explored by \citet{liu06}, who proposed a scenario where protons are accelerated via stochastic acceleration in the turbulent flow in the inner 20 $\rm R_S$ (with a power of $10^{37}$ erg~s$^{-1}$) and interact within 3 pc of \sgra. Magnetic reconnection has also been proposed as a mechanism to accelerate particles in the RIAF \citep{rodrig19}. An important consequence of these models is that time variability of the acceleration is washed out because of propagation timescales. Hence no variability of \hessGC is to be expected below at least year timescales \citep{liu06}, consistent with current observational results. \citet{ballan07} has showed that a likely site of hadronic interactions is the dense molecular circum-nuclear disk (see \Sec\ref{subsubsec:SgrAcomplex}). 
If the TeV emission is produced by ions accelerated in the accretion flow of \sgra, it is therefore likely to be as extended as the CND, up to 30$''$ radius \citep{linden12}. 
While current measurements do not reach the required angular resolution, it is expected that CTAO will be able to resolve it \citep{cheren19}.  

Accounting for the bright and steep spectrum of 4FGL~J1745.6--2859 and the fainter but harder one of \hessGC requires energy dependent propagation or temporal injection effects. \citet{cherny11} note that if particles diffuse in the inner GC in a similar manner as CR in the Galaxy, the mean free path of multi-TeV protons will reach pc scales making their propagation ballistic rather than diffusive. An expected consequence is a spectral hardening in the VHE range where the accelerator spectrum is not modified by diffusion, while the GeV emission is brighter and steeper because of particle confinement. Following this idea, \citet{cherny11} obtain a scenario that can account for most of the \fermi/LAT and \hess emissions, assuming that particles steadily injected over 10 kyr account for the GeV spectrum while the VHE one requires more recently injected particles, possibly connected with the century old flares seen through X-ray reflection (see \Sec\ref{subsec:DXNT}). Following this idea, \citet{guo13} propose a model where most of the particles would be injected during a recent acceleration episode linked to a past \sgra flare. 
IC emission of the electron population would account for the \Fermi/LAT emission while hadronic emission account for the VHE signal. A consequence of such a model would be a long-term decrease of the GC source flux which has not been observed so far.  This idea is supported by the results of \citet{malysh15} showing that the spectrum of the source below 200 MeV is not consistent with a pure hadronic contribution.

\paragraph{The PWN candidate \GCPWN.}\label{para:GRCS_interpretation_PWN}

Discovered with \chandra by \citet{wang06a}, \GCPWN is a cometary non-thermal X-ray structure located only 7$''$ away from \sgra. It displays clear spectral cooling from the head to the tail, making it a likely ram-pressure confined PWN. It is detected by \nustar above 40 keV demonstrating the presence of very high energy electrons in the nebula \citep{mori15}.  With an X-ray synchrotron luminosity of $\sim 10^{34} \rm~erg~s^{-1}$, it is a plausible origin of the TeV emission provided inverse Compton emission is the dominant radiative process to account for the higher TeV luminosity. \citet{hinaha07} have modeled its emission with a magnetic field of about 100 $\mu$G and a far-IR radiation energy density of up to 5000 eV cm$^{-3}$. 
A key question for the association of \hessGC and \GCPWN is therefore the intensity of the magnetic field that produces the X-ray emission. The measurement of radio pulses associated with the magnetar SGR J1745--2900 suggests that the ambient magnetic field could be as large as the mG level \citep[][see also the discussion in \Sec\ref{subsubsec:otherpop}]{eatoug13}. 
Such a high field would make synchrotron the dominant radiative process and strongly suppress the expected gamma-ray emission \citep{kistle15}. Irrespective of this aspect, inverse Compton emission from \GCPWN fails short of explaining the high luminosity of the GeV source 4FGL~J1745.6--2859 \citep{cafard21}.

\paragraph{Sgr A East.}\label{para:GRCS_interpretation_sgraeast}

It is a bright radio supernova remnant shell filled with intense X-ray plasma emission located behind \sgra. It has been discussed in detail in \Sec\ref{subsec:SGCE}. It was initially discussed as the counterpart of the EGRET GC gamma-ray source \citep{melia98, pohl05}. Given its age and its interaction with the nearby 50 km~s$^{-1}$ cloud, Sgr A East is reminiscent of GeV gamma-ray bright evolved SNRs in the Galaxy, which could make it a plausible gamma-ray emitter \citep{acero16b}. 
Yet, based on the incompatibility of the measured size and position of \hessGC with Sgr~A East morphology, \citet{acero10} have discarded it as a plausible counterpart of the TeV source (see also \Fig\ref{fig:gamma-gc-position}). Similarly, while the extension upper limit of 4FGL J1745.6--2859 is consistent with the SNR size, \citet{cafard21} have argued that its position is markedly offset from the barycenter of the radio emission hence making the association with the GeV source unlikely as well.  

\paragraph{Collection of sources.}\label{para:GRCS_interpretation_collective}

The central cluster should contain a large number of neutron stars, among which millisecond pulsars (MSP). Relativistic leptons accelerated above the TeV in the winds of these pulsars should produce VHE emission through inverse Compton scattering over the intense radiation field in the region. \citet{bedsob13} estimate that energetic leptons from a population of around 100 such pulsars could explain the VHE GC source flux. Because of the strong losses, the diffusion distance of these leptons is limited to a few pc at most, consistent with the measured constraint on the source extension. Cumulative emission from the magnetospheric emission of the MSPs could account for the measured GeV flux, but their typical spectrum falls off quickly above few GeV. Whether the combination of the two components can account for the spectrum measured in the 10--100 GeV range remains to be tested \citep{cafard21}.

\subsection{GC HE and VHE gamma-ray discrete sources } \label{subsec:GRDS}

In addition to the GC source, there are 4 \Fermi/LAT sources within 1$^\circ$ from \sgra and 26 within 2$^\circ$. While the majority is unidentified, a few are associated with extragalactic objects as well as a couple of foreground pulsars \citep{abdoll20}. Similarly, the \hess catalog contains 4 objects in the region.  

We briefly review below three of these objects, two pulsar wind nebulae (\Sec\ref{subsubsec:G09+01} and \ref{subsubsec:G013-011}) and a bright extended region of GeV--TeV emission: HESS~J1745--303 and 4FGL~J1745.8--3028e (\Sec\ref{subsubsec:GeVTeVsource}).

\subsubsection{G0.9+0.1 and its pulsar wind nebula} \label{subsubsec:G09+01} 

G0.9+0.1 was first identified as a composite radio supernova with 20-cm VLA observations \citep{helbec87}. The diameter of the radio shell is 8$'$ while the one of the compact radio nebula is 2$'$. Non-thermal X-ray emission from the nebula was observed with \BeppoSAX, confirming it is a PWN \citep{meregh98, sidoli00}. \chandra resolved the emission of the G0.9+0.1 PWN and revealed an arc and jet-like feature as well as a plausible pulsar candidate \citep{gaensl01}.  Radio pulsations were detected with the GBT \citep{camilo09}. PSR J1747--2809 has a period of 52 ms, a spin-down age of 5.3 $\rm kyr$ and a spin-down luminosity of $\dot{E} = 4.3 \times 10^{37}\rm~erg~s^{-1}$, one of the largest in the Galaxy. The large dispersion measure observed suggests that the pulsar is at the GC distance or beyond.  This is confirmed by interstellar X-ray absorption measurement ($N_H \sim 10^{23}\rm ~ cm^{-2}$) \citep{sidoli00, porque03a}. The electron density model of \citet{yao17} yields a distance of 8.14~kpc for this pulsar \citep{manche05}. 
Assuming a distance of 8.5~kpc, the PWN radio luminosity is $\sim 1.2 \times 10^{35}\rm~erg\ s^{-1}$ \citep{dubner08} and its 2--10 keV luminosity is $L_X \sim 4.7 \times 10^{34}\rm\ erg\ s^{-1}$ \citep{porque03a}, both values consistent with a young nebula powered by PSR~J1747--2809.

\begin{figure*}
	\centering
    \includegraphics[width=0.8\textwidth]{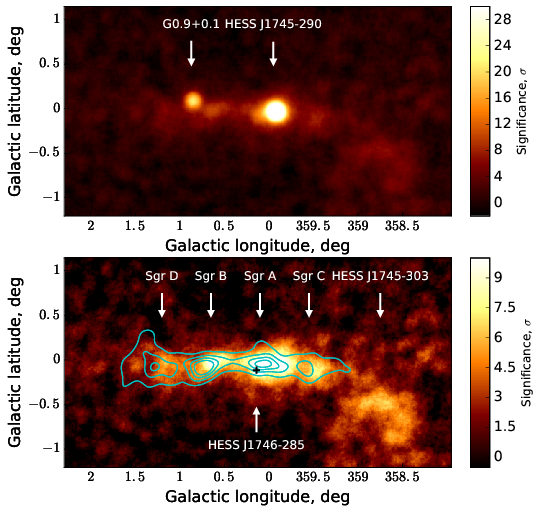}
	\caption{Upper panel: Significance map of the CMZ seen by \hess above 400 GeV. The emission is dominated by two point sources: \hessGC, the source at the GC position, and G0.9+0.1 a pulsar wind nebula. Lower panel: Same map after subtraction of the two point sources. The residual diffuse emission is distributed following the dense gas distribution as traced by the CS molecule (cyan contours from \citealt{tsuboi99}). Figure from \citet{hessco18b}\AG{, reproduced with permission © ESO}.}
	\label{fig:hess-CMZ-image}
\end{figure*}

VHE emission from G0.9+0.1 was detected by \hess (see top panel of \Fig\ref{fig:hess-CMZ-image}). HESS J1747--281 was found to be spatially coincident with the PWN and point-like, excluding the SNR shell as a possible origin of the $\gamma$-rays \citep{aharon05}.  The spectrum was fitted with a power-law of index $2.4 \pm 0.11$. Assuming a distance of 8.5~kpc, the gamma-ray luminosity above 200 GeV is $L_\gamma \sim 2 \times 10^{34}\rm~erg\ s^{-1}$.  These results were confirmed by subsequent observations with \MAGIC \citep{ahnen17, magicc20} and \Veritas \citep{archer16, adams21} who found compatible source parameters, but no coincident HE source was found by the \Fermi/LAT \citep{abdoll20}. The VHE emission is well explained by inverse Compton emission of energetic electrons in the PWN \citep[see][and references therein]{fiori20}.

\subsubsection{Radio Arc gamma-ray source}\label{subsubsec:G013-011}

There is evidence of the existence of a gamma-ray source located in the vicinity of the GC Radio Arc based on EGRET data \citep{pohl05}.  The \Fermi/LAT has confirmed the existence of such a source: 4 FGL J1746.4--2852 is located within the arc at $l=0.15^\circ \pm 0.02^\circ$ and $b=-0.12^\circ \pm 0.02^\circ$. Its spectrum is significantly curved and becomes soft 
above a few GeV. Its luminosity is $L(>1\ \rm GeV) \sim 10^{35}\rm\ erg\ s^{-1}$ \citep{abdoll20}.

IACTs have also detected gamma-ray emission in the vicinity of the GC Radio Arc: VER J1746--289 for \Veritas \citep{archer16}, HESS J1746--285 for \hess \citep{hessco18b} and the ``Arc" for \MAGIC \citep{ahnen17}. 
This emission lies on top of a molecular complex, its measured characteristics, both morphological and spectral, are therefore strongly sensitive to the energy range and to the diffuse emission model (see \Sec\ref{subsec:GRDE}) used to extract them (if any) because of the different levels of contamination.
The source spectrum measured by \hess is found to be rather hard, a power-law of index $\sim 2$ and its luminosity is $L(0.35$--$10~{\rm TeV}) \sim 7 \times 10^{33} \rm ~ erg ~ s^{-1}$ \citep{hessco18b}. Its position at $l= 0.14^\circ \pm 0.01^\circ$ and $b=-0.11^\circ \pm 0.01^\circ$  is found to be coincident with the PWN candidate G0.13--0.11 (note that there is an additional positional uncertainty due to the contribution of the underlying diffuse gamma-ray emission and the way it is modeled).

G0.13--0.11, an X-ray thread on the edge of the GC Radio Arc and the 6.4 keV line emitting molecular cloud G0.11--0.11, was first detected with \chandra \citep{yusefz02a}.  
Its emission consists in a filamentary emission  aligned with the Radio Arc and a point like emission, both having very hard non-thermal spectra. G0.13--0.11 was interpreted as the leading edge of a pulsar wind nebula, produced by a pulsar moving in a strong magnetic field environment \citep{wang02b}. Additional \chandra and \nustar observations have confirmed this scenario and revealed emission extending up to the hard X-ray domain \citep{zhang20}. The overall luminosity of G0.13--0.11 is $(12.0 \pm 0.3) \times 10^{33}\rm\ erg\ s^{-1}$ in the 2--79 keV range. 
The point source (CXOGCS J174621.5--285256) has a very hard spectrum of index $\Gamma =  1.4 \pm 0.5$, which makes it a plausible pulsar candidate albeit no pulsations were observed so far. The filament is very elongated with a dimension of $40''\times 2''$ or $1.6\rm\ pc \times 0.07\ pc$ at the GC distance. The X-ray flux decreases with distance from the pulsar candidate consistent with electrons radiating their energy as they propagate away from the pulsar. Yet, the spectral index along the filament is in the range $1.5$--$1.8$ without significant evidence of spectral cooling with distance from the point source \citep{zhang20}.
The dimensions of the filament allows to place constraints on the magnetic field: the length of the filament imposes that the field as to be below 300 $\rm \mu G$ while its thickness implies $B \ge 20\ \rm \mu G$ \citep{wang02b}.

Recently, a high degree of polarization of the 3--6 keV emission of G0.13--0.11  has been measured ($57 \pm 18$\%) with \IXPE.
The polarization angle is approximately perpendicular to the non-thermal radio filaments \citep{churaz24}. This proves that the X-ray emission is synchrotron emission of electrons propagating along magnetic field lines traced by the filaments.

While the GeV luminosity is much larger than the X-ray and VHE gamma-ray luminosities, the latter are comparable. With the intense radiation fields produced by the nearby Quintuplet cluster, the energetic electrons powering the nebula could produce the observed VHE gamma-ray emission via inverse-Compton scattering. Yet the contribution of the nebula to the GeV emission is likely too low to explain its flux \citep{hessco18b}. 

\subsubsection{HESS~J1745--303 and 4FGL~J1745.8--3028e} \label{subsubsec:GeVTeVsource}

The region around (l, b) = ($358.5^\circ$, $-1^\circ$) is a complex $\gamma$-ray emitting region with several confused sources with complex morphologies from $\sim$ 100 MeV to the multi-TeV region.  An unidentified EGRET source, 3EG~J1744--301, was first found in this region  \citep{hartma99}.  A VHE counterpart, HESS~J1745--303,  was discovered in good spatial coincidence with the EGRET source \citep{aharon06a}. \citet{aharon08a} have resolved its complex morphology into several hotspots.  And the source was finally resolved in two distinct objects in the HGPS. The first, HESS~J1745--303, is centered on (l=$358.65^\circ$, b=$-0.56^\circ$) and has a radius of $0.18^\circ$. The second, HESS~J1746--308, is centered on l=$358.44^\circ$, b=$-1.11^\circ$ and has a radius of $0.16^\circ$. Their spectra are steep with power-law indices respectively of $2.56 \pm 0.06$ and $3.27 \pm 0.22$ \citep{hessco18a}. \Fermi/LAT observations of the GeV source, 4FGL~J1745.8--3028e, found its morphology is well modeled by a disk of radius 0.53$^\circ$ centered on  (l=$358.44^\circ$, b=$-1.11^\circ$) which overlaps both VHE sources \citep{abdoll20}. Its spectrum is modeled by a power-law of index $2.20$, consistent with a spectral softening occurring between the GeV and TeV domains. A second \Fermi/LAT source, 2FHL~J1745.1--3035, has been revealed in the catalog of objects detected above 50 GeV \citep{ackerm16}. Being centered on (l$=358.53^\circ$, b=$-0.78^\circ$), it lies on the edges of HESS~J1745--303, but its very hard spectral, with an index of $1.2 \pm 0.4$ suggests an additional distinct object \citep{marche24}.    

The brightest of the VHE component, HESS~J1745--303, is found to partly overlap the large radio shell SNR G359.1--0.5. At the GC distance, its angular size would result in a unrealistic physical size of 60 pc. G359.1--0.5 must therefore lie in front of the GC. Recent observations suggest a distance of $\sim 4$ kpc \citep{eppens20, suzuki20}. The  presence of OH masers indicates the shell interacts with molecular gas. The interaction of CR accelerated at the shock with the dense gas might therefore produce the observed VHE $\gamma$-rays. 
The absence of any clear X-ray counterpart in \xmm \citep{aharon08a}  and \suzaku \citep{bamba09} supports such a scenario. 
The latter paper also revealed an excess of 6.4 keV line emission suggesting the presence of dense gas. Yet, no clear correlation is found between the TeV emission  and CO clouds in the vicinity of the remnant \citep{eppens20, suzuki20}.  Recently, \citet{marche24} have discovered a very-hard compact, yet slightly extended (5$''$ in \chandra images) non-thermal X-ray source spatially coincident with 2FHL~J1745.1--3035. They argue this could be a young PWN system, where the compact nebula would power the hard \Fermi source while the softer and more extended component (HESS~J1745--303 and 4FGL~J1745.8--3028e) could be explained by an older diffuse hadronic component.

\subsection{Diffuse HE and VHE gamma-ray emission from the CMZ} \label{subsec:GRDE}

\subsubsection{An excess of multi-TeV cosmic rays in the CMZ}

In 2006, the H.E.S.S. Collaboration reported on the detection of a diffuse emission of VHE gamma rays extending for about 2$^{\circ}$ along the Galactic disk and roughly centered around the GC \citep{aharon06b}. 
The diffuse emission is quite faint in terms of brightness. Thus, its detection required a relatively long exposure (55 hours of observations) and the subtraction of the two known discrete VHE sources in the region: \hessGC (see \Sec\ref{subsec:GRCS}) and G0.9+0.1 (see \Sec\ref{subsec:GRDS}).
As shown in the bottom panel of \Fig\ref{fig:hess-CMZ-image}, at Galactic longitudes in the range $|l| \lesssim 1^{\circ}$, the morphology of the VHE diffuse emission (taken from \citealt{hessco18b}) follows quite closely the distribution of dense gas in the CMZ, as traced by CS line emission \citep{tsuboi99}.
This spatial correlation points towards a hadronic origin of the VHE emission, i.e., gamma rays are generated in the decay of neutral pions produced by interactions between CR nuclei and ambient matter in the CMZ.
While multi-TeV CR electrons might also produce VHE radiation through inverse Compton scattering \citep{yusefz07} or non-thermal Bremsstrahlung emission \citep{yusefz13}, the very short radiative loss time of such particles (mainly due to synchrotron emission, see e.g.~\citealt{hessco16})  would prevent them to propagate away from their sources and spread over a region as large as the CMZ. 
A leptonic origin for the VHE emission seems then disfavored.

If the emission is indeed hadronic, the intensity and spectrum of CR nuclei in that region can be derived from gamma-ray data and from the knowledge of the gas mass distribution in the CMZ.
The energy density of multi-TeV CRs measured in this way exceeds by a factor of 3 to 9 the one measured in the Solar neighbourhood. 
Moreover, the spectrum of CRs in the CMZ is significantly harder (slope of the differential energy distribution close to 2.3, \citealt{aharon06b}) than that of the local interstellar spectrum (slope close to 2.7, see e.g.\ \citealt{gabici19}).
These observations can be explained by invoking the presence of an additional population of energetic particles emerging in the TeV domain above the sea of CRs that pervades the entire Galactic disk.

\subsubsection{Origin of the diffuse emission: impulsive cosmic-ray source?}

The diffuse gamma-ray emission was initially detected at photon energies exceeding $\sim 400$~GeV \citep{aharon06b}.
Such photons are produced in interactions of CR nuclei of energy larger than $\sim 4$~TeV/nucleon \citep[e.g.][]{aharon04a}, 
and in order to explain the observed diffuse gamma-ray flux such particles must carry an overall energy of $\sim 10^{49}$~erg.
This estimate increases to $\sim 10^{50}$~erg if the CR spectrum is extrapolated down to the GeV domain.
For this reason, the H.E.S.S. Collaboration initially suggested that the additional CR component in the CMZ might be accelerated by a single supernova explosion, provided that about 10\% of the typical supernova explosion energy (which is $\sim 10^{51}$~erg) is converted into CR nuclei.

At the distance of the GC, the angular extent of the diffuse gamma-ray emission corresponds to a physical extent of few hundred parsecs.
The large physical extent suggests that observations are best explained by assuming that CRs were produced in the past at an accelerator located somewhere in the GC region, and that they then escaped the production site and diffused away, filling a volume much larger than the one of the accelerator itself.
The injection of CRs must have happened recently enough, otherwise the accelerated particles would have diffused well beyond the boundaries of the CMZ, and would be nowadays too diluted to produce any appreciable gamma-ray emission.
This result is truly remarkable, as it was the first time that the presence of a CR accelerator in the Galaxy was revealed indirectly through the emission produced by runaway CRs, a strategy suggested long ago by \citet{ahaato96}.

The morphology of the diffuse gamma-ray emission provides tight constraints on where and when the hypothetical supernova explosion might have taken place.
With this respect, in \citet{aharon06b} it was noticed that the spatial correlation between gamma-ray emission and gas distribution degrades at Galactic longitudes $l > 1^{\circ}$, where a peak in the gas distribution at $l \sim 1.3^{\circ}$ does not correspond to an enhancement in the gamma-ray flux (see \Fig\ref{fig:hess-CMZ-image}).
If the supernova exploded $t$ years ago, and was located close to the GC, the CRs that were injected into the CMZ had time to diffuse and spread over a region of size $R_{diff} \sim \sqrt{2 a ~D~ t}$, where $D$ is the spatial diffusion coefficient of multi-TeV CRs, and $a$ accounts for the dimensionality of diffusion ($a = 1$ corresponds to 1D diffusion along a magnetic flux tube, and $a = 3$ to isotropic 3D diffusion).
This expression can be inverted to derive the supernova explosion time, which is $t \sim 3 \times 10^3 (a ~ \eta)^{-1}$~yr, where it was assumed that CRs are spread over a region of apparent size of $1^{\circ}$ around the GC, and that $D = 10^{30} \eta$~cm$^2$/s, where the numerical factor represents the characteristic diffusion coefficient in the Galactic disk for multi-TeV CRs \citep[e.g.][]{gabici19}, and $\eta$ accounts for possible deviations from this value.

A natural candidate CR accelerator in the region is the supernova remnant Sgr A East, located within the inner few parsecs around the GC (see \Sec\ref{subsubsec:SgrAEast} and \citealt{maeda02}).
The age of the remnant was initially estimated to be of the order of $\approx 10^4$~yr \citep{maeda02,ehlero22} which would require a suppressed diffusion coefficient ($\eta \ll 1$).
This might be indeed plausible, given the highly turbulent environment in the GC region.
However, according to other studies the estimated age should be reduced to $\approx 10^3$~yr \citep{rockef05,fryer06,zhang23}, which would be consistent with an ordinary value ($\eta \sim 1$) of the diffusion coefficient.
As we will see below, the hypothesis of a single impulsive source originating the CRs responsible for the VHE diffuse emission is now disfavored, and therefore we do not comment further on the discrepancies in the estimated age of Sgr A East.

\subsubsection{The morphology of the emission is best explained by a continuous cosmic-ray source at the GC}

Following \citet{aharon06b}, further observations of the CMZ in VHE gamma rays were carried out by the Cherenkov telescope arrays  H.E.S.S. \citep{hessco16,hessco18b}, MAGIC \citep{magicc20}, and VERITAS \citep{archer16}.
In particular, the most recent results obtained by H.E.S.S. and presented in \citet{hessco16} and \citet{hessco18b} were based on much deeper exposures obtained after 226 and 259 hours of observations, respectively.
Therefore, a more accurate reconstruction of the morphology of the gamma-ray emission became possible.

Based on such improved study of the morphology of the emission, an alternative interpretation of the deficit of VHE emission observed at the peak in mass distribution located at $l \sim 1.3^{\circ}$ was presented in \citet{hessco16}.
In that work, the spatial distribution of multi-TeV CRs was inferred from the estimate of the gas mass integrated along the line of sight and from the VHE emission from a number of regions distributed along the CMZ.
The excess in the energy density of multi-TeV CRs found in \citet{aharon06b} was confirmed, and the morphology of the gamma-ray emission could be reproduced by a density of CRs declining away from the GC roughly as $\sim 1/R$, $R$ being the distance from the GC
(the best $1/r^{\alpha}$ fit to data is obtained for $\alpha = 1.1$, \citealt{hessco16}).
Within this framework, the deficit of VHE emission from the region located at $l \sim 1.3^{\circ}$ is not due to the inability of CRs to reach such a distance after escaping from a source located in the GC, but rather to a gradual decline of their energy density.

A $\sim 1/R$ decline is suggestive of a quasi-continuous injection of CRs over an extended period of time taking place very close to the GC.
If multi-TeV CRs are injected at a constant rate $\dot{Q}_{CR}$ at the GC ($R = 0$) and then undergo isotropic and homogeneous spatial diffusion ($a = 3$), their number density as a function of the radial coordinate and at a time $t$ is indeed $n \sim (\dot{Q}_{CR}/4 \pi D R) \propto 1/R$ for $R < R_{diff}$.
The continuous injection of CRs has to last a time $\Delta t$ larger than the CR diffusion time, up to at least $l \sim 1.3^{\circ}$, which implies $\Delta t > 2 \times 10^3 \eta^{-1}$~yr.
The rate at which CR nuclei have to be injected in order to explain the VHE flux is $\dot{Q}_{CR}(>10~{\rm TeV}) \sim 4 \times 10^{37} \eta$~erg~s$^{-1}$.
Interestingly, within this framework a strict upper limit on the value of both the diffusion coefficient and the injection rate can be obtained by recalling that the diffusive approximation breaks down when the diffusion time $\tau_{\rm diff} \sim R^2/6 D$ becomes comparable or smaller than the crossing time $\tau_{\rm cross} \sim R/c$, where $R$ is the size of the region under examination. 
Imposing $\tau_{\rm diff} > \tau_{\rm cross}$ gives $\eta \lesssim 3$ which implies $\dot{Q}_{CR}(>10~{\rm TeV}) \lesssim 3 \times 10^{38}$~erg/s \citep{hessco16}.

The most detailed analysis of the diffuse VHE emission from the CMZ was presented in the most recent publication by the \citet{hessco18b}, from which \Fig\ref{fig:hess-CMZ-image} is taken.
Thanks to the improved statistics, individual molecular cloud complexes within the CMZ are now clearly resolved (Sgr B2, C and D, as indicated by the labels in the figure).

A template-based likelihood fit analysis of the data confirmed that most of the diffuse emission correlates with the distribution of very dense gas as traced by the CS emission, and that a strong (negative) gradient is present in the spatial distribution of CRs away from the GC.
This further supports the evidence for a CR accelerator located at the GC.
However, the fit to the data requires the existence of an additional large-scale component, that does not correlate with and is more extended in latitude than the CS gas template.
One possibility is that such component might be generated by the interaction of CRs with molecular hydrogen in a diffuse phase, characterised by a comparatively lower density ($\sim 100~\rm cm^{-3}$), which is not mapped by CS emission but could account for about a third of the entire molecular matter in the region \citep{dahmen98}.
This shows that accurate estimates of the global energy requirement to explain the gamma-ray emission from the CMZ can be obtained only by combining information on the mass distribution coming from different tracers.

\begin{figure}[ht]
	\centering
    \includegraphics[width=0.5 \textwidth]{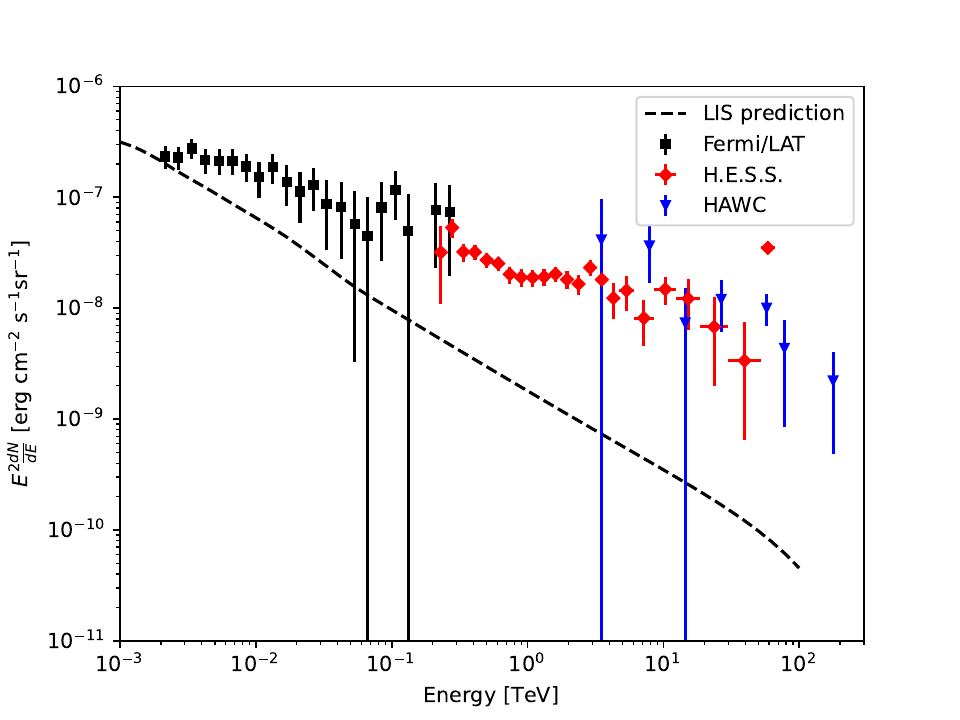}
	\caption{Broadband SED
	of the gamma-ray diffuse emission at the GC extracted in a 0.45$^\circ$ annulus centered around \sgra as reported by \citet{gagger17}, by \citet{hessco16} and by \citet{albert24}. The extrapolated contribution from \hessGC has been subtracted from the HAWC flux points. The dashed line shows the predicted SED assuming the CR density measured in the local interstellar (LIS) medium.
	}
	\label{fig:hess-fermi-sed}
\end{figure}

\subsubsection{A cosmic-ray PeVatron at the GC?}

The measurement of the VHE gamma-ray spectrum of the CMZ attracted much interest as it was suggested that its extension to very large energies could imply the presence of a CR PeVatron, i.e., an accelerator capable of accelerating particles up to the PeV domain, in the GC \citep{hessco16}.
The acceleration of particles up to PeV energies is a central issue in cosmic-ray physics, as explanations of the observed spectrum of CRs require that Galactic accelerators must be able to accelerate protons to at least several PeVs, but the nature of such accelerators is not known \citep[e.g.][]{gabici19}.

The spectrum of the diffuse emission was found to be quite hard, and could be described well by a pure  power law of photon index $\sim 2.3$ \citep{aharon06b}.
Its extension up to photon energies of several tens of TeV suggests that the spectrum of the CR protons responsible for the emission likely extends up to at least a fraction of PeV, and possibly even beyond that. Otherwise, deviations from a pure power-law behaviour should have been observed in the gamma-ray spectrum \cite{hessco16,hessco18b}.

Extending the observations to larger photon energies is crucial in order to better characterise the spectral shape of the gamma-ray emission and to constrain the maximum energy of the parent CR protons.
This was recently done thanks to the data collected over 7 years by HAWC, which extend the observed spectral range to photon energies exceeding 100~TeV \citep{albert24}.
The HAWC Collaboration reported on the detection of a source, named HAWC~J1745--290, coincident with the GC region and characterised by a maximum significance above the background at the 6.5~$\sigma$ level.
Observations can be explained by a point-like gamma-ray source (68\% confidence level upper limit on the size of 0.48$^{\circ}$) exhibiting a steep power-law spectrum of photon index $\sim 2.9$.

A comparison between HAWC and \hess observations of the CMZ is not straightforward, mainly due to the very different angular resolutions of the two instruments, but also to the different energy ranges probed.
Indeed, two \hess sources (\sgra and the Radio Arc, see \Sec\ref{subsec:GRCS} and \ref{subsec:GRDS}, respectively) are found within the upper-limit angular extension of HAWC~J1745-290.
For this reason, the gamma-ray emission from the CMZ was estimated by the HAWC Collaboration by subtracting the flux of these two H.E.S.S. sources (extrapolated up to $\gtrsim 100$~TeV photon energies) from that of HAWC~J1745-290.
When this is done, the \hess and HAWC fluxes are well compatible at $\gtrsim 10$~TeV photon energies (see \Fig\ref{fig:hess-fermi-sed}), but the spectrum estimated by HAWC is markedly steeper than the one measured by \hess within a very similar region around the GC.
As HAWC is more sensitive at higher photon energies, the difference in the measured slopes suggests that the gamma-ray spectrum of the CMZ is not described by a pure power law, but is rather curved.
The presence of a curvature in the spectrum is also supported by a recent preliminary analysis of the most recent \hess data (presented at the 8th Heidelberg International Symposium on High-Energy Gamma-Ray Astronomy).
Even though not conclusive, these latest results put into question the existence of a CR PeVatron in the GC region.
The presence or not of a spectral steepening will also impact dramatically on the expectations of future detections of the GC region with neutrino telescopes such as Baikal-GVD \citep{belola22} or KM3NeT \citep{adrian16}, which are most sensitive in the multi-TeV/PeV energy domain.
Neutrinos are produced in the very same hadronic interactions that produce the observed VHE photons and, in the absence of a spectral suppression, the GC region should be detected after 5 to 10 years of exposure \citep{celli17,guo17}.

\subsubsection{What is the nature of the accelerator?}\label{subsubsec:gamma-acc}

In any case, whether or not the maximum energy of CRs in the CMZ reaches the nominal value of 1 PeV, gamma-ray data tell us that a powerful CR accelerator is present in that region.
Remarkably, as discussed above, the morphology of the gamma-ray emission requires an injection of CRs located very close (less than $\approx$~10~pc) to the GC \citep{hessco16}.
This disfavors a number of potential accelerators, including the Arches and Quintuplet star clusters (located just outside of the region of interest, see e.g. \citealt{genzel10} and \Sec\ref{subsubsec:SFR}), the radio filaments \cite[see \Sec\ref{subsec:PONT} and][]{yusefz13}, or hypothetical discrete sources distributed throughout the entire CMZ \citep[e.g.][]{wommer08}.
The SNR Sgr A East is placed within the inner 10 pc but, as said above, it would most likely inject CRs in an impulsive event, rather than in a continuous (or quasi-continuous) manner, as required by the brightness profile of gamma rays.
The YNC is also located very close to the GC, and should therefore considered a possible candidate for particle acceleration (see \citealt{gabici24} for a review on CR acceleration in stellar clusters).
The recent detection of UHE gamma rays from the direction of stellar clusters reported by LHAASO (notably the OB association Cygnus OB2, \citealt{lhaaso24}, and the young star forming region W43, \citealt{cao25}) provides support to the idea that the YNC might power the diffuse very-high-energy gamma-ray emission observed from the CMZ.
Intriguingly, another plausible candidate is the central SMBH and its immediate surroundings. 
The explanation of gamma-ray observations would require the conversion of a fraction of the energy associated to accretion/ejection phenomena around the SMBH into relativistic particles.

Observations of the CMZ in the VHE domain do not suffice to constrain the global energetic associated with CRs.
This is because for CR spectra steeper than $E^{-2}$ the bulk of the energy is carried by $\approx$~GeV particles. 
Therefore, HE (GeV photon energies) observations of the CMZ are needed.
An analysis of \Fermi data showed that the spectrum of the gamma-ray emission from the CMZ region can be described reasonably well by a single power law extending from the GeV to the multi-TeV domain \citep[see \Fig\ref{fig:hess-fermi-sed} and][]{gagger17}.

Remarkably, as indicated in \Fig\ref{fig:hess-fermi-sed}, at multi-GeV photon energies, the flux is just a factor of $\lesssim 2$ larger than what one would expect from the interaction between the Galactic CR nuclei that pervade the entire Galaxy (assumed to have an intensity roughly equal to the one measured locally) and the dense ambient gas \citep{schcua24,raviku25}. 
However, CRs with a spectral energy distribution similar to the local (near Earth) one would produce a gamma-ray emission from the CMZ characterised by a spectrum much steeper than the observed one.
This fact prompted \citet{gagger17} to propose that at least part of the gamma-ray emission from the CMZ could be explained if the Galactic CRs have a harder spectrum close to the GC.
Indeed, a progressive hardening of the Galactic CR spectrum towards the GC is  required to explain \fermi observations of the Galactic diffuse emission \citep{acero16a,yang16,pothas18}, but including this effect would not eliminate the need of a CR source located at the GC, but only reduce slightly its power \citep{gagger17}.

Various estimates of the CR power associated to the (unknown) central source exist, and they mostly converge to values of $\sim 10^{38}$--$10^{39}$~erg~s$^{-1}$ \citep{crocke11,hessco16,yoasth14,macias15,dorner24,raviku25}.
This is a few orders of magnitude larger than the current bolometric luminosity of \sgra ($\sim 10^{36}$~erg~s$^{-1}$), indicating that the SMBH environment (if responsible for the acceleration of particles) might be much more effective in producing CRs rather than photons.
On the other hand, the required CR power is much smaller than the mechanical power needed to inflate the giant \fermi and \erosita bubbles \citep{su10,predeh20}, which has been estimated to be $\sim 10^{41}$~erg~s$^{-1}$, and significantly smaller than the mechanical power of $\lesssim 10^{40}$~erg~s$^{-1}$ needed to excavate the X-ray chimneys (\citealp{ponti19} and \Sec\ref{subsec:chimneys}), i.e., the exhaust channels through which matter and energy flows from the GC into the bubbles \citep[see][for a review]{sarkar24}.
This implies that, while the nature of the particle accelerator is still unknown, the energy reservoir for non-thermal particles is certainly not an issue.

In fact, as pointed out by \citet{jouvin17}, it is probably more interesting to compare the required CR power with some sort of minimum available mechanical power in the region.
A few percent of the massive star formation of the entire Galaxy takes place in the GC region (see \Sec\ref{sec:Intro}). 
The supernova explosion rate is correspondingly enhanced and is estimated to be, with large uncertainty, equal to $\approx 4 \times 10^{-4}$~yr$^{-1}$ within few hundred parsec from the GC \citep{crocke11,ponti15b,jouvin17}.
Assuming that each supernova releases $10^{51}$~erg in form of mechanical energy and that, as required by CR observations in the Galaxy, $\sim 10\%$ of such energy is converted into CRs \citep[e.g.][]{gabici19}, one gets a CR power of $\sim 10^{40}$~erg~s$^{-1}$, which actually exceeds the power required to explain the gamma-ray emission from the CMZ \citep{jouvin17,jouvin20}.
Therefore, instead of asking what is the origin of the CR excess in the CMZ, one should rather wonder why is the excess so small.
Remarkably, a fast removal of CRs from the CMZ region, either due to a star-formation induced wind \citep{su10,croaha11} or to AGN-like activity events at the SMBH \citep{su10,guomat12}, would provide a natural link between the GC and the giant bubbles detected by \fermi and \erosita.

\subsection{Gamma-ray emission from the Galactic bulge} \label{subsec:GRDM}

The \Fermi GeV excess and the 511~keV line emission 
are both peaked within the Galactic center region considered in this review 
but their roughly spherically symmetric signals extend much further out, likely covering the entire bulge of our Galaxy. 
Since the origin of these features are not yet understood and may be related with the activity at the GC,
we briefly summarize their characteristics and mention the models evoking a possible connections with the center and between them.

\subsubsection{The \Fermi GeV excess}

After subtracting all known discrete sources and diffuse components from the \Fermi/LAT data, \citet{goohoo09} first revealed an excess of emission at GeV energies in the inner degrees of our Galaxy, the so-called \Fermi GeV excess.  The need for a extra component to fit the \Fermi/LAT data has remained true for all independent analyses performed ever since, leading to a very significant detection of this excess \citep[see][for a review]{murgia20}. However, models of the gamma-ray sky used in these analyses are subject to many uncertainties \citep[see e.g.][]{calore15,ackerm17} and the precise description of the \Fermi GeV excess, in particular of its morphology, is therefore still debated. 

The most robust characteristic is its spectrum. The excess is significantly detected between 1 and 10~GeV with a peak at $\sim1$--$3$~GeV \citep[][]{hoosla13, gormac13, calore15, daylan16, abazaj14, ajello16}.
The presence of the excess above and below the 1--10~GeV energy range is more uncertain \citep[see discussion in][]{murgia20}. The \Fermi GeV excess extends at least 10$^\circ$ away from the dynamical center of the Galaxy and its spectrum is almost uniform across this extent \citep[e.g.][]{calore15,daylan16}. Nevertheless, there is no consensus on the spatial morphology of the \Fermi GeV excess. Several descriptions have been proposed, and the main competing models are either a spherically symmetric morphology \citep[][and references therein]{murgia20,cholis22,mcderm23} or a bulge-like morphology \citep{bartel18,macias18}, they are also pointing to two different origins for the \Fermi GeV excess.

Dark matter annihilation was the first interpretation proposed for the excess \citep{goohoo09}. This scenario has since been largely explored and remains viable. Dark matter particles with a mass $\sim60$~GeV annihilating with a velocity-averaged cross section of $\sim4\times10^{-26}\rm~cm^{3}~s^{-1}$ would produce a diffuse spherically-symmetric signal having the same spectral properties than the excess \citep[][and references therein]{murgia20,dimwin21}. On the other hand, the cumulative emission of unresolved point sources distributed as the old stellar population of the bulge could also explain the \Fermi GeV excess \citep[e.g.][]{calore21,list21,lee16,bartel16}. After \citet{abazaj11} showed that the spectra of several globular clusters match the one of the \Fermi GeV excess, millisecond pulsars, which are gamma-ray emitters known to be present in these clusters and that could also be present in the Galactic bulge, became natural candidates \citep[e.g.][]{gormac13,abakap12,brakoc15,gautam22}. Despite on-going extensive searches for bulge MSPs, this population has not been discovered yet, possibly due to the difficulty of detecting fast pulsation for distant sources with high dispersion measures \citep[see e.g.][]{calore16}. DM and MSPs are currently the two main competing scenarios, but other interpretations involving cosmic-ray outbursts from the central region have also been proposed \citep[see][for more details]{murgia20}. 

\subsubsection{The 511 keV emission line}\label{subsec:511}

The celestial emission from electron-positron annihilation \citep[see][for a review]{prantz11} 
produces a pure gamma-ray line at 511~keV (the electron rest mass energy)
or a line mixed with a lower-energy continuum, when the process goes through formation of positronium. 
It was detected from outside the solar system as early as the 1970s from the general direction of the GC, 
by low-energy resolution balloon-borne instruments \citep[][]{johnso72}
and then studied with high resolution Ge detectors \citep[][]{levent78}.
Many experiments have tried to constrain the properties of the emission
and to derive the sources of positron production, their power and the location of the annihilation. 
They established that the radiation is steady, mainly diffuse and extending over the Galactic bulge.
The most precise results are from the \INTEGRAL SPI 
\citep[][]{knodls05, sieger16, churaz20, sieger23}. 
The emission appears centered at the GC, but shifted by $\approx$1$^\circ$ towards negative longitudes, 
and extends with a Gaussian shape of $\approx$ 8$^\circ$ FWHM width 
(even if two Gaussian components of widths $\approx$ 20$^\circ$ and 6$^\circ$ 
better fit the data) encompassing the GB.
An additional point-like component, centered on \sgra, 
could also be present in the data, but with low significance.
A Galactic disk component, 24$^\circ$ FWHM wide in latitude and without longitudinal asymmetries, was instead clearly, and for the fist time, unambiguously detected.
The 511 keV fluxes are respectively (8.9--10.1), (13.1--20.1) and (0.6--1.2) in units of 10$^{-4}$~ph~cm$^{-2}$ s$^{-1}$
for the bulge, disk, and central source, respectively. 
The flux ratio between the bulge and the disk results in a luminosity ratio of $1.0 \pm 0.1$,
which is a very high ratio compared with what is seen at other wavelengths.
High-resolution spectroscopy of the line  
shows that it is well centered at 511~keV with a width of 2.6~keV 
(but it could be a mixture of a broad line of 5~keV FWHM and a narrow one of 1~keV width) 
and a large positronium fraction, compatible with 1.0. 
This suggests that the annihilation of positrons occurs mostly at low ($\approx$ keV) energies,
and mainly in a warm ISM with gas temperature of 7000--40000~K and ionization state of 2–25$\%$.
These data imply a (model-dependent) estimate of the e$^+$ annihilation rate of 2--$3\times10^{43}$ e$^+$ s$^{-1}$ 
in both bulge and disk components \citep[e.g.][]{sieger23}. 

Positrons are typically produced at mildly or highly-relativistic energies 
and they are expected to propagate through the ISM 
on timescales of 0.1--10 Myr along path lengths of 0.1--10 kpc from their initial sources \citep[][]{sieger23}. 
Therefore the measured morphology is certainly a smeared-out image of the initial source distribution 
due to the propagation effects and the distribution of the gas where they annihilate.
The observed flux of MeV $\gamma$-rays from the inner Galaxy also constrains 
the initial energy of the bulk of the positrons to less than a few MeV,
otherwise the continuum emission from in-flight annihilation, i.e.\ before positron thermalization, 
would exceed the observed flux.
Due to these elements, the large uncertainties of the measurements and the source and background model dependence of the results, 
the origin of the emission is still not understood \citep[][]{prantz11, sieger23},
and the large GB/GD ratio is particularly hard to explain. 
The main sources of e$^+$ considered 
include $\beta^+$-radioactivity from isotopes generated by nucleo-synthesis processes in stars, both in steady and in explosive phases 
(SNs, hypernovae, novae), but also particle interactions or photon pair productions taking place close to compact objects 
such as pulsars, magnetars, XRBs or GNs, 
and even processes involving DM. 

The case for \sgra to be a contributor to this emission has been recently re-discussed by \citet{jean17}
who studied the propagation of emitted positrons from the SMBH into the bulge.
This has been first suggested in the 1980s, but was considered as unlikely given the present very-low level of activity of \sgra \citep{riegle81}.
When the possibility of large past activity of the BH emerged, 
a model for pair production in the hot RIAF  
plasma of \sgra, assumed to accrete at larger rate in the past, 
was proposed by \citet{totani06} 
who claimed that this process could explain the required e$^+$ production rate. 
Alternatively, a large amount of accelerated particles can be produced by the SMBH when accretion increases drastically 
during stellar TDE that are expected on time scales of 10$^{4-5}$ yr, 
depending on the star mass \citep{cheng06, cheng07}. 
In the latter work, the authors found that TDEs of low-mass stars by the SMBH, 
would liberate $\approx 10^{52}$~erg in positrons through p-p interactions and following pion decays, at a rate of one per 10$^5$~yr. 
The repeated events over 10 million years and the following propagation of e$^+$ would allow to fill the bulge with low-energy positrons
that then annihilate in its warm ionized medium, giving the observed 511 keV line emission and associated continuum.
Indeed, considering the different propagation effects in the zones encountered from the GC to the GB, \citet{jean17} showed that an outburst of 10$^{52-53}$ erg of MeV positrons from \sgra, occurring 10$^{5-6}$ years ago, can explain the latest measurements of the 511 keV line emission in the broad bulge.

Models invoking DM to produce the 511 keV emission, as for the SMBH models, allow for the large GB/GD ratio of the signal, 
due to the expected DM concentration towards the GN \citep[see][for reviews and references]{boehm09, prantz11}. 
Viable such models for the GB 511 keV line 
involve either de-excitation of weakly interactive massive particles (1 GeV -- 1 TeV), 
possibly the same than those invoked for the \Fermi GeV excess, 
that have been excited to energies above the ground state of 
$\approx$ few MeV,
or annihilation of light DM particles ($< 10$~MeV). 
In this case, the disk component could be explained by nucleo-synthetic products or other stellar contributions. 
However, detailed predictions are scarce because the properties of the DM particles are unknown.
~\\

In spite of the large amount of data collected on the two GB components of the Galactic gamma-ray emission, 
the GeV excess and the 511 keV line emission,
their origins are still a puzzle and 
the link between them remains unclear.
New generations of instruments may help elucidate this relation in the future (\Sec\ref{sec:Con}).


\begin{figure*}
    \centering
    \includegraphics[width=\textwidth]{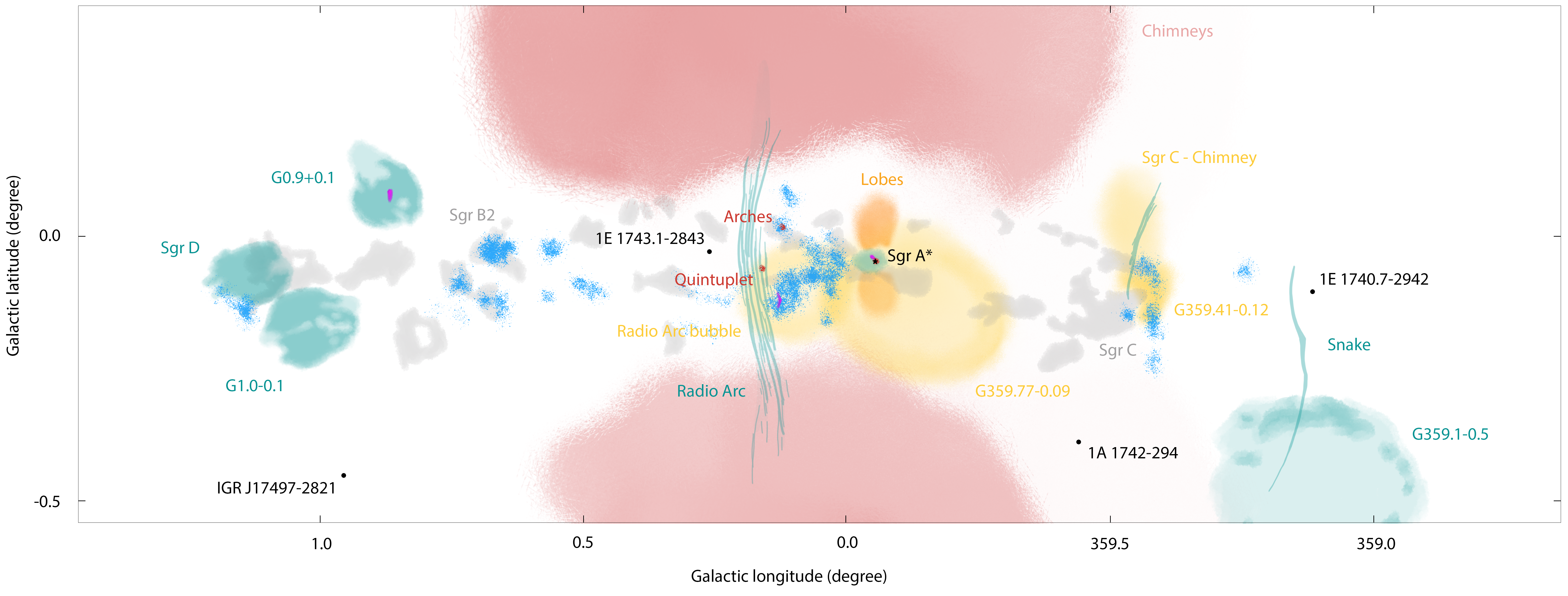}
    \caption{Map of Galactic center features discussed in this review. The location of \sgra and of a selection of X-ray point sources is shown in black (star and points). Diffuse X-ray thermal structures are drawn in yellow, orange and pink (e.g.\ Radio Arc bubble, Lobes and Chimneys, see \Sec\ref{subsec:DXSC}). Reflection features detected at 6.4~keV are shown in patchy blue (see \Sec\ref{subsec:DXNT}). The three young stellar clusters are shown in red (YSC, Arches and Quintuplet clusters, see \Sec\ref{subsec:POTH}). Three PWNe possibly associated with gamma-ray counterparts are highlighted in purple (G359.95--0.04, G013--0.11 and the one within G0.9+0.1, see \Sec\ref{subsec:PONT} and \ref{sec:GRCR}). To help the comparison with maps at other wavelength presented in \Sec\ref{sec:Intro}, radio SNR (emitting in X-rays, e.g.\ Sgr~A East and Sgr~D) plus a selection of radio non-thermal filaments (e.g.\ Radio Arc) are drawn in teal, and a catalog of molecular clouds from the CMZ is shown in light grey (e.g.\ Sgr~B2 and Sgr~C). See \Fig\ref{fig:conclu2} for a close-up of the inner region.
    }
    \label{fig:conclu}
\end{figure*}

\begin{figure}
    \centering
    \includegraphics[width=1\columnwidth]{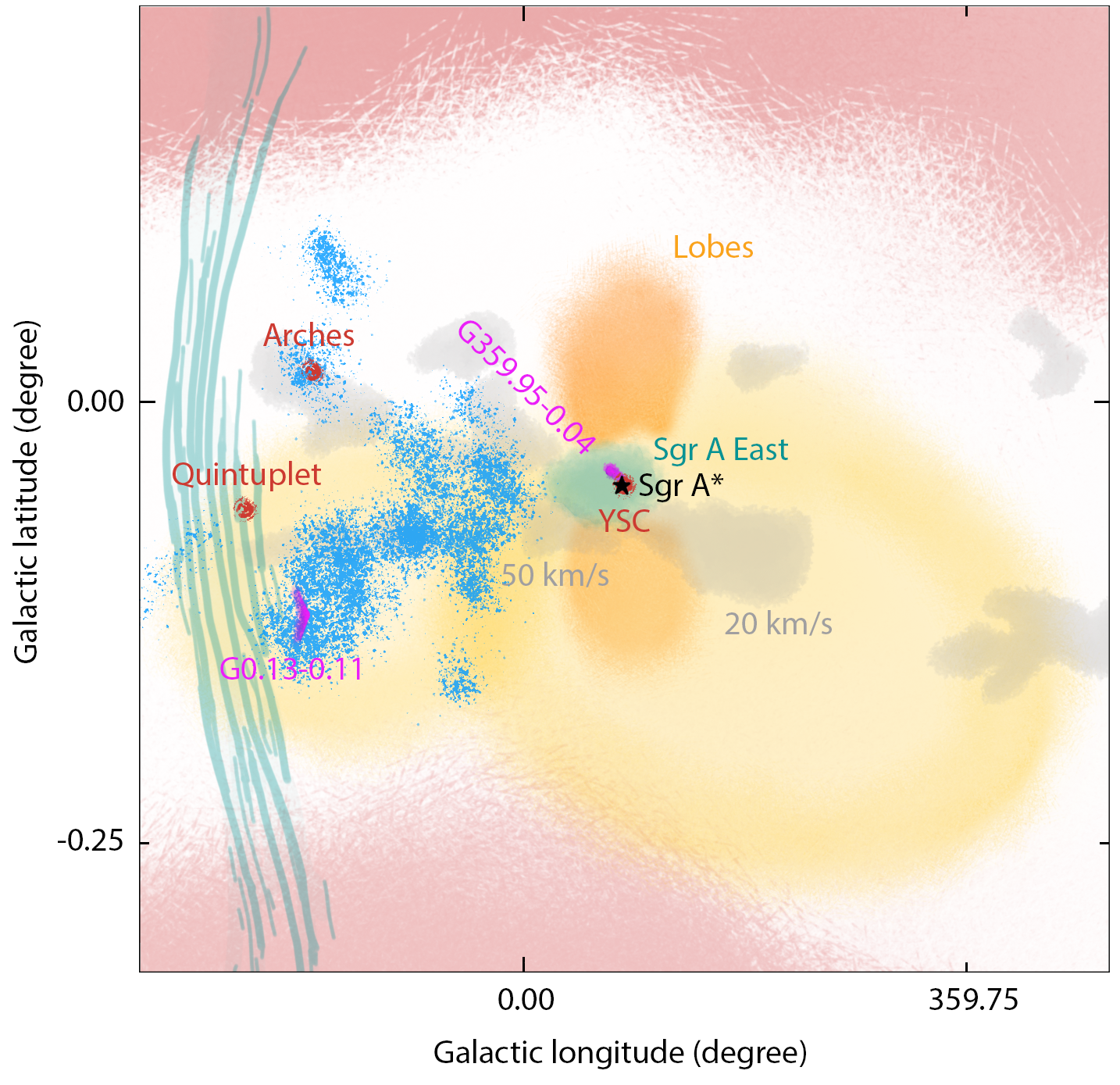}
    \caption{Map of the inner Galactic center features, see caption of \Fig\ref{fig:conclu} for more details.
    }
    \label{fig:conclu2}
\end{figure}

\section{Conclusions and perspectives} \label{sec:Con}
The HE emission from the GC is prominent and very complex. 
It is produced by thermal and non-thermal processes in many different discrete and diffuse sources, 
superposing, and sometimes interacting, with each other in the narrow sky region of the CMZ.
Figures \ref{fig:conclu} and \ref{fig:conclu2} present 
a summary finding chart of the main features discussed in this review
and in particular of the HE sources studied in the last 25 yr with the present generation of X-ray and gamma-ray observatories, with references to the sections where they are described.

\subsection{Summary and open questions}

\sgra is currently characterised by a very weak ($\sim 10^{-11} L_E$), persistent thermal X-ray emission, likely powered by the accretion flow fed by the YNC stellar winds.
This makes it the best prototype of a very low accretion rate BH system,
but the detailed characteristics of the plasma flow and, especially, 
the form of the outflows that seem to dominate it, are not yet understood.
The X-ray emission increases by up to a couple of orders of magnitudes during flares, which happens at a daily rate and last at most few hours.
Their variability timescale implies that they are produced very close ($< 20~R_S$) to the BH horizon, and their spectrum is non-thermal.
Detected also at IR frequencies, their broad-band emission may be due to synchrotron emission of a population of relativistic electrons possibly generated at transient magnetic reconnection events.
Their origin is still not understood. 

Although rather quiet today, the GC was very active in the past.
A direct proof of that came from the discovery of variability of the non-thermal diffuse X-ray emission (iron 6.4 keV line) from the GC MCs.
The emission is dominated by reflection of past X-ray outbursts emitted less than 1 kyr ago most likely by \sgra.
Each outburst had a duration in the range 1-10 yr and released an energy of $\approx 10^{46}$--$10^{48}$~erg.
Also in this case, the origin of these outbursts is not known.
If connected to \sgra, explaining the bursts would require a large (but short) accretion (Low Luminosity AGN - like) of material into the SMBH.

Several signatures of more ancient eruptions and/or outflows originated in the very center of the CMZ have been revealed by the deep X-ray surveys of the region and by their extensions in the polar directions.
They include, from the very center outwards, two symmetrical polar 15-pc lobes, two not axi-symmetric 50-pc superbubbles 
(one towards the Radio Arc the other in the SW direction) and finally two 150-pc polar chimneys. 
The energy involved in these outflows ranges from 10$^{50}$ up to 10$^{52}$~erg for the larger structures.
 Their origin is not fully clear, with both SF events and BH activity as possible explanations.
 
 Other hints for the presence of very energetic events taking place in the GC region, and not necessarily related to the presence of the SMBH, come from the extremely large gas ionisation rates measured within the CMZ, and from the possible presence of a hot ($kT \approx 7$~keV) diffuse plasma in the region.
 The origin of the large ionisation rates remains puzzling.
Ionisation by LECRs seems to be the most plausible scenario, but it requires an exceedingly large energy input in form of CR particles ($\approx 10^{40} - 10^{41}$~erg/s). 
The presence of a hot plasma might explain the residual 6.7~keV line diffuse X-ray emission, which is mostly but probably not entirely due to unresolved mCv and non-mCv.
Maintaining this hot plasma would require an energy input of the order of $10^{53}$~erg to be refilled on a time scale of about $10^4$~yr if not bounded to the GC, and a strong toroidal magnetic field to confine it within the CMZ.
We note that the energy input might be partly spread over the entire GC region

Given all these evidences for the availability of a large energy reservoir, it is somewhat odd that the excess of high-energy CRs in the GC region, deduced from the TeV gamma-ray diffuse emission of the CMZ, is very modest.
This implies that CRs are likely evacuated very rapidly from the CMZ, due to either a wind driven by star formation or AGN activity at the SMBH.
The morphology of the diffuse gamma-ray emission indicates that a powerful accelerator of CRs is present within the inner 10~pc of the Galaxy, injecting continuously energetic particles for at least thousands of years.
The source, whose nature is still debated, was believed to be a CR proton PeVatron, i.e. an extreme Galactic particle accelerator, but recent data cast doubts on this hypothesis.

\subsection{Future observations of the GC region}

New X-ray observables will be essential for progress. In particular, high resolution micro-calorimeter spectroscopy (as provided by \Xrism and \Athena)
will be decisive for characterizing the present and past emission of \sgra. 

Plasma line diagnostics of its quiescent emission will further constrain the physics of the accretion flow on the SMBH, although achieving this will require good angular resolution observations as well to disentangle \sgra from the neighbouring plasma components.
In addition, while X-ray polarimetry appears a very promising tool to constrain properties of \sgra echoes, high resolution spectroscopy of illuminated clouds will further constrain the geometry of the reflections and will enable velocity measurements.
This will provide robust associations with ISM tracers and will be key to determine the number, timing and energetics of past outbursts. 

High resolution spectroscopy will also improve our understanding of diffuse thermal emissions in the CMZ.  Measurements of the diffuse He-like Fe line emission will probe its signatures of rotation around the GC constraining the respective roles of unresolved stellar systems and the hot ISM phase.  More generally, detailed line diagnostics will allow measurements of bulk motions, gas expansion velocities, and turbulences along the LoS, thereby setting constraints on the physical parameters of the soft X-ray diffuse sources. 

Understanding the mechanisms causing \sgra flares and their radiation, will require further coordinated MWC involving new generation instruments from sub-mm to NIR coupled with next generation X-ray telescopes providing large effective area and high angular resolution.

The next generation gamma-ray observatories, such as CTAO, will provide a precise location and extension of the GC gamma-ray source which, in addition to long term monitoring of its emission will be crucial to determine its nature. Emission from \sgra flares might be detectable thanks to improved sensitivity around 100 GeV. Determining the location and nature of the CR accelerator in the CMZ will be made possible by the improved energy resolution and energy coverage of the diffuse gamma-ray emission. 

A new generation MeV mission would also be essential to measure the GC source spectrum below the pion bump to constrain the radiation processes at play as well as characterize the positron annihilation bulge emission properties. It could also observe nuclear line signatures of low energy CR interactions in the CMZ and help solve the mystery of the very large ionisation rate in the region. With this respect, progresses might also come from alternative and more robust ways to estimate the cosmic-ray ionisation rate, based on the study of infrared H$_2$ rovibrational lines from {\it dense} clouds. This is feasible in the local ISM with JWST and might be applicable to the CMZ as well.

\subsection{Link to the Galaxy: SMBH activity or Star Formation?} \label{subsec:Con-LinkG}

One of the key question is of course what level of feedback the GC exerts on
the MW and whether HE observations of the GC can help to answer this question.
To do so, it is necessary to establish the link between the GC features 
and the large Galactic polar structures which are the strongest signs of nuclear outflows.

The \Fermi bubbles (FB)
are gigantic elliptical lobes of HE gamma-ray emission extending from the central regions in both polar directions
for 8--10~kpc along the rotation axis of the MW \citep[][]{su10}. 
The symmetry of their morphology and their sharp edges imply they are powered from the central regions
and the mechanical energy of the process is estimated to $10^{54}$~erg 
over timescales of the order of $10^{6}$~yr or more. 
They are enveloped by the thermal X-ray \eRosita bubbles, that extend for 12--14~kpc,
implying larger energies ($10^{56}$~erg), timescales (10~Myr) and power \citep[$10^{41}$~erg~s$^{-1}$,][]{predeh20}.

The HE GC features most connected to the FEB are certainly the soft X-ray chimneys,
which rise to their bases, and have been suggested to trace the channels through which energy 
is injected from the GC in the FEB outflows. However the estimated power of the chimneys is
at least one order of magnitude lower than the lower estimate of the FB power.
It is possible that repeated events occurred and what is seen is the carved residual of the last event
while the FEB accumulated the energy of several events along a long period.
The smaller structures are difficult to reconnect to the giant polar outflows producing the FEB, 
but the 15-pc lobes could represent part of the same process giving rise to the chimneys \citep{ponti19}.
As discussed in \Sec\ref{subsubsec:gamma-acc}, both the central gamma-ray source and the accelerator responsible for the 
gamma-ray diffuse emission are rather under-luminous compared to the powers needed to fill the FEB
and even to explain the chimneys.
However as the source and processes generating the CR that produce the gamma-rays are not identified yet,
it is difficult to conclude on their relation with the outflows. 
As discussed in \Sec\ref{subsec:GRDE}, the existing estimates of the power needed by the central accelerator to produce
the diffuse gamma-rays in the CMZ are 2--3 orders of magnitude lower than the FB power.
Since the expected CR power in the CMZ should be of the order of the FB power,  
it has been proposed that processes (either SF winds or AGN-like outflows)
rapidly remove CR out of the GC, possibly feeding the FB.
Searches for an observational connection between the GC TeV emission and the bases of the FEB are in progress
with \HESS and certainly foreseen with CTAO. 

Moving on even larger scales, giant ($\approx$~200~kpc) X-ray halos have been recently revealed by \eRosita around Milky Way-like galaxies \citep{zhang24a}, 
and the presence of a gamma-ray halo of similar size has been reported, based on \Fermi observations, for Andromeda \citep{karwin19}.
A similar halo might exist also around the Milky Way, and phenomena related to the GC activity might play a role in explaining its characteristics \citep{recchi21}.

As discussed in \Sec\ref{subsec:GRDM}, 
if the large \sgra activity could release a fraction of the energy to power the production of $10^{52}$~erg in MeV 
positrons, this event could also generate the bulge component of the Galactic 511~keV emission, 
which GC SF processes may have difficulty to explain.  
A prototype of \textit{COSI} has recently flown on a balloon and provided results on the bulge 511 keV emission
that indicate a more extended morphology (width of $\approx$ 30\deg) 
than the one derived from \INTEGRAL data \citep{kieran20}. 
Although these results need to be confirmed, 
improved designs of devices such as \textit{COSI} may provide new important results that can contribute to
resolve the mystery of the Galactic 511 keV emission and clarify its relation with the GC
and the activity of the SMBH.

As a final comment, we remark that if all the central features, their energies and timescales,
are attributed to the SMBH, the history of the \sgra activity in the past can be reconstructed.
From the present $< 10^{-8}$~L$_E$ (quiescent and flaring state), 
it was at $\approx 10^{-5}$~L$_E$ about 300~yr ago (MC echoes),  
liberated $\approx 10^{-3}$~L$_E$ about 10$^{4}$~yr before (15-pc lobes),  
then $\approx 10^{-2}$~L$_E$ about 10$^{5}$~yr ago (chimneys)
and finally $\approx 10^{-1}$~L$_E$ more than 1~Myr ago (FB).
~\\

Clearly more data and more theoretical efforts 
are needed to understand what is going on in the center of the MW and its link with the whole Galaxy,
and HE observations are crucial in these exploratory programs.
The collection of new observatories progressively set in operation over the next 10--15~yr,
working in synergy with those operating at low frequencies and in the multi-messenger domain, 
and in particular with SKA, ALMA, EHT, GRAVITY+, \textit{JWST}, KM3Net and \textit{LISA} between others, 
will hopefully clarify a number of key issues as discussed above.

\newpage 

\appendix 
\section{Abbreviations, acronyms and conventions} \label{subsec:AnnA} 

In this manuscript we use a number of abbreviations and acronyms,
which we list here with their definition.
Observatories are named using their acronyms (known in the community)
and for the HE ones we explicitly give the main references in the text (\Sec\ref{sec:SEAR}).
These contain further references, in particular for the instruments.
Space observatory names are in italics, and instruments are reported after a separation slash: \textit{Observatory}/Instrument.

\noindent
\begin{longtable}{l l}
AB & coronally Active Binary \\
ADAF & Advection Dominated Accretion Flow \\
BH & Black Hole\\
CCD & Charge Coupled Device \\
CDF & Cumulative Distribution Function \\
CMZ & Central Molecular Zone \\ 
CND & Circum-Nuclear Disk or Ring \\
CR & Cosmic Rays\\
(m)CV & (magnetic) Cataclysmic Variable\\
DM & Dark Matter\\
EW & Equivalent Width \\
FB & Fermi Bubbles \\
FEB & Fermi eRosita Bubbles \\
FoV & Field of View \\
FWHM & Full Width Half Maximum \\
GB & Galactic Bulge \\
GC & Galactic Center \\ 
GD & Galactic Disk (intended as thin GD) \\
(A)GN & (Active) Galactic Nucleus \\
GR & General Relativity \\
GRXE & Galactic Ridge X-ray Emission \\
HE & High-Energy \\ 
HED/P/W & Half Energy Diameter / Power / Width\\
(M)HD & (Magneto) Hydro Dynamics \\
(E)IC & (External) Inverse Compton \\
IMBH & Intermediate Mass Black Hole \\
IMF & Initial Mass Function \\
IP & Intermediate Polars \\
ISM & Interstellar Medium \\
(N/M/F)IR  & (Near/Mid/Far) Infrared  \\
LECRe/p & Low Energy CR electrons/protons \\
LMXB & Low Mass X-ray Binary\\
LoS & Line of Sight \\
MAD & Magnetically Arrested Disk \\
(G)MC & (Giant) Molecular Cloud \\
MSP & Millisecond Pulsar\\
MW & Milky Way (the Galaxy) \\
MWC & Multi-Wavelength Campaign \\
NB & Nuclear Bulge \\
NS & Neutron Star\\
NSC & Nuclear Star Cluster\\
NSD & Nuclear Stellar Disc\\
NTF & radio Non-Thermal Filament \\
PSF & Point Spread Function \\
PWN & Pulsar Wind Nebula \\
RIAF & Radiative Inefficient Accretion Flow \\
SANE & Standard and Normal Evolution \\
SED & Spectral Energy Distribution \\
SF(R) & Star Formation (Rate) \\
SMBH & Supermassive Black Hole \\
SMD & Stellar Mass Distribution \\ 
SN(R) & Supernova (Remnant) \\
SSC & Syncrotron Self inverse Compton \\
SYN & Synchrotron \\
TDE & Tidal Disruption Event \\
UHE & Ultra High Energy (gamma-ray) \\
UV & Ultra-Violet \\ 
VHE & Very High Energy (gamma-ray)\\
YNC & Young Nuclear Cluster \\
x1 / x2 & orbit types induced by the bar \\
XRB & X-Ray Binary\\
XVP & \chandra X-ray Visionary Project \\
WD & White Dwarf \\ 
WR & Wolf Rayet \\ 
2D / 3D & two / three Dimension \\
\end{longtable}

\begin{acknowledgements}
The authors acknowledge the Centre National d'\'Etudes Spatiales (CNES) for financial support.
SG acknowledges support from Agence Nationale de la Recherche (grant ANR-21-CE31-0028).
We thank the referees for the careful reading of the manuscript and 
the valuable suggestions that have improved the paper.
This paper employs a list of Chandra data-sets, obtained by the Chandra X-ray Observatory, 
contained in the Chandra Data Collection (CDC) 369~\href{https://doi.org/10.25574/cdc.369}{doi:10.25574/cdc.369}.
\end{acknowledgements}

\bibliographystyle{aa}
\bibliography{heegc.bib}    
 
\end{document}